\documentclass[%
 preprint, 
 amsmath,amssymb,
 aps, physrev,
 showkeys,
]{revtex4-2}

\usepackage{graphicx}
\usepackage{dcolumn}
\usepackage{bm}

\usepackage{amsmath}
\newtheorem{theorem}{Theorem}
\newtheorem{proof}{Proof}[section]
\usepackage{mathrsfs}
\usepackage{float}
\usepackage{multirow}
\usepackage{subfigure}
\usepackage{tabularx}
\usepackage{booktabs}

\begin{document}

\preprint{APS/123-QED}

\title{\textbf{Analytical solutions of layered Poiseuille flows in the diffuse interface model} 
}%

\author{Jun Lai}
\email{Contact author: junlai@pku.edu.cn}
 \affiliation{%
 	School of Naval Architecture and Ocean Engineering, Guangzhou Maritime University, Guangdong 510725, P.R. China
 }%
\author{Yiming Qi}%
\affiliation{%
 State Key Laboratory for Turbulence and Complex Systems, College of engineering, Peking University, Beijing 100871, P.R. China
}%

\author{Lian-Ping Wang}
\email{Contact author: wanglp@sustech.edu.cn}
\affiliation{
 Guangdong Provincial Key Laboratory of Turbulence Research and Applications, Center for Complex Flows and Soft Matter Research and Department of Mechanics and Aerospace Engineering, Southern University of Science and Technology, Shenzhen 518055, P.R. China
}%
\affiliation{
Guangdong-Hong Kong-Macao Joint Laboratory for Data-Driven Fluid Mechanics and
Engineering Applications, Southern University of Science and Technology, Shenzhen
518055, P.R. China
}%


\date{\today}

\begin{abstract}
Based on the two-phase macroscopic governing equations in the phase field model, the governing equations and analytical solutions for the steady-state layered Poiseuille flows in the diffuse interface (DI) model are derived and analyzed. Then, based on three dynamic viscosity models commonly used in the literature, the corresponding analytical solutions of the velocity profile are obtained. Under the condition of high dynamic viscosity ratio, the analytical solution of DI model may be significantly different from that of the sharp interface (SI) model, and the degree of deviation depends on the dynamic viscosity model and the interface thickness. Therefore, the numerical simulation of layered Poiseuille flow with DI model should be compared with the analytical solution of DI model with the same dynamic viscosity model.  A direct comparison of the numerical solution results with the SI analytical solution could
misinterpret the model error with the numerical error. In addition, the direct numerical simulation data and the DI analytical solutions agree well, which validates the theoretical results.
Finally, a new set of symmetrical dynamic viscosity models is proposed and recommended for the simulation of two-phase flows in the DI model, which makes both the viscosity profiles and velocity profiles close to the SI model. 
\end{abstract}

\maketitle


\section{\label{sec:Intro}Introduction}

Since the birth of the Navier-Stokes (NS) equations, the study of analytical solutions has been one of the major efforts. Unfortunately, problems with analytical solutions are rare, and much rarer for multiphase flows, which exist only under very simple physical and boundary conditions~\cite{2016Multiphase}. 
The layered Poiseuille flow is one of such problems.
The flow field of the layered Poiseuille flow is a simple laminar flow in which the streamlines are parallel to each other. 
On the other hand, layered Poiseuille flow is one of the benchmark problems for the two-phase simulations~\cite{2007Lattice,2008Lattice,LECLAIRE20122237,zu2013phase,WANG2015404,2016Multiple,ren2016Improved,2017Improved,PhysRevE.97.033309,zhang2018discrete}.
Therefore, the precise theoretical results of this problem are of great significance both in advancing the theory of fluid dynamics and in examining the accuracy of numerical algorithms.

The phase field (PF) model uses the convection and diffusion of a scalar (order paremeter $\phi$) to describe the evolution of two-phase interface, is one of the commonly used approaches to perform the interface-resolved simulation of two-phase flows.
This model has been applied both in simple flow configurations~\cite{zu2013phase,zhang2018discrete,lai2022simulation,Bachini_Krause_Nitschke_Voigt_2023,Kou_Salama_Wang_2023,Ba_Liu_Li_Yang_2024} and in complex turbulent flows~\cite{scarbolo2013turbulence,komrakova2015numerical,roccon2017viscosity,lai2022systematic}.
Therefore, it is very important to be able to examine the accuracy of numerical simulations of the phase field model.
We point out that the key  is to recognize that the phase field model is a diffusion interface (DI) model, and the analytical solution
has to incorporate this fact. 

At present, the analytical solution of layered Poiseuille flow in the literature is based on the assumption of the sharp interface (SI) model.
When the DI model is applied to simulate this problem, it is always compared with the SI analytical solution~\cite{2007Lattice,2008Lattice,LECLAIRE20122237,zu2013phase,WANG2015404,2016Multiple,ren2016Improved,2017Improved,PhysRevE.97.033309,zhang2018discrete}. 
However, the analytical results under different interface model may be different.
The DI model should be considered when developing a more precise theory under the PF model.

In this paper, we give the analytical solution of steady-state layered Poiseuille flows under the DI model.
For this purpose, we first need to combine the literature and examine the two aspects of this problem.
One aspect is the specific flow field of layered Poiseuille flows, since the layer number may be different.
The other one is the most important physical parameter for the layered Poiseuille flow, {\it i.e.}, the dynamic viscosity $\mu$, which affects the velocity profile significantly.

For the flow field, two kinds of the layered Poiseuille flows inside a channel are mainly simulated in the literatures. One is the double-layer Poiseuille flow~\cite{2007Lattice,LECLAIRE20122237,zu2013phase,2017Improved,PhysRevE.97.033309}, the other one is  the three-layer Poiseuille flow~\cite{2007Lattice,2008Lattice,WANG2015404,2016Multiple}. The solution methods and results are similar for these two physical problems with the DI model. In this paper, we choose the double-layer Poiseuille flow to analyze, for convenience. Our analytical method can be extended to the three-layer Poiseuille flow.

For the description of the dynamic viscosity $\mu$, 
it is typically a function of the order paremeter $\phi$. 
There are two main approaches to calculate $\mu$. We can formulate the kinematic viscosity $\nu$ (or relaxation time $\tau$ in the mesoscopic methods) first~\cite{2010Lattice,PhysRevE.87.023304,WANG201541,2017Improved,PhysRevE.97.033309}, then obtain dynamic viscosity $\mu$ by multiplying density and kinematic viscosity ($\mu=\rho\nu$). Another approach is to calculate the dynamic viscosity directly~\cite{HE1999642,2000Lattice,DING20072078,zu2013phase,chen2019simulation,2019A}, and the kinematic viscosity $\nu$ (or relaxation time $\tau$) is obtained after that. In this paper, we use the second approach.
The analysis of the first approach is similar to that of the second approach, except that the dynamic viscosity models are different.

In this paper, we derive the governing equations and analytical solutions of double-layer Poiseuille flows in the phase field model.
The rest of the paper is organized as follows. In Section~\ref{sec :gover}, the theory of two-phase flow is presented, including the phase field model, the governing equations, and the mixture dynamic viscosity model, which is the key variable in the layered Poiseuille flow.
In Section~\ref{sec: SI}, the equation and analytical solution of layered Poiseuille flow in the sharp interface (SI) model is shown, for better discussion and comparison.
In Section~\ref{sec: DI}, the governing equations and analytical solutions of layered Poiseuille flow with the diffuse interface (DI) model are derived and discussed.
In Section~\ref{sec: Simulations}, our analytical results are compared to the PF numerical solutions.
In Section~\ref{sec: Guidelines}, the choice of the viscosity models and the interfacial thickness are discussed based on the analytical solutions.
In Section~\ref{sec: NewMu}, new viscosity models are proposed and the corresponding velocity profiles are shown.
The summary and conclusions are given in Section~\ref{Conclusion}.
Furthermore, the convective term in the interface equation, the relationship between the continuity equation and divergence-free velocity condition,
some derivation details, special cases,
information at the interface,
properties of the new dynamic viscosity models
are shown and discussed in Appendices~\ref{sec: convective}-\ref{sec: property of the second set},
making the results in this paper more self-contained and rigorous.

	\section{Theory of Two-Phase Flow}\label{sec :gover}

\subsection{Phase field method}\label{subsec :PFM}

The phase field model is a diffuse interface (DI) method, which is widely applied to simulate the fluid-fluid two-phase flow problems~\cite{Anderson1998DIFFUSEaa,zhang2018discrete,chen2019simulation,liang2014phase,church2019high,zhang2019interface}.
In the phase field model, different constant values of the order parameter $\phi$ represent different phases.
For instant, $\phi_A=1$ and $\phi_B=0$ represent phases $A$ and $B$.
The free-energy functional ${\cal F}(\phi,\nabla \phi)$ can be expressed as~\cite{2002Molecular,swift1996lattice,liu2003phase,jacqmin1996energy,yue2004diffuse}
\begin{equation}
{\cal F}(\phi,\nabla \phi)=\int_{V}\left[\psi(\phi)+\frac{\kappa}{2}|\nabla \phi|^{2}\right] d V,\label{Fphi}
\end{equation}
where $V$ is the volume of the considered system. 
$\frac{\kappa}{2}|\nabla \phi|^{2}$, which tends to mix the two phases, is the free-energy density of the interfacial region.
$\psi(\phi)$, which tends to separate the two phases, is the free-energy density of the bulk fluids. It has a double-well form
\begin{equation}
\psi(\phi)=\beta\left(\phi-\phi_{A}\right)^{2}\left(\phi-\phi_{B}\right)^{2}.
\end{equation}
$ \kappa $ and $\beta$ are constant coefficients which can control the surface tension force $\sigma$ and the interfacial thickness parameter $W$ together,
\begin{equation}
\sigma=\frac{\left|\phi_{A}-\phi_{B}\right|^{3}}{6} \sqrt{2 \kappa \beta},\quad
W=\frac{1}{\phi_{A}-\phi_{B}} \sqrt{\frac{8 \kappa}{\beta}}.
\end{equation}
\par 
The chemical potential $\mu_{\phi}$ is the variation of the free-energy functional ${\cal F}(\phi,\nabla \phi)$ with respect to the order parameter,
\begin{equation}
\begin{array}{l}{\mu_{\phi}=\frac{\delta {\cal F}}{\delta \phi}=4 \beta\left(\phi-\phi_{A}\right)\left(\phi-\phi_{B}\right)
	\left( \phi - { {\phi_{A} + \phi_{B} } \over 2} \right)-\kappa \nabla^{2} \phi}\end{array}.
\end{equation}
For a ﬂat surface at equilibrium, $\mu_{\phi}=0$. Then $\phi$ will be
\begin{equation}\label{phi}
\phi(y)=\frac{\phi_{A}+\phi_{B}}{2}+\frac{\phi_{A}-\phi_{B}}{2} \tanh \left(\frac{2 y}{W}\right),
\end{equation}
where $y$ is the signed distance normal to the interface. This distribution is used to initialize $\phi$ in the simulations.

The evolution equation of $\phi$ (interface equation) is the conservative Allen-Cahn (AC) equation~\cite{H2016Comparative,ren2016Improved,2017Improved,PhysRevE.97.033309} or Cahn-Hilliard (CH) equation~\cite{cahn1958free,cahn1959free,DING20072078,WANG201541,WANG2015404,H2016Comparative}, combined with a convection term. The convection term is $\nabla \cdot(\phi \boldsymbol{u})$~\cite{WANG2015404,ren2016Improved,zhang2018discrete,chen2019simulation,liang2014phase,zhang2019interface,H2016Comparative} or $\boldsymbol{u}\cdot\nabla \phi$~\cite{DING20072078,CHIU2011185,ABELS2012THERMODYNAMICALLY,yue2004diffuse,PhysRevE.87.023304,2017Error}, 
where $\boldsymbol{u}$ is the fluid velocity. 
$\boldsymbol{u}\cdot\nabla \phi$ should be the correct physical description, while $\nabla \cdot(\phi \boldsymbol{u})$ is the simplified one under certain assumptions (see Appendix~\ref{sec: convective}).
Other newly proposed evolution equations of $\phi$~\cite{2016A,zhang2019interface,AFlux} are generally modified on the basis of these two equations (AC eq. and CH eq.), and some are essentially linear combinations of them~\cite{2016A}.
In any case, the order parameter describing the evolution of two-phase flow satisfies the convection-diffusion equation.
Therefore, it has the following general form,
\begin{subequations}
	\begin{equation}\label{evolphi1}
	\frac{\partial \phi}{\partial t}+\nabla \cdot(\phi \boldsymbol{u})=J
	\end{equation}      
	or
	\begin{equation}\label{evolphi2}
	\frac{\partial \phi}{\partial t}+\boldsymbol{u}\cdot\nabla \phi=J.
	\end{equation}      
\end{subequations}
$t$ is the evolution time. The flux term should have a conservative form
\begin{equation}
J=\nabla \cdot \boldsymbol{J}_\phi,
\end{equation}
where 
\begin{subequations}\label{Jphi}
	\begin{equation}\label{JphiAC}
	\boldsymbol{J}_\phi=M_{AC}(\nabla \phi-\theta \boldsymbol{n})
	\end{equation}
	and
	\begin{equation}\label{JphiCH}
	\boldsymbol{J}_\phi=M_{CH} \nabla \mu_{\phi}
	\end{equation}
\end{subequations}
for conservative AC equation and CH equation, respectively. The scalar $\theta=\frac{-4\left(\phi-\phi_{A}\right)\left(\phi-\phi_{B}\right)}{W\left(\phi_{A}-\phi_{B}\right)}$ is the function of $\phi$ and interfacial thickness parameter $W$. The vector $\boldsymbol{n}=\frac{\nabla\phi}{|\nabla\phi|}$ is unit vector normal to the two-phase interface. $M_{AC}$ and $M_{CH}$ are the mobilities corresponding to the two equations. For the static flow field at an equilibrium state, $J$ should be zero because the LHS of Eqs.~(\ref{evolphi1}) and~(\ref{evolphi2}) are zero.
In fact, substituting Eq.~\eqref{phi} into Eqs.~\eqref{Jphi}, it is easy to demonstrate that
\begin{equation}\label{Jphi eq}
\boldsymbol{J}_\phi=\boldsymbol{0}.
\end{equation}
Then $J=0$ as expected.

The density is the function of order parameter, {\it i.e.}, $\rho=\rho\left( \phi\right) $. Hence, the continuity equation and divergence-free velocity field in phase-field model are generally incompatible (see Theorem~\ref{contrad} in Appendix~\ref{sec: conti}). For incompressible model, the divergence of velocity is zero in the whole flow field~\cite{DING20072078,H2016Comparative,PhysRevE.97.033309,lai2022simulation},
\begin{equation}\label{eqmass1}
\nabla \cdot \boldsymbol{u}=0,
\end{equation}
then the continuity equation cannot hold near the interface,
\begin{equation}\label{eqmass2}
\frac{\partial \rho}{\partial t}+\nabla \cdot(\rho \boldsymbol{u})=\frac{d\rho}{d\phi} J.
\end{equation}
On the other hand, if the evolution of the fluids satisfy the continuity equation~\cite{2017Improved,lai2022systematic},
\begin{equation}\label{eqmass3}
\frac{\partial \rho}{\partial t}+\nabla \cdot(\rho \boldsymbol{u})=0,
\end{equation}
then the velocity divergence-free condition cannot hold near the interface,
\begin{subequations}\label{eqmass4}
	\begin{equation}\label{eqmass4.1}
	\nabla \cdot \boldsymbol{u}=\frac{d\rho/d\phi}{\phi d\rho/d\phi-\rho}J
	\end{equation}
	or
	\begin{equation}\label{eqmass4.2}
	\quad \nabla \cdot \boldsymbol{u}=\frac{d\rho/d\phi}{-\rho}J 
	\end{equation}
\end{subequations}
corresponding to Eqs.~(\ref{evolphi1}) and~(\ref{evolphi2}), respectively. It is called the quasi-incompressible model~\cite{PhysRevE.93.043303,zhang2018discrete,chen2019simulation}. 
It can be seen that the incompressible model cannot always satisfy the continuity equation near the interface, and the quasi-incompressible model is designed to satisfy the continuity equation in the whole domain.
Usually, the density is selected to be a linear function, $
\rho=\frac{\phi-\phi_{B}}{\phi_{A}-\phi_{B}} \rho_{A}+\frac{\phi_{A}-\phi}{\phi_{A}-\phi_{B}} \rho_{B}
$, where $\rho_{A}$ and $\rho_{B}$ are the densities of the two ﬂuids. Then $\frac{d\rho}{d\phi}=\frac{\rho_{A}-\rho_{B}}{\phi_{A}-\phi_{B}}$, $\frac{d\rho/d\phi}{\phi d\rho/d\phi-\rho}=-\frac{\rho_{A}-\rho_{B}}{\phi_{A}\rho_{B}-\phi_{B}\rho_{A}}$,
which give us the explicit expressions for the RHS of Eqs.~\eqref{eqmass4.1} and~\eqref{eqmass4.2}.

The momentum equation is 
\begin{equation}
\frac{\partial(\rho \boldsymbol{u})}{\partial t}+\nabla \cdot(\rho \boldsymbol{u} \boldsymbol{u})=-\nabla p+\nabla \cdot\left[\mu\left(\nabla \boldsymbol{u}+ \boldsymbol{u}\nabla\right)\right]+ \boldsymbol{F},\label{EqMom}
\end{equation}
where $p$ is the pressure, the dynamic viscosity $\mu=\mu\left( \phi\right) $. The total force is $\boldsymbol{F}=\boldsymbol{F}_s+\boldsymbol{F}_b$, where $\boldsymbol{F}_{s}=-\phi\nabla\mu_{\phi}$~\cite{2005A,WANG201541} or $\mu_{\phi}\nabla\phi$~\cite{DING20072078,2017Improved,PhysRevE.97.033309,2005A} is the interfacial force, $\boldsymbol{F}_{b}$ is other body force such as gravity.

The above three equations (interface equation, mass equation, and momentum equation) constitute the governing equations of two-phase flow.

\par  

\subsection{The mixture dynamic viscosity models}\label{Vis}     

The formulation of dynamic viscosity profoundly affects the velocity profile in the layered Poiseuille flow.
In order to simulate the evolution of a two-phase flow under the diffusive interface representation, it is necessary to model the dynamic viscosity at the interface.


The mixture dynamic viscosity model refers to the relationship between the effective dynamic viscosity and the order parameter.
Typically it is empirically defined.
Three models of mixture dynamic viscosity have been used in DI model to express the smooth transition of the dynamic
viscosity. For convenience, we denote M1~\cite{zu2013phase,chen2019simulation} for the inverse model, M2~\cite{HE1999642,DING20072078,2017Improved,2019A} for the linear model, M3~\cite{2000Lattice} for the logarithmic model, respectively. 
The dynamic viscosity for phase $A$ and phase $B$ alone is denoted by  $\mu_{A}$ and $\mu_{B}$, respectively.
Then, the variation of $\mu$ across the smooth interface is assumed to be a function of  $\phi$, namely,

Form 1
\begin{equation}
\mu=\mu(\phi;\phi_{A},\phi_{B},\mu_{A},\mu_{B}).  \label{Form11}  
\end{equation}
For better understanding of the structure of viscosity models, we can rewrite them as

Form 2
\begin{equation}\label{Form21} 
F\left(\mu\right)=F\left(\mu_{A}\right) \frac{\phi-\phi_{B}}{\phi_{A}-\phi_{B}}+F\left(\mu_{B}\right) \frac{\phi-\phi_{A}}{\phi_{B}-\phi_{A}},    
\end{equation}
where $F(x)$ is an invertible function. For M1, M2, M3, $F(x)=1/x,x,\ln (x/x_0)$, respectively. Form 2 shows a linear relation between $\phi$ and $F(\mu)$.
Different models written in Form 1 and Form 2 are summarized in Table~\ref{Form12}.

\begin{table}[htbp]
	\centering
	\caption{Different expressions of the three dynamic viscosity models}
	\label{Form12}
	\begin{tabular}{ccc}
		\toprule
		\multirow{2}{*}{} & \multicolumn{1}{c}{Form 1} & \multicolumn{1}{c}{Form 2}   \\
		\midrule
		M1
		&$\mu=\frac{\mu_{A} \mu_{B}\left(\phi_{A}-\phi_{B}\right)}{\left(\phi-\phi_{B}\right) \mu_{B}+\left(\phi_{A}-\phi\right) \mu_{A}}$&$\frac{1}{\mu}=\frac{1}{\mu_A} \frac{\phi-\phi_{B}}{\phi_{A}-\phi_{B}}+\frac{1}{\mu_B} \frac{\phi-\phi_{A}}{\phi_{B}-\phi_{A}}$\\
		M2          &$\mu=\mu_{A} \frac{\phi-\phi_{B}}{\phi_{A}-\phi_{B}}+\mu_{B} \frac{\phi-\phi_{A}}{\phi_{B}-\phi_{A}}$&$\mu=\mu_{A} \frac{\phi-\phi_{B}}{\phi_{A}-\phi_{B}}+\mu_{B} \frac{\phi-\phi_{A}}{\phi_{B}-\phi_{A}}$\\
		M3        &$\mu=\mu_{A}^\frac{\phi-\phi_{B}}{\phi_{A}-\phi_{B}}\mu_{B}^ \frac{\phi-\phi_{A}}{\phi_{B}-\phi_{A}}$&$\ln\left( \frac{\mu}{\mu_0}\right) =\ln\left( \frac{\mu_A}{\mu_0}\right)\cdot \frac{\phi-\phi_{B}}{\phi_{A}-\phi_{B}}+\ln\left( \frac{\mu_B}{\mu_0}\right)\cdot \frac{\phi-\phi_{A}}{\phi_{B}-\phi_{A}}$\\
		\bottomrule
	\end{tabular}\\
\end{table}

Now let us take a closer look at the dynamic viscosity.
$\mu$ is a physical variable, but $\phi_{A}$ and $\phi_{B}$ are different order parameters, which can be set to any different constants in
the phase field model.
Therefore, $\mu$ should only depend on a normalized value of $\phi$.
One can introduce $\tilde{\phi} \equiv (\phi - \phi_B)/ (\phi_A - \phi_B)$, then Form 2 is
$F(\mu) = F (\mu_A)  \tilde{\phi} + F (\mu_B) (1- \tilde{\phi} )$ and Form 1 is $\mu = \mu ( \tilde{\phi} ; \mu_A , \mu_B )$.

In fact, for a flat surface at equilibrium state, we can substitude the expression of $\phi$ (Eq.~(\ref{phi})) into the expressions of $\mu$ and obtain a new form of viscosity models, which are not the functions of $\phi_{A}$ or $\phi_{B}$.
Corresponding to Form 1 and Form 2, there are 

Form 3
\begin{equation}\label{Form31}
\mu=\mu(y;W,\mu_{A},\mu_{B})
\end{equation}
and

Form 4
\begin{equation}\label{Form41} 
F\left(\mu\right)=\frac{F\left(\mu_A\right)+F\left(\mu_B\right)}{2}+\frac{F\left(\mu_A\right)-F\left(\mu_B\right)}{2} \operatorname{tanh}\left(\frac{2 y}{W}\right),
\end{equation} 
respectively.
The function $F$ in Form 4 is the same as that in Form 2.
The results for different models are in Table~\ref{Form34}.

\begin{table}[htbp]
	\centering
	\caption{The three dynamic viscosity models for a flat surface at equilibrium state}
	\label{Form34}
	\begin{tabular}{ccc}
		\toprule
		\multirow{2}{*}{} & \multicolumn{1}{c}{Form 3} & \multicolumn{1}{c}{Form 4}   \\
		\midrule
		M1
		&$\mu=\frac{2 \mu_{A} \mu_{B}}{\left(\mu_{A}+\mu_{B}\right)+\left(\mu_{B}-\mu_{A}\right) \operatorname{tanh}\left(\frac{2 y}{W}\right)}$&$\frac{1}{\mu}=\frac{\frac{1}{\mu_{A}}+\frac{1}{\mu_{B}}}{2}+\frac{\frac{1}{\mu_{A}}-\frac{1}{\mu_{B}}}{2} \operatorname{tanh}\left(\frac{2 y}{W}\right)$\\
		M2          &$\mu=\frac{1}{2}\left[\left(\mu_{A}+\mu_{B}\right)+\left(\mu_{A}-\mu_{B}\right) \operatorname{tanh}\left(\frac{2y}{W}\right)\right]$&$\mu=\frac{\mu_{A}+\mu_{B}}{2}+\frac{\mu_{A}-\mu_{B}}{2} \operatorname{tanh}\left(\frac{2 y}{W}\right)$\\
		M3        &$\mu=\sqrt{\mu_A\mu_B}\cdot
		\left(\frac{\mu_{A}}{\mu_{B}}\right)^{\frac{1}{2}\operatorname{tanh}\left(\frac{2y}{W}\right)}$&$\ln\left(\frac{\mu}{\mu_0} \right) =\frac{\ln\left(\frac{\mu_{A}}{\mu_0} \right)+\ln\left(\frac{\mu_{B}}{\mu_0} \right)}{2}+\frac{\ln\left(\frac{\mu_{A}}{\mu_0} \right)-\ln\left(\frac{\mu_{B}}{\mu_0} \right)}{2} \operatorname{tanh}\left(\frac{2 y}{W}\right)$\\
		\bottomrule
	\end{tabular}\\
\end{table}

\begin{figure}[]
	\centering
	\includegraphics[width=0.5\columnwidth,trim={0cm 0cm 0cm 0cm},clip]{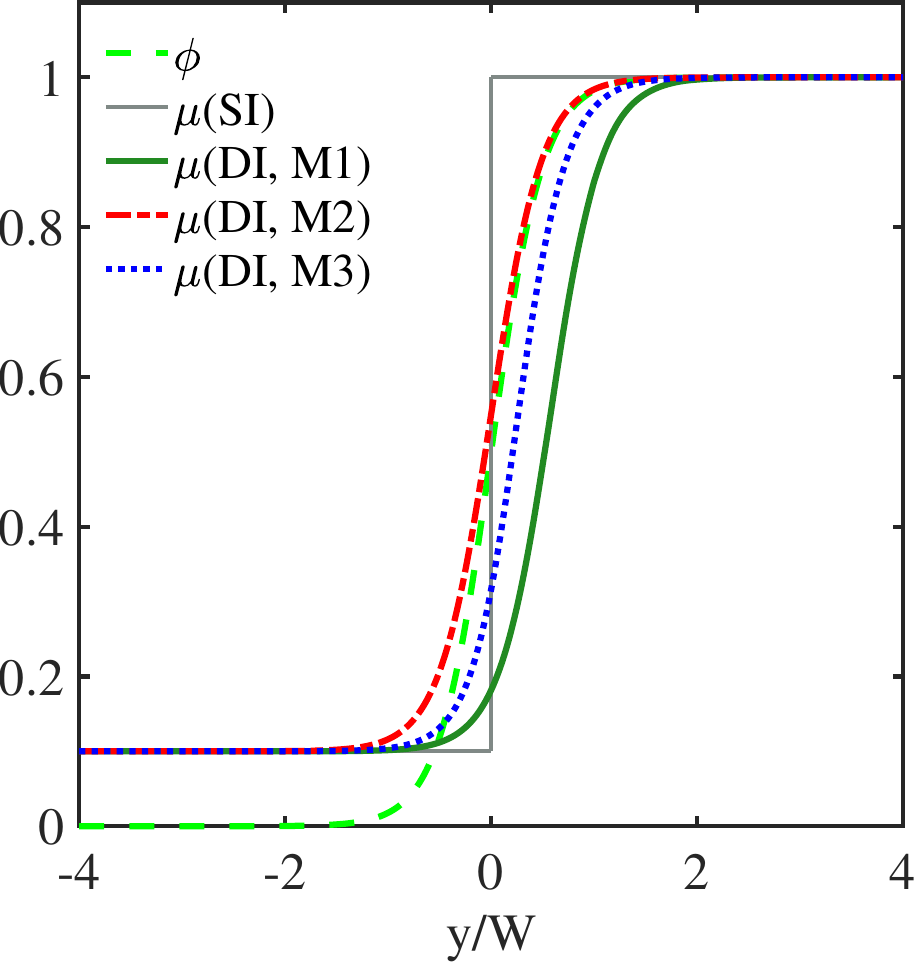}
	\centering
	\caption{The profiles of order parameter, dynamic viscosity in the sharp interface (SI) model, and the three dynamic viscosity models in the diffuse interface (DI) model. The two-phase interface is at $y=0$. Here, the fixed dynamic viscosities for the two phases are $0.1$ and $1$, respectively.
	}
	\label{fig:mu3models}
\end{figure}

Eq.~\eqref{Form41} or Table~\ref{Form34} shows that all the models have the same form as the expression of $\phi$ (Eq.~(\ref{phi})), for a flat surface at equilibrium state. 
In fact, the distribution of $\mu$ are similar as the distribution of $\phi$ for any flow field due to Eq.~\eqref{Form21} or Table~\ref{Form12}.
The profiles of the three models are shown in Fig.~\ref{fig:mu3models}. It can be seen that only M2 is symmetry with respect to the two-phase interface ($\mu(y_0)+\mu(-y_0)=(\mu_{A}+\mu_{B})/2$). M1 is closer to the sharp interface (SI) model in the region of low viscosity, but it differs greatly from the SI model in the region of high viscosity.

For these three viscosity models, we can conclude four properties of the above viscosity $\mu=\mu(\phi;\phi_{A},\phi_{B},\mu_{A},\mu_{B})$:
(1) (Single phase property) $\mu(\phi=\phi_{A})=\mu_{A}$, $\mu(\phi=\phi_{B})=\mu_{B}$, 
$\min\{\mu_{A},\mu_{B}\}<\mu(\min\{\phi_{A},\phi_{B}\}<\phi<\max\{\phi_{A},\phi_{B}\})<\max\{\mu_{A},\mu_{B}\}$;		
(2) (Equal viscosity property) If $\mu_{A}=\mu_{B}=\mu_{0}$, then $\mu\equiv const.\equiv\mu_{0}$; 
(3) (Sharp interface limit property) If $W\rightarrow 0^+$, then $\mu$ is constant on each side of the interface, like the sharp interface model ({\it i.e.}, $\mu\equiv\mu_{A}, y>0; \mu\equiv\mu_{B}, y<0$);
(4) ($\phi$ independent property) If we substitute the initial equilibrium expression $\phi=\phi(y;W,\phi_{A},\phi_{B})$ into $\mu$, then $\phi_{A}$ and $\phi_{B}$ will disappear, and $\mu$ would become $\mu=\mu(y;W,\mu_{A},\mu_{B})$. 

For the specific general expression in Form 2 and Form 4 (Eqs.~(\ref{Form21}) and~(\ref{Form41})), we can also prove these properties because $F$ is invertible.
Therefore, from the point of view of model design, we can design infinitely many hybrid viscosity models given the specific expression of $F(x)$.

\section{Analytical solution of steady-state layered Poiseuille flow in the sharp-interface (SI) model}\label{sec: SI}

For comparison, we first show the well-known analytical result of steady-state sharp-interface (SI) layered Poiseuille flow.

\begin{figure}[]
	\centering
	\includegraphics[width=0.8\columnwidth,trim={0cm 0cm 0cm 0cm},clip]{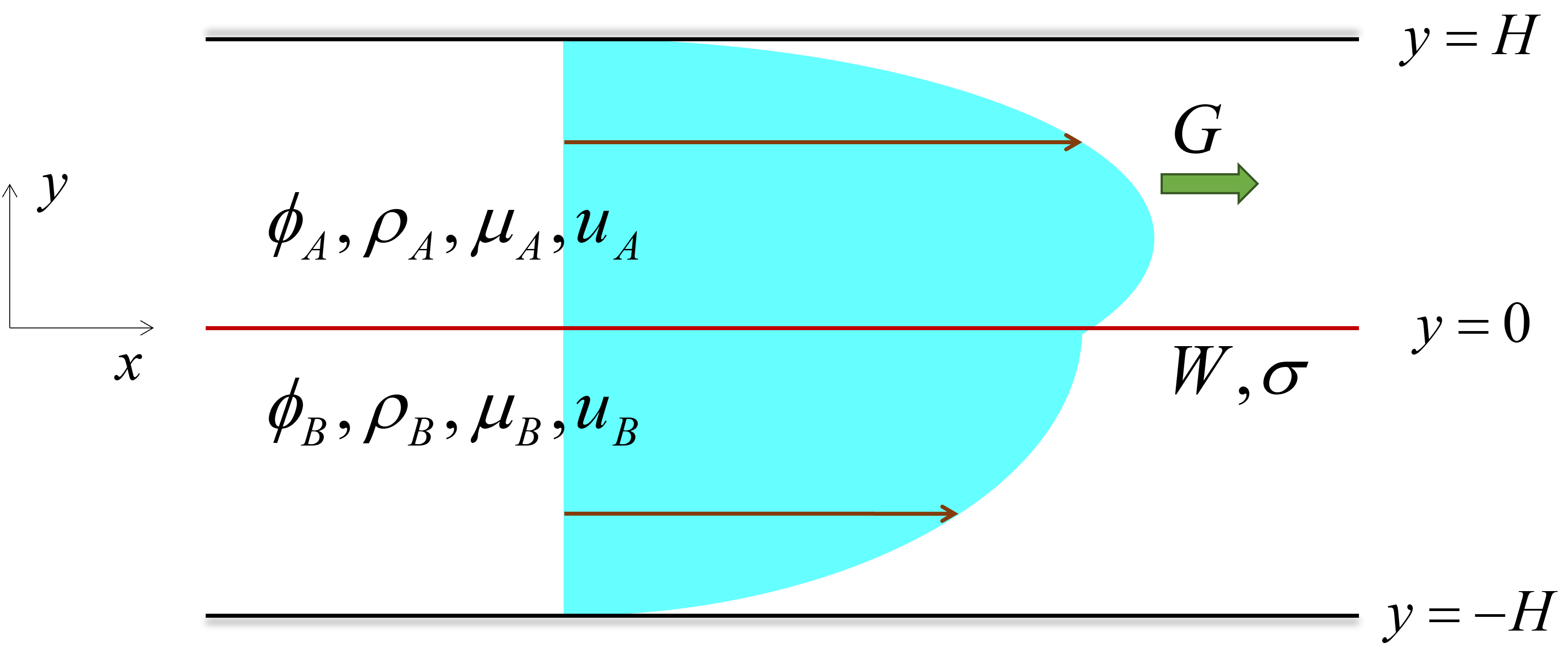}
	\centering
	\caption{Schematic diagram of layered Poiseuille flow.
	}
	\label{fig:layeredsketch}
\end{figure}

As shown in Fig.~\ref{fig:layeredsketch}, the layered Poiseuille flow is two immiscible fluids driven by pressure gradient or other body forces in the flow direction between two infinite plates. The interface of the fluids is in the middle of the plates. 
$A$ and $B$ represent the fluids on the top and on the bottom, respectively. The flow direction is $x$. The height between the two plates is $2H$ in $y$ direction. Then, for SI assumption (the interfacial thickness is $0$, the physical property in any part of the fluid is the same as that in single-phase fluid) and no-slip boundary conditions, the governing equation of steady-state layered Poiseuille flow is

\begin{equation}\label{PoiseuilleSI}
-G=\frac{d}{d y}\left(\mu\frac{d u_x}{d y}\right)=\left\{\begin{aligned}
&{\mu_A\frac{d^2 u_A}{d y^2},\quad 0< y< H,} \\
&{\mu_B\frac{d^2 u_B}{d y^2},\quad -H< y< 0,}
\end{aligned}\right.
\end{equation}
with boundary conditions 
\begin{subequations}\label{SIBC}
	\begin{equation}\label{SIBCu}
	u_x=\left\{\begin{array}{ll}
	{u_A=0,} & {y=H,} \\
	{u_A=u_B,} & {y=0,} \\
	{u_B=0,} & {y=-H,}
	\end{array}\right.
	\end{equation}
	\begin{equation}\label{SIBCtau}
	\tau=\mu_A\frac{d u_A}{d y}=\mu_B\frac{d u_B}{d y},\quad y=0,
	\end{equation}
\end{subequations} 
where $G$ is the magnitude of driving force per unit volume. 
$\mu$ and $u_x$ are the dynamic viscosity and the velocity magnitude in the streamwise direction, respectively.
$\mu_A$, $\mu_B$, $u_A$ and $u_B$ are the dynamic viscosities and the velocity magnitudes of phases $A$ and $B$, respectively.
$\tau$ is the viscous stress.

The solution of Eqs.~\eqref{PoiseuilleSI} and~\eqref{SIBC} is~\cite{2007Lattice,LECLAIRE20122237,zu2013phase,PhysRevE.97.033309,zhang2018discrete}
\begin{equation} \label{uSI}
u_x=\left\{\begin{aligned}
& u_A=-\frac{GH^2}{2\mu_{A}}\left(\frac{y}{H}-1\right)\left(\frac{y}{H}+\frac{2\mu_A}{\mu_A+\mu_B}\right), & {0<y\le H,} \\ 
& u_A=u_B=u_c=\frac{GH^2}{\mu_A+\mu_B}, & {y=0,} \\ 
& u_B=-\frac{GH^2}{2\mu_{B}} \left(\frac{y}{H}+1\right)\left(\frac{y}{H}-\frac{2\mu_B}{\mu_A+\mu_B}\right), & {-H\le y<0,}
\end{aligned}\right.
\end{equation}
where $u_c$ is the velocity at the interface. Define dimensionless parameters
\begin{equation}
u^{*}=\frac{u_{x}}{GH^2/\left(\mu_A+\mu_B\right)}, \quad y^{*}=\frac{y}{H},\quad \mu^{*}=\frac{\mu_{B}}{\mu_{A}},
\end{equation}
then the solution has the following dimensionless form
\begin{equation} \label{uSInonD}
u^{*}=\left\{\begin{aligned}
&\frac{1-y^*}{2}\left[2+\left(\mu^*+1\right)y^*\right], & {0<y^{*}\le 1,} \\ 
& 1, & {y^{*}=0,} \\ &\frac{1+y^*}{2\mu^*}\left[2\mu^*-\left(\mu^*+1\right)y^*\right], & {-1\le y^{*}<0.}\end{aligned}\right.
\end{equation}

\section{Theoretical analysis of steady-state layered Poiseuille flow in the diffuse-interface (DI) model} \label{sec: DI}

In this section, 
we first discuss the general form of equations and solutions, and then substitute three commonly used dynamic viscosity models to obtain specific results.
That is to say, 
our results on the governing equations and analytical solutions are applicable to the general form of dynamic viscosity models, including the models in Section~\ref{sec: NewMu}.

\subsection{Governing equation of layered Poiseuille flow in the DI model}\label{subsec: Governing equation}

The physical definition of layered Poiseuille flow in DI model
is almost the same as that in SI model. The only difference is the assumption of the interface. 
For DI, the macro variables are  assumed to transit smoothly across the interface. Therefore, the interfacial thickness is a positive number, not $0$. 

First, we simplify the macroscopic governing equations in Subsection~\ref{subsec :PFM}.
It is obvious that $\frac{\partial}{\partial t}=0,  u_{y}=u_{z}=0,  \frac{\partial}{\partial x}=0$, which means the LHS of interface Eqs.~(\ref{evolphi1}) and~(\ref{evolphi2}), mass Eqs.~(\ref{eqmass1}),~(\ref{eqmass2}),~(\ref{eqmass3}), and~(\ref{eqmass4}), and momentum Eq.~(\ref{EqMom}) are all $0$. 

This physical problem is at equilibrium state with a ﬂat surface, then the chemical potential $\mu_{\phi}=0$ and the expression of $\phi$ ({\it i.e.}, Eq.~(\ref{phi})) should be satisfied. 
As a result, $J=0$ for AC equation and CH equation due to Eq.~\eqref{Jphi eq}. The RHS of interface equations and mass equations are all zero for AC model and CH model. Therefore, mass equations and interface equations are trivial for layered Poiseuille flow in DI model, which means that the velocity is always divergence-free, the mass is always conserved locally, the distribution of order parameter is always the same as Eq.~(\ref{phi}) in this problem.

We only need to consider the RHS of the momentum Eq.~(\ref{EqMom}). Since $\mu_{\phi}=0$, it is obvious that $\boldsymbol{F}_{s}=\boldsymbol{0}$ for the two expressions of $\boldsymbol{F}_{s}$. That means there is no influence of surface tension, which is consistent with the physics of flat interface. Since $\mu=\mu\left( \phi\right)=\mu\left( \phi\left(y \right) \right) $, $\boldsymbol{u}=u_x\left(y \right)\boldsymbol{e}_x  $, we have $\nabla \cdot\left(\mu  \boldsymbol{u}\nabla\right)=\boldsymbol{0}$. Assume $\boldsymbol{G}$ is the driving force, {\it i.e.}, $\boldsymbol{G}=-\nabla p+\boldsymbol{F}_b=G\boldsymbol{e}_x$. Therefore, Eq.~(\ref{EqMom}) with the boundary condition is
\begin{equation}\label{general vector}
\left\{\begin{array}{ll}
{\nabla \cdot\left(\mu\nabla \boldsymbol{u}\right)+\boldsymbol{G}=\boldsymbol{0},} & {-H< y< H,} \\
{\boldsymbol{u}=\boldsymbol{0},} & {y=\pm H,}\end{array}\right.
\end{equation}
or the scalar form in $x$ direction (the terms in $y$ and $z$ directions are always $0$) is
\begin{equation}\label{general scalar}
\left\{
\begin{aligned}
&\frac{d}{dy}\left(  \mu \frac{du_x}{dy} \right)  + G = 0, & -H< y< H, \\
& u_x=0, & y=\pm H.
\end{aligned}
\right.
\end{equation}
Eq.~(\ref{general vector}) and~(\ref{general scalar}) are the general governing equations of steady-state layered Poiseuille flow in the vector form and the scalar form, respectively. It does not depend on the property of interface (SI or DI), the evolution equation of $\phi$ in DI model (AC equation or CH equation or other modified equations), the property of the velocity field (incompressible or quasi-incompressible), the form of interfacial force ($\mu_{\phi}\nabla\phi$ or $-\phi\nabla\mu_{\phi}$), {\it etc}. Therefore, the governing equations are universal to this physical problem. 
The results show that the fluid density $\rho$ does not affect the flow characteristics of steady-state layered Poiseuille flows in theory. The main physical parameter that affects the flow is the dynamic viscosity $\mu$.

For the dynamic viscosity models satisfying the Equal viscosity property (2) in subsection~\ref{Vis},
if the viscosities of the two phases are the same, then the viscosity is constant in the whole region. 
As a result,
\begin{equation}\label{momsamemu}
-G=\mu \frac{d^2u_x}{dy^2}.
\end{equation}
Therefore, in this case the following three equations will be the same for steady-state layered Poiseuille flow: (1) the governing equation for two-phase flow in DI model; (2) the governing equation for two-phase flow in SI model; (3) the governing equation for single-phase flow.
As a result, the solutions and the properties of these three physical problems are the same. All the results of single-phase flow can be used for same-viscosity two-phase flow directly. In this case, the interfacial thickness parameter $W$ has no influence on the result.

Generally, the viscosities of the two phases are not necessarily the same. Since $\mu=\mu(\phi)$, the momentum equation would be
\begin{equation}
\frac{d^{2} u_{x}}{d y^{2}}+\frac{\theta}{\mu} \frac{d \mu}{d \phi} \frac{d u_{x}}{d y}=-\frac{G}{\mu} .\label{mom1}
\end{equation}
Now the parameters $\phi_{A}$ and $\phi_{B}$ are in the expressions of $\theta$ and $d\mu/d\phi$, meaning the profile of $u_x$ may change with $\phi_{A}$ and $\phi_{B}$, which is not physical. To solve this problem, substituting Eqs.~(\ref{Form21}) and~(\ref{phi}) into  Eq.~(\ref{mom1}), we have
\begin{equation}\label{mom3}
\frac{d^{2} u_{x}}{d y^{2}}+  \frac{1-\tanh \left(\frac{2 y}{W}\right)}{W}\frac{F\left(\mu_{A}\right)-F\left(\mu_{B}\right)}{\mu \frac{d}{d\mu} F(\mu)} \frac{d u_{x}}{d y}=-\frac{G}{\mu} .
\end{equation}
Eq.~(\ref{mom3}) is not the function of order parameters, but the function of some physical parameters and one model parameter $W$.
Therefore, the expansion form of the governing equation for steady-state layered Poiseuille flow in the DI model is
\begin{equation}\label{mom2}
\left\{
\begin{aligned}
&\frac{d^{2} u_{x}}{d y^{2}}+f \frac{d u_{x}}{d y}
=g, & -H< y< H, \\
& u_x=0, & y=\pm H.
\end{aligned}
\right.
\end{equation}
Since $\mu$ has the expression of Form 3, {\it i.e.}, Eq.~(\ref{Form31}), $f$ and $g$ have the following forms,
\begin{subequations}\label{eqfg}
	\begin{equation}
	f=\frac{1-\tanh \left(\frac{2 y}{W}\right)}{W}\frac{F\left(\mu_{A}\right)-F\left(\mu_{B}\right)}{\mu\frac{d}{d\mu} F(\mu)}=f\left(y ; W, \mu_{A}, \mu_{B}\right),
	\end{equation}
	\begin{equation}
	g=-\frac{G}{\mu}=g\left(y ; W, \mu_{A}, \mu_{B},G\right),
	\end{equation}
\end{subequations}
where $F(x)=1/x,x,\ln (x/x_0)$ for M1, M2, M3, respectively.
Specifically, the expressions of $f$ and $g$ for the three models are shown in Table~\ref{Tabfg}.

\begin{table}[htbp]
	\centering
	\caption{Expressions of coefficients $f,g$ in Eq.~\eqref{mom2} under different viscosity models}
	\label{Tabfg}
	\begin{tabular}{ccc}
		\toprule
		\multirow{2}{*}{} & \multicolumn{1}{c}{$f$} & \multicolumn{1}{c}{$g$}   \\
		\midrule
		M1
		&$\frac{2\left(\mu_{A}-\mu_{B}\right)\left[1-\operatorname{tanh}^{2}\left(\frac{2 y}{W}\right)\right]}{W\left[\left(\mu_{B}+\mu_{A}\right)+\left(\mu_{B}-\mu_{A}\right)\operatorname{tanh}\left(\frac{2 y}{W}\right)\right]}$&$\frac{-G}{2 \mu_{A} \mu_{B}}\left[\left(\mu_{B}+\mu_{A}\right)+\left(\mu_{B}-\mu_{A}\right) \operatorname{tanh}\left(\frac{2 y}{W}\right)\right]$\\
		M2          &$\frac{2\left(\mu_{A}-\mu_{B}\right)\left[1-\operatorname{tanh}^{2}\left(\frac{2 y}{W}\right)\right]}{W\left[\left(\mu_{A}+\mu_{B}\right)+\left(\mu_{A}-\mu_{B}\right) \operatorname{tanh}\left(\frac{2 y}{W}\right)\right]}$&$\frac{-2 G}{\left(\mu_{A}+\mu_{B}\right)+\left(\mu_{A}-\mu_{B}\right) \operatorname{tanh}\left(\frac{2 y}{W}\right)}$\\
		M3        &$\frac{1}{W}\left[1-\operatorname{tanh}^{2}\left(\frac{2 y}{W}\right)\right]
		\ln\left(\frac{\mu_{A}}{\mu_{B}}\right)$&$\frac{-G}{\sqrt{\mu_A\mu_B}}
		\left(\frac{\mu_{A}}{\mu_{B}}\right)^{-\frac{1}{2}\operatorname{tanh}\left(\frac{2y}{W}\right)}$\\
		\bottomrule
	\end{tabular}\\
\end{table}

In order to reduce the unknowns and facilitate derivation and discussion, Eq.~\eqref{mom2} and~\eqref{eqfg} shall be made dimensionless.
Define
\begin{equation}
u^{*}=\frac{u_{x}}{GH^2/\left(\mu_A+\mu_B\right)}, \quad y^{*}=\frac{y}{H},\quad W^{*}=\frac{W}{H},\quad \mu^{*}=\frac{\mu_{B}}{\mu_{A}},
\end{equation}
\begin{equation}
f^{*}=\frac{f}{H^{-1}}, \quad g^{*}=\frac{g}{G/\left(\mu_A+\mu_B\right)},
\end{equation}
then the dimensionless form of Eq.~(\ref{mom2}) is
\begin{equation}\label{mom4}
\left\{
\begin{aligned}
&\frac{d^{2} u^{*}}{d y^{* 2}}+f^{*} \frac{d u^{*}}{d y^{*}}=g^{*}, & -1< y^{*}< 1, \\
& u^{*}=0, & y^{*}=\pm 1,
\end{aligned}
\right.
\end{equation}
where
\begin{subequations}
	\begin{equation}
	f^{*}=f^{*}\left(y^{*} ; W^{*}, \mu^{*}\right),
	\end{equation}
	\begin{equation}
	g^{*}=g^{*}\left(y^{*} ; W^{*}, \mu^{*}\right).
	\end{equation}
\end{subequations}
The specific expressions of $f^{*}$ and $g^{*}$ are in Table~\ref{Tabfgstar}.

\begin{table}[htbp]
	\centering
	\caption{Expressions of coefficients $f^{*},g^{*}$ in Eq.~\eqref{mom4} under different viscosity models}
	\label{Tabfgstar}
	\begin{tabular}{ccc}
		\toprule
		\multirow{2}{*}{} & \multicolumn{1}{c}{$f^{*}$} & \multicolumn{1}{c}{$g^{*}$}   \\
		\midrule
		M1
		&$\frac{2\left(1-\mu^{*}\right)\left[1-\operatorname{tanh}^{2}\left(\frac{2 y^{*}}{W^{*}}\right)\right]}{W^{*}\left[\left(\mu^{*}+1\right)+\left(\mu^{*}-1\right) \operatorname{tanh}\left(\frac{2 y^{*}}{W^{*}}\right)\right]}$&$-\frac{\mu^{*}+1}{2 \mu^{*}}\left[\left(\mu^{*}+1\right)+\left(\mu^{*}-1\right) \operatorname{tanh}\left(\frac{2 y^{*}}{W^{*}}\right)\right]$\\
		M2          &$\frac{2\left(1-\mu^{*}\right)\left[1-\operatorname{tanh}^{2}\left(\frac{2 y^{*}}{W^{*}}\right)\right]}{W^{*}\left[\left(1+\mu^{*}\right)+\left(1-\mu^{*}\right) \operatorname{tanh}\left(\frac{2 y^{*}}{W^{*}}\right)\right]}$&$\frac{-2 \left(1+\mu^{*}\right)}{\left(1+\mu^{*}\right)+\left(1-\mu^{*}\right) \operatorname{tanh}\left(\frac{2 y^{*}}{W^{*}}\right)}$\\
		M3        &$\frac{\ln\mu^*}{W^*}\left[\operatorname{tanh}^{2}\left(\frac{2 y^*}{W^*}\right)-1\right]$&$-\left(1+\mu^*\right)\cdot
		\left(\mu^*\right)^{\frac{1}{2}\left[\operatorname{tanh}\left(\frac{2 y^*}{W^*}\right)-1\right]}$\\
		\bottomrule
	\end{tabular}\\
\end{table}

Since $\mu^{*}>0$, $W^{*}\neq 0$, it is easy to prove that $\left(\mu^{*}+1\right)+\left(\mu^{*}-1\right) \operatorname{tanh}\left(\frac{2 y^{*}}{W^{*}}\right)\neq 0, \left(1+\mu^{*}\right)+\left(1-\mu^{*}\right) \operatorname{tanh}\left(\frac{2 y^{*}}{W^{*}}\right)\neq 0$ for any $y^{*}$. So $f^{*}$ and $g^{*}$ in these three models are always meaningful based on Table~\ref{Tabfgstar}.
To further check the suitability of the above expressions, we examine two special cases in Appendix~\ref{sec: App f*g*}.


\subsection{Analytical solutions of layered Poiseuille flow in the DI model}\label{subsec: Analytical solutions}

We have several methods to obtain the analytical solutions of layered Poiseuille flow.

First, integrating the momentum equation in Eq.~(\ref{general scalar}) twice, and substituting the boundary conditions, we have (See Appendix~\ref{sec: App first solution})
\begin{subequations}\label{uPoiseuilleMethod1}
	\begin{equation}\label{integrate1}
	u_x=\frac{G}{\int_{-H}^{H} \frac{1}{\mu} d Y}\cdot\left\{ 
	\int_{-H}^{H} \frac{1}{\mu} d Y\int_{y}^{H} \frac{Y}{\mu} d Y+ \int_{-H}^{H} \frac{Y}{\mu} d Y\int_{H}^{y} \frac{1}{\mu} d Y
	\right\}
	\end{equation}
	or
	\begin{equation}\label{integrate2}
	u_x=G \left\{ 
	\left[I_y(H)-I_y(y) \right]  
	+\frac{  I_y(H) - I_y(-H)}{I(H) - I(-H)}\left[I(y)-I(H) \right] \right\},
	\end{equation}
\end{subequations}
where $I(y) = \int_a^y \frac{dY}{\mu(Y)}$ is a certain primitive function of $\frac{1}{\mu(y)}$, $I_y(y) = \int_a^y \frac{YdY}{\mu(Y)}$ is a certain primitive function of $\frac{y}{\mu(y)}$.
From the discussions in Subsection~\ref{subsec: Governing equation}, the velocity profile of a layered Poiseuille flow mainly depends on the dynamic viscosity.
From the analytic form, Eq.~\eqref{uPoiseuilleMethod1}, we further know that this velocity profile depends only on the two integrals $I(y)$ and $I_y(y)$ related to the dynamic viscosity.
Therefore, when the analytical solutions of the three viscosity models need to be deeper analyzed, the similarities and differences of $I(y)$ and $I_y(y)$ in different models should be compared. Here the results of these two integrals are derived for reference only (see Appendix~\ref{sec: int mu}).

Second, the general solution of Eq.~(\ref{mom2}) is
\begin{equation}\label{uPoiseuilleMethod3}
u=\int e^{-\int f d y}\left(\int g e^{\int f d y} d y+C\right) d y+D,
\end{equation}
where $\int h(y) dy$ is a certain primitive function of $h(y)$.
$C$ and $D$ are constants determined by the boundary conditions. 
Similarly, the general solution of Eq.~(\ref{mom4}) is
\begin{equation}\label{analyticalustar}
u^{*}=\int e^{-\int f^{*} d y^{*}}\left(\int g^{*} e^{\int f^{*} d y^{*}} d y^{*}+C^*\right) d y^{*}+D^*,
\end{equation}
where $\int h^*(y^*) dy^*$ is a certain primitive function of $h^*(y^*)$.
$C^*$ and $D^*$ are constants, which are determined by the boundary conditions. 

By substituting three different viscosity models, the dimensionless analytical solutions of the velocity profiles can be obtained as follows
(details are given in Appendix~\ref{CDI}).


\begin{equation}\label{uM1star}
u^*_{M1}=-\left(\mu^{*}+1\right)\cdot\left\{
\begin{aligned}
&\frac{2y^{*2}-\left(\mu^*-1\right)y^*-\left(\mu^*+1\right)}{4\mu^*}-\frac{\left(\mu^*-1\right)W^*}{4\mu^*}I_1\left(y^*;W^{*}\right)
\\&
+\frac{(\mu^*-1)W^*y^*}{8\mu^*}
\cdot
\left\{
\begin{array}{l}
2\ln \left(1+e^{\frac{4y^*}{W^*}}\right)
\\-\ln \left(2+2\operatorname{cosh}\left(\frac{4}{W^{*}}\right)\right)
\\
+2I_1\left(1;W^{*}\right)
\end{array}
\right\}
\\&
-\left\{
\begin{array}{l}
\frac{(\mu^*-1)^2W^{*2}}{16\mu^*(\mu^*+1)}
\ln \left(\frac{\operatorname{cosh}\frac{2y^* }{W^*}}{\operatorname{cosh}\frac{2 }{W^*}}\right)
\\\cdot
\left[\ln \left(2+2\operatorname{cosh}\left(\frac{4}{W^{*}}\right)\right)+\frac{2}{W^*}-2I_1\left(1;W^{*}\right)
\right]
\end{array}\right\}
\end{aligned}
\right\},
\end{equation}
where
\begin{equation}
\begin{aligned}
I_1\left(y;W\right):=&
\int_0^y \ln \left(1+e^{\frac{4x}{W}}\right) dx
\\=&\left\{\begin{array}{ll}
{\frac{2y^{2}}{W}+\frac{\pi^{2}W}{48}+\frac{W}{4}Li_2\left(-e^{-\frac{4 y}{W}}
	\right),} & {y\ge 0,} \\ {-\frac{\pi^{2}W}{48}-\frac{W}{4}Li_2\left(-e^{\frac{4 y}{W}}
	\right),} & {y\le 0.}\end{array}\right.
\end{aligned}
\end{equation}
The definition of polylog function is
\begin{equation}
Li_s\left(z
\right):=\sum_{m=1}^{\infty}
\frac{z^m}{m^s}.
\end{equation}


\begin{equation}\label{uM2star}
u^*_{M2}=\frac{
	\left\{\begin{array}{l}
	16\left(y^{*2}-1\right)
	\\+2\left(\mu^*-1\right) W^{*}\cdot
	\left\{\begin{array}{l}
	4y^*\ln\left(1+\frac{e^{\frac{4y^*}{W^*}}}{\mu^*}\right)
	\\+\left(y^*-3\right)\left(y^*+1\right)\ln\left(1+\frac{e^{\frac{4}{W^*}}}{\mu^*}\right)
	\\
	-\left(y^*+3\right)\left(y^*-1\right)\ln\left(1+\frac{e^{-\frac{4}{W^*}}}{\mu^*}\right)
	\\
	+2\left(y^*+1\right)I_2\left(1;W^*,\mu^*\right)
	\\
	-4I_2\left(y^*;W^*,\mu^*\right)
	\end{array}\right\}
	\\
	+\left(\mu^*-1\right)^2 W^{*2}\cdot
	\left\{\begin{array}{l}
	\left(y^*-1\right)
	\ln\left(1+\frac{e^\frac{4}{W^*}}{\mu^*}\right)
	\ln\left(1+\frac{e^{\frac{4y^*}{W^*}}}{\mu^*}\right)
	\\-\left(y^*+1\right)
	\ln\left(1+\frac{e^{-\frac{4}{W^*}}}{\mu^*}\right)
	\ln\left(1+\frac{e^{\frac{4y^*}{W^*}}}{\mu^*}\right)
	\\
	+2\ln\left(1+\frac{e^\frac{4}{W^*}}{\mu^*}\right)
	\ln\left(1+\frac{e^{-\frac{4}{W^*}}}{\mu^*}\right)
	\\
	+I_2\left(1;W^*,\mu^*\right)\ln\left(1+\frac{e^{\frac{4y^*}{W^*}}}{\mu^*}\right)
	\\
	-I_2\left(y^*;W^*,\mu^*\right)\ln\left(1+\frac{e^{\frac{4}{W^*}}}{\mu^*}\right)
	\\
	+\left\{\begin{array}{l}
	I_2\left(y^*;W^*,\mu^*\right)
	\\-I_2\left(1;W^*,\mu^*\right)
	\end{array}\right\}
	\ln\left(1+\frac{e^{-\frac{4}{W^*}}}{\mu^*}\right)
	\end{array}\right\}
	\end{array}\right\}}
{-\frac{4\mu^*}{\mu^*+1}\cdot
	\left[8+(\mu^*-1) W^* \ln \left(\frac{\mu^*+e^{\frac{4}{W^*}}}{\mu^*+e^{-\frac{4}{W^*}}}\right)\right]
},
\end{equation}
where
\begin{equation}
\begin{aligned}
&I_2\left(y;W,\mu\right):=
\int_{-1}^y \ln \left(1+\frac{1}{\mu}e^{\frac{4x}{W}}\right) dx
\\=&\left\{\begin{array}{ll}
{-\frac{W}{4}\left[Li_2\left(-\frac{e^{\frac{4 y}{W}}}{\mu}
	\right)-Li_2\left(-\frac{e^{-\frac{4 }{W}}}{\mu}
	\right)\right],} & {
	\left\{\begin{array}{l}\frac{W}{4}\ln\mu\ge 1 \text { or }\\ -1\le y\le\frac{W}{4}\ln\mu<1,\end{array}\right\}} \\ 
{\left\{\begin{array}{l}\frac{2y^{2}}{W}-y\ln\mu+\frac{W}{8}\ln^2\mu+\frac{\pi^2 W}{24}\\
	+\frac{W}{4}
	\left[
	Li_2\left(-\mu e^{-\frac{4 y}{W}}
	\right)
	+Li_2\left(-\frac{e^{-\frac{4 }{W}}}{\mu}
	\right)
	\right]
	\end{array}\right\}
	,} & { -1<\frac{W}{4}\ln\mu\le y\le 1,}
\\ {\left\{\begin{array}{l}\frac{2}{W}\left(y^{2}-1\right)-\left(y+1\right)\ln\mu\\
	+\frac{W}{4}
	\left[
	Li_2\left(-\mu e^{-\frac{4 y}{W}}
	\right)-Li_2\left(-\mu e^{\frac{4 }{W}}
	\right)
	\right]
	\end{array}\right\}
	,} & { \frac{W}{4}\ln\mu\le - 1.}\end{array}\right.
\end{aligned}
\end{equation}


\begin{equation}\label{uM3star}
u_{M3}^*=\frac{\mu^{*}+1}{\sqrt{\mu^{*}}}\cdot
\frac{
	\left\{\begin{array}{l} 
	\int_{-1}^{1}\left(\mu^{*}\right)^{\frac{1}{2}\tanh\left(\frac{2 Y^{*}}{W^{*}}\right)}  Y^{*} d Y^{*}\cdot
	\int_{1}^{y^*}\left(\mu^{*}\right)^{\frac{1}{2}\tanh\left(\frac{2 Y^{*}}{W^{*}}\right)}   d Y^{*}
	\\
	+
	\int_{-1}^{1}\left(\mu^{*}\right)^{\frac{1}{2}\tanh\left(\frac{2 Y^{*}}{W^{*}}\right)}   d Y^{*}
	\cdot
	\int_{y^*}^{1}\left(\mu^{*}\right)^{\frac{1}{2}\tanh\left(\frac{2 Y^{*}}{W^{*}}\right)} Y^{*}  d Y^{*}
	\end{array}\right\}
}
{\int_{-1}^{1}\left(\mu^{*}\right)^{\frac{1}{2}\tanh\left(\frac{2 Y^{*}}{W^{*}}\right)}   d Y^{*}}.
\end{equation}

Again, to check the suitability of the above expressions, we examine two special cases in Appendix~\ref{sec: App Analytical solutions}.

In the layered Poiseuille flow, the most important physical quantity is the velocity magnitude. Other physical quantities (such as viscous stress, flow rate, maximum velocity magnitude, {\it etc.}) can be calculated from the velocity field. The results of viscous stress are shown in Appendix~\ref{sec: App Viscouse stress}.

Further, in order to investigate the rationality of the information at the two-phase interface, the velocity and viscous stress at the interface are derived and discussed in Appendix~\ref{sec: App Information at the interface}. The results show that when the interface thickness is very small, the velocity and viscous stress in DI model and SI model are consistent, indicating that different dynamic viscosity models discussed in this paper are compatible.

\section{Numerical validation of the analytical solutions for the layered Poiseuille flow}\label{sec: Simulations}

In this section, we utilize numerical simulations to validate the correctness of the analytical solution.
We apply two methods to simulate this flow.
The governing equations of these methods are the Navier-Stokes equation coupled with Allen-Cahn/Cahn-Hilliard equation in the phase field model.
They are the ACNS system~\cite{lai2022simulation} and CHNS system~\cite{lai2022systematic}.
We use the mesoscopic approach, namely, the discrete unified gas kinetic scheme (DUGKS)~\cite{Guo2013,Guo2015,Guo2021}.
The numerical details can be found in our previous studies~\cite{lai2022simulation,lai2022systematic}.
The only difference is that we do not apply the weighted essentially non-oscillatory (WENO) schemes in the present simulation, due to the simple profile of layered Poiseuille flow.
The two simulation methods are denoted by PF(AC)-DUGKS and PF(CH)-DUGKS.

The schematic diagram is shown in Fig.~\ref{fig:layeredsketch}.
The computational domain and physical parameters in this section are consistent with those in
Zu and He~\cite{zu2013phase}.
The computational domain is $xyz=100\times100\times2$,
with meshgrid $dx=dy=dz=1$.
The height of half channel is $H=50$.
$y=\pm H$ are two plates, with no-slip boundary conditions.
The boundary conditions in $x$ and $z$ directions are periodic.
The densities of the two phases are the same,
$\rho_A = \rho_B = 1.0$.
The interfacial thickness parameter and the surface tension are
$W=4.0$ and $\sigma=5\times 10^{-5}$.
CFL number is $CFL=0.5$.
The dynamic viscosity of phase $A$ is $\mu_A = 1.0$, and the viscosity ratio is $\mu^*=\mu_A/\mu_B=100$.
The velocity of the center line is $u_c=5\times 10^{-5}$,
corresponding to a driving force
$G=u_c\left( \mu_B+\mu_A\right) /H^2$.

\begin{figure}[]
	\centering
	\includegraphics[width=0.6\columnwidth,trim={0cm 0cm 0cm 0cm},clip]{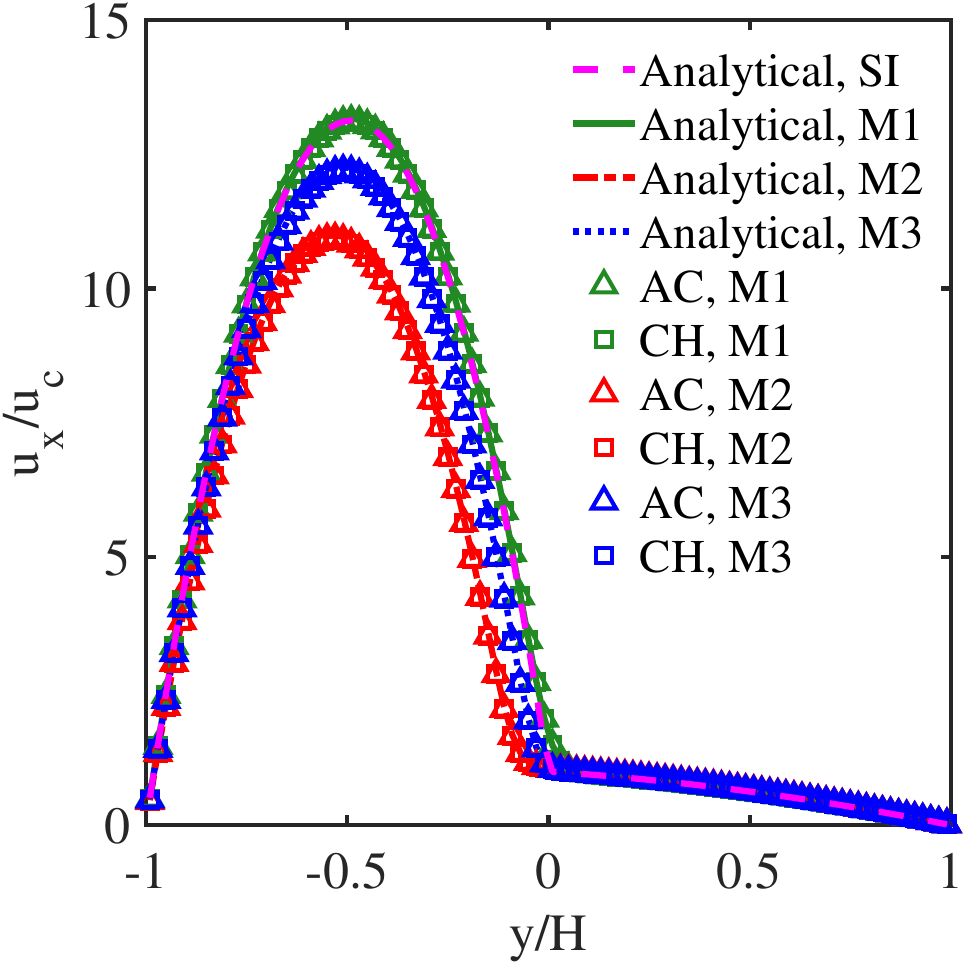}
	\centering
	\caption{Comparison of numerical simulation results with velocity profiles of analytical solutions. “SI” is the SI model. “M1”,“M2”, and “M3” represent different dynamic viscosity models in the DI model. “Analytical” is the analytical results we derive in Subsection~\ref{subsec: Analytical solutions}. “AC” and “CH” represent the simulation results obtained by PF(AC)-DUGKS method and PF(CH)-DUGKS method, respectively.}
	\label{fig:layeredACHmu100}
\end{figure}

\begin{table}[htbp]
	\centering
	\caption{Comparison of the normalized maximum velocity obtained by different methods}
	\label{tab: umax}
	\begin{tabular}{cccc}
		\toprule
		\multirow{2}{*}{} & \multicolumn{1}{c}{Analytical} & \multicolumn{1}{c}{PF(AC)-DUGKS} & \multicolumn{1}{c}{PF(CH)-DUGKS}  \\
		\midrule
		SI
		&$13.13$&$-$&$-$\\
		M1
		&$13.16$&$13.12$&$13.12$\\
		M2          &$10.99$&$10.93$&$10.93$\\
		M3        &$12.26$&$12.20$&$12.20$\\
		\bottomrule
	\end{tabular}\\
\end{table}

We apply PF(AC)-DUGKS method and PF(CH)-DUGKS method to simulate the layered Poiseuille flow with three different dynamic viscosity models. The comparison of the velocity profiles between the numerical results and the analytical solutions is shown in Fig.~\ref{fig:layeredACHmu100}.
It can be seen that the numerical results are consistent with the corresponding analytical solutions, regardless which numerical method is used (PF(AC)-DUGKS or PF(CH)-DUGKS).

The maximum velocity magnitude obtained by different methods are shown in Table~\ref{tab: umax}.
As long as the dynamic viscosity model is fixed, the relative error between PF-DUGKS and the maximum velocity magnitude obtained by the analytical solution does not exceed $1\%$.
The difference between the numerical results and the analytical solution of SI model is mainly 
due to the difference of dynamic viscosity model and the thickness of the interface.
To reduce the difference, we can reduce $W/H$ or change the viscosity model.
The velocity profile of M1 is closer to the analytical solution in the SI model, because the dynamic viscosity of M1 is closer to that of the SI model near the interface region with low dynamic viscosity, as shown at the $y<0$ region in Fig.~\ref{fig:mu3models}.
At the interface region with high viscosity, there is a significant difference between M1 and the SI model (the $y>0$ region in Fig.~\ref{fig:mu3models}). However, the region with high viscosity has small velocity and thus has little influence on the overall velocity profile (the $y>0$ region in Fig.~\ref{fig:layeredACHmu100}).
This explains why the velocity profile obtained by the M1 model is closer to the SI analytical solution.

Our result is consistent with that of Zu and He~\cite{zu2013phase}, but they did not provide the explanation.
We believe that the main reason is that the dynamic viscosity profile of M1 is closer to the SI model in the region of low viscosity, and this region plays a decisive role in the velocity profile of the layered Poiseuille flow.
Using this perspective, it is easy to understand that the velocity profile of M2 shows the largest deviation from the SI model, among the three models (M1, M2, and M3).
However, this does not mean that M2 is the worst, only that in the particular flow of layered Poiseuille flow, the M2 results happen to be the least desirable.
Of these three dynamic viscosity models, only M2 is symmetric, which could be better from this point of view.

\begin{figure}
	\centering    
	\subfigure[The profiles of $\phi$ and $\mu$ at $W^*=0.4$]{
		\begin{minipage}[t]{0.48\linewidth}
			\centering
			\includegraphics[width=0.85\columnwidth,trim={0cm 0cm 0cm 0cm},clip]{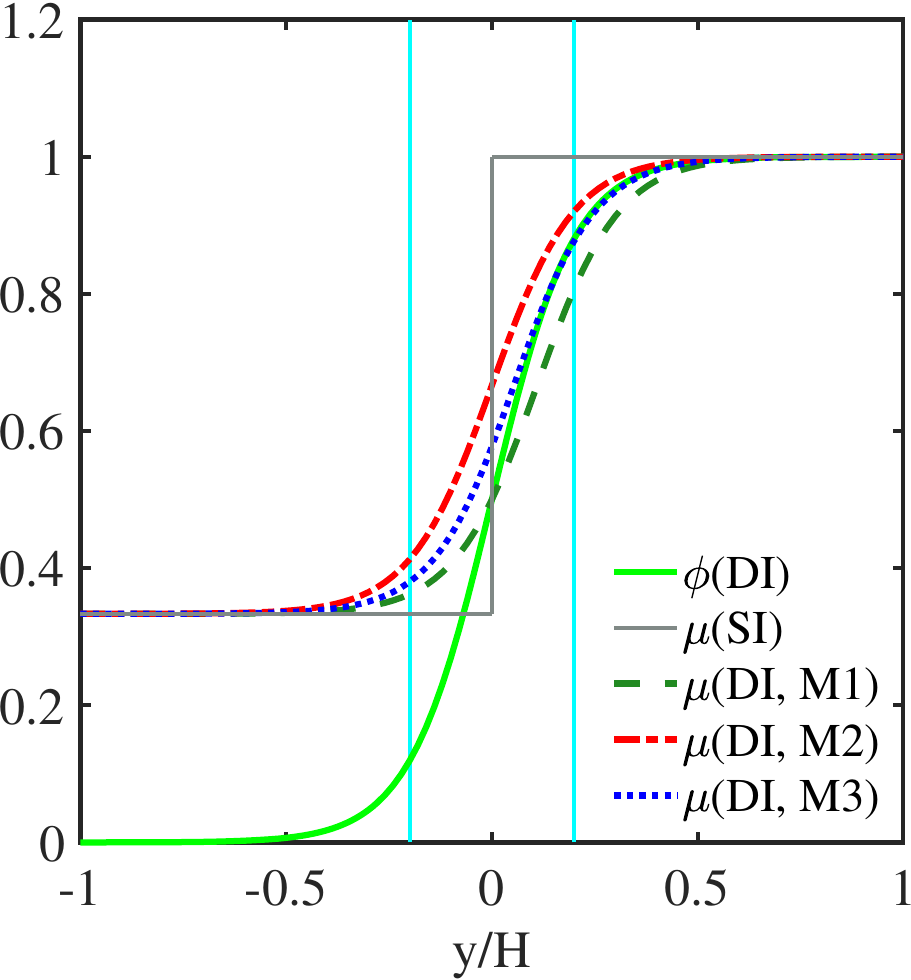}
			\label{fig:3modelsmu3W04mu}
		\end{minipage}
	}
	\subfigure[The profiles of $\phi$ and $\mu$ at $W^*=0.2$]{
		\begin{minipage}[t]{0.48\linewidth}
			\centering
			\includegraphics[width=0.85\columnwidth,trim={0cm 0cm 0cm 0cm},clip]{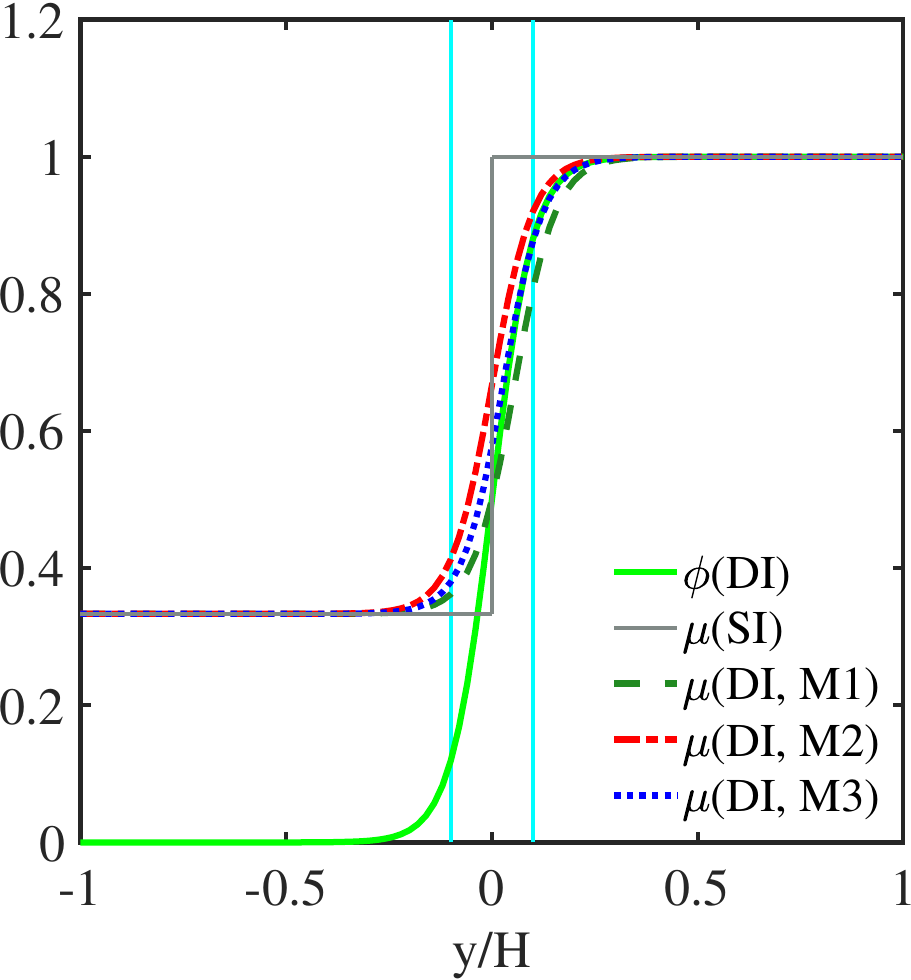}
			\label{fig:3modelsmu3W02mu}
		\end{minipage}
	}\\
	\subfigure[The profiles of $u^*$ at $W^*=0.4$]{
		\begin{minipage}[t]{0.48\linewidth}
			\centering
			\includegraphics[width=0.9\columnwidth,trim={0cm 0cm 0cm 0cm},clip]{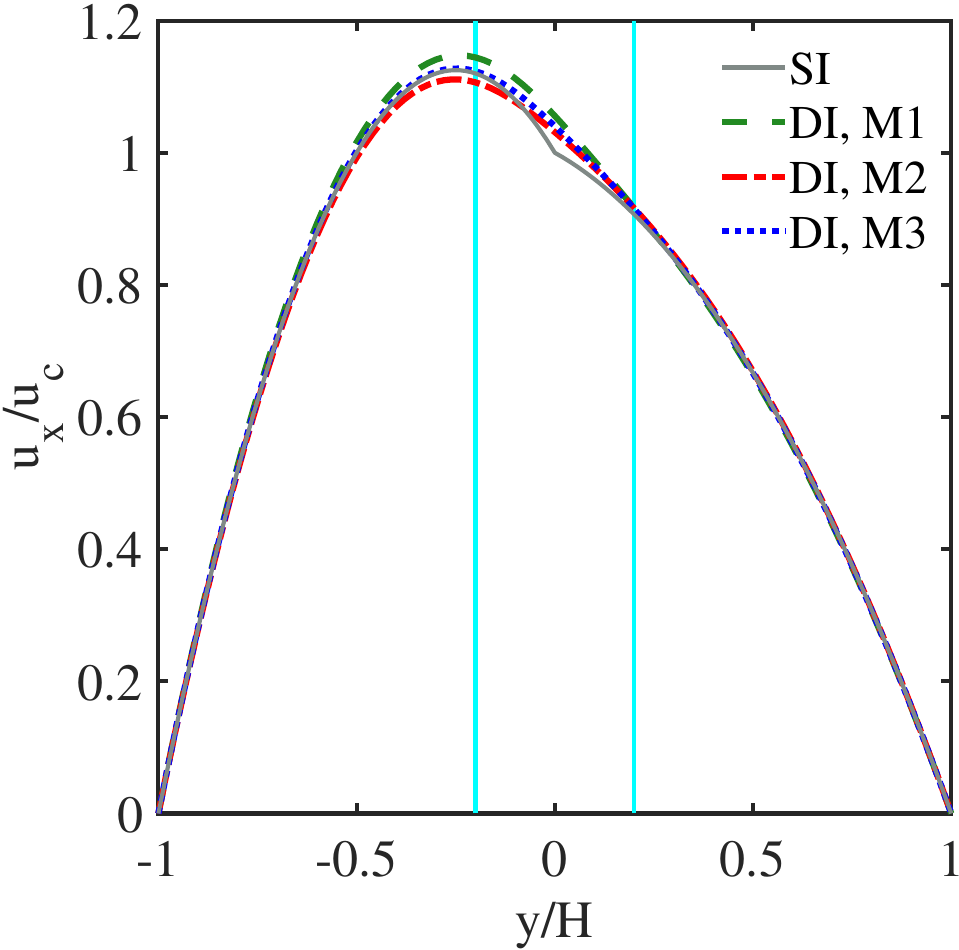}
			\label{fig:3modelsmu3W04u}
		\end{minipage}
	}
	\subfigure[The profiles of $u^*$ at $W^*=0.2$]{
		\begin{minipage}[t]{0.48\linewidth}
			\centering
			\includegraphics[width=0.9\columnwidth,trim={0cm 0cm 0cm 0cm},clip]{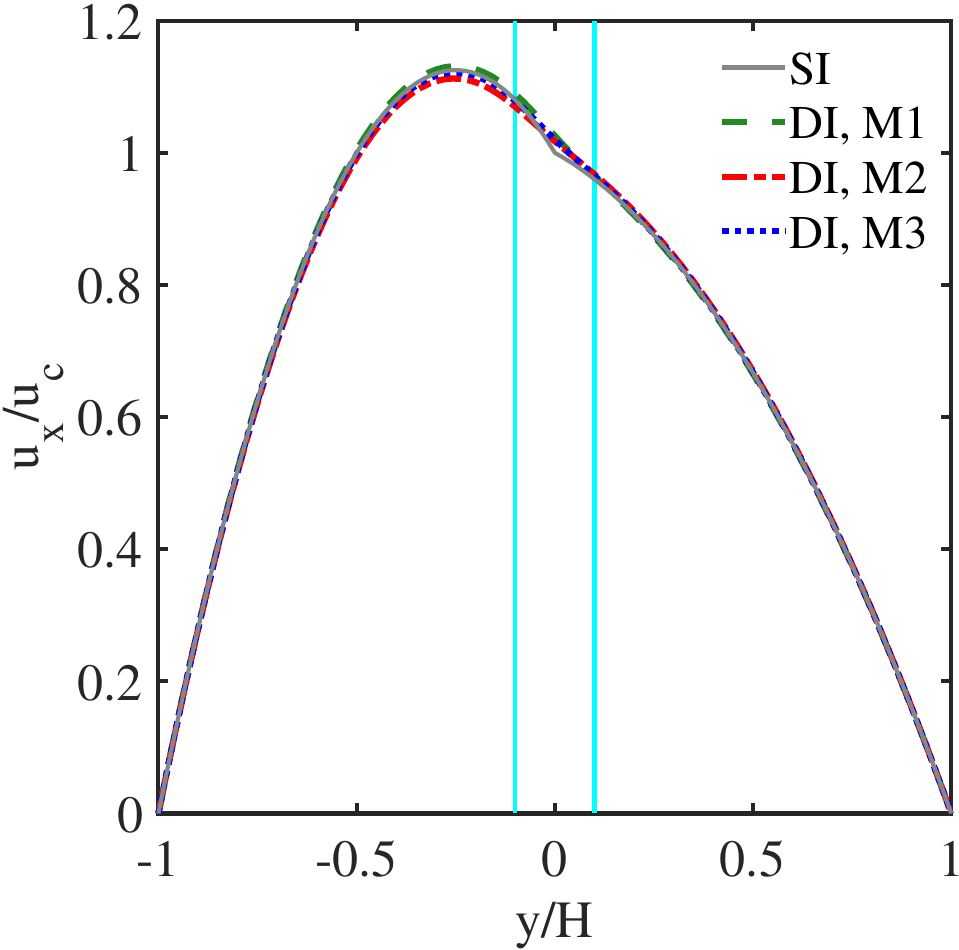}
			\label{fig:3modelsmu3W02u}
		\end{minipage}
	}
	\centering
	\caption{(a) and (b) are the profiles of $\phi$, $\mu$ at different $W^*$. (c) and (d) are the corresponding analytical velocity profiles. The width of the two light blue lines represents the value of $W^*$. $\mu_{A}$ is set to $1$ without losing generality. $\mu^*=3$ in this case.}
	\label{fig:3modelsmu3}
\end{figure}

\begin{figure}
	\centering    
	\subfigure[The profiles of $\phi$ and $\mu$ at $W^*=0.12$]{
		\begin{minipage}[t]{0.48\linewidth}
			\centering
			\includegraphics[width=0.85\columnwidth,trim={0cm 0cm 0cm 0cm},clip]{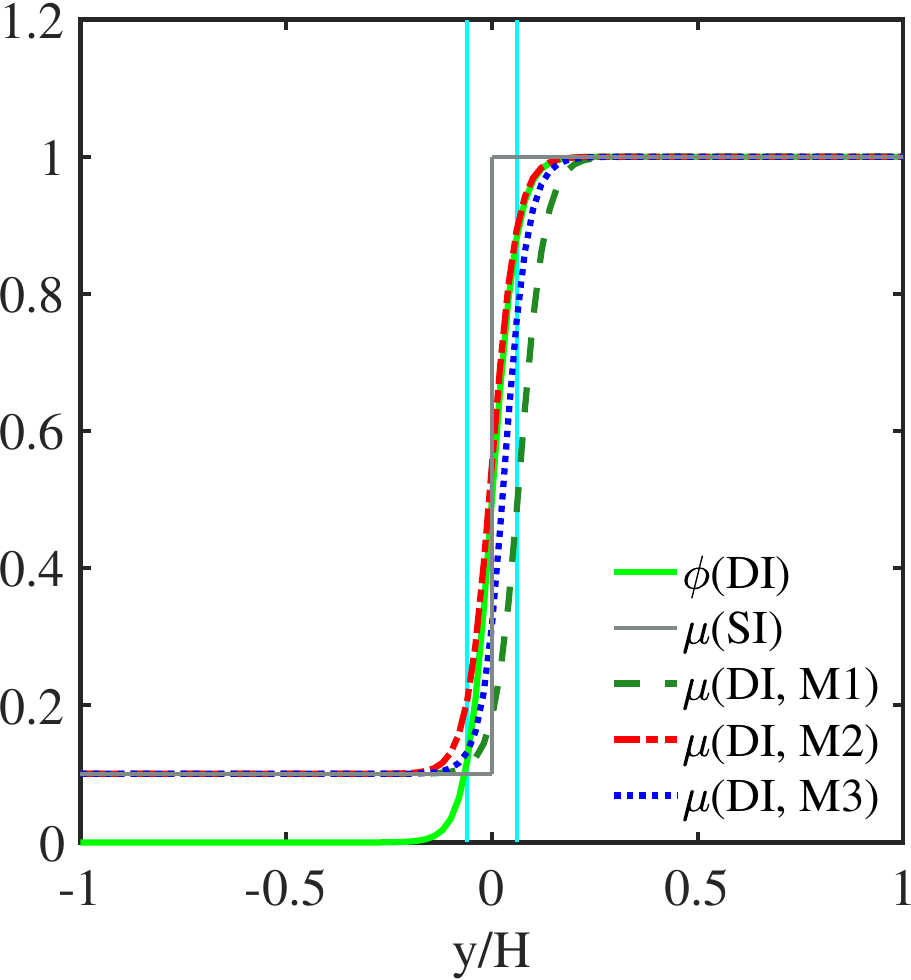}
			\label{fig:3modelsmu10W012mu}
		\end{minipage}
	}
	\subfigure[The profiles of $\phi$ and $\mu$ at $W^*=0.06$]{
		\begin{minipage}[t]{0.48\linewidth}
			\centering
			\includegraphics[width=0.85\columnwidth,trim={0cm 0cm 0cm 0cm},clip]{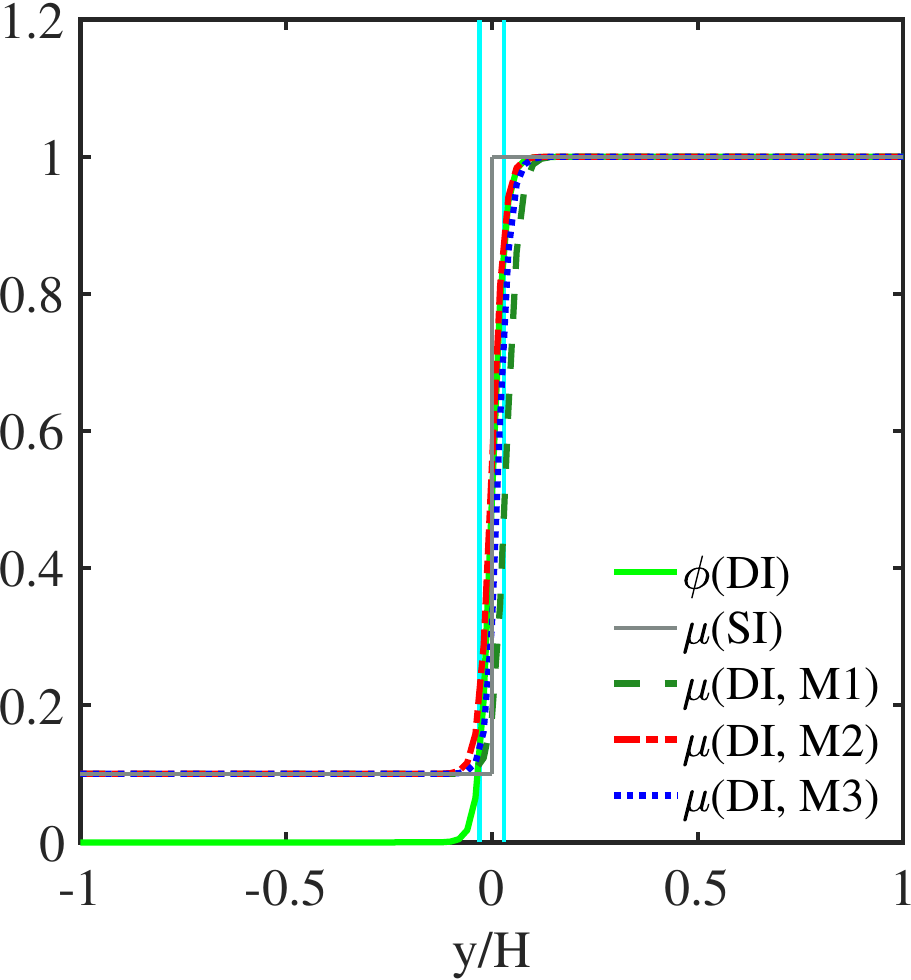}
			\label{fig:3modelsmu10W006mu}
		\end{minipage}
	}\\
	\subfigure[The profiles of $u^*$ at $W^*=0.12$]{
		\begin{minipage}[t]{0.48\linewidth}
			\centering
			\includegraphics[width=0.9\columnwidth,trim={0cm 0cm 0cm 0cm},clip]{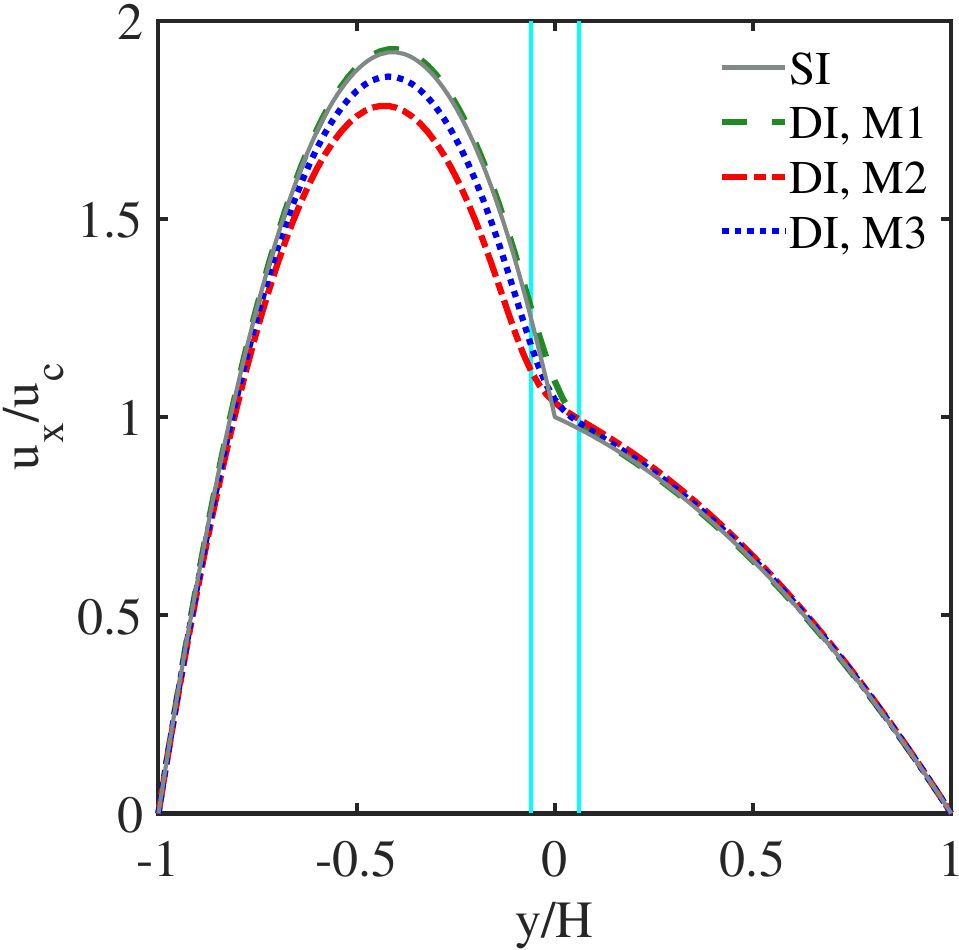}
			\label{fig:3modelsmu10W012u}
		\end{minipage}
	}
	\subfigure[The profiles of $u^*$ at $W^*=0.06$]{
		\begin{minipage}[t]{0.48\linewidth}
			\centering
			\includegraphics[width=0.9\columnwidth,trim={0cm 0cm 0cm 0cm},clip]{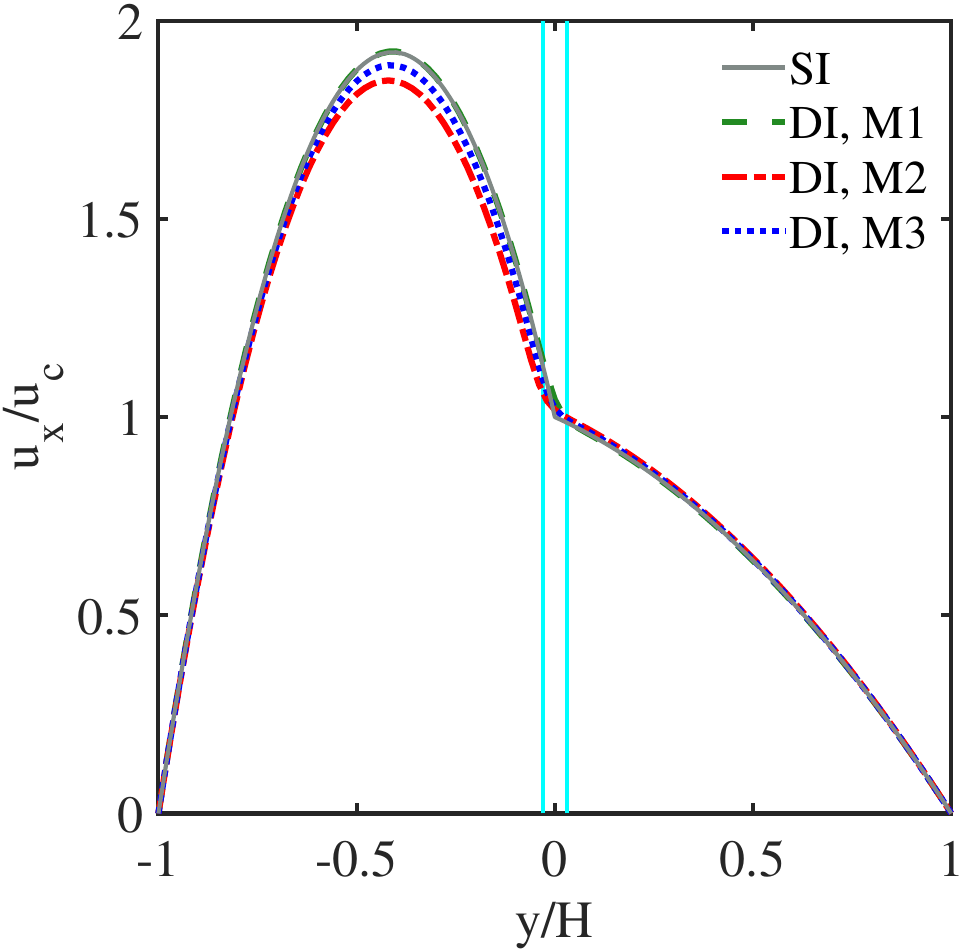}
			\label{fig:3modelsmu10W006u}
		\end{minipage}
	}
	\centering
	\caption{(a) and (b) are the profiles of $\phi$, $\mu$ at different $W^*$. (c) and (d) are the corresponding analytical velocity profiles. The width of the two light blue lines represents the value of $W^*$. $\mu_{A}$ is set to $1$ without losing generality. $\mu^*=10$ in this case.}
	\label{fig:3modelsmu10}
\end{figure}

\begin{figure}
	\centering    
	\subfigure[The profiles of $\phi$ and $\mu$ at $W^*=0.06$]{
		\begin{minipage}[t]{0.48\linewidth}
			\centering
			\includegraphics[width=0.85\columnwidth,trim={0cm 0cm 0cm 0cm},clip]{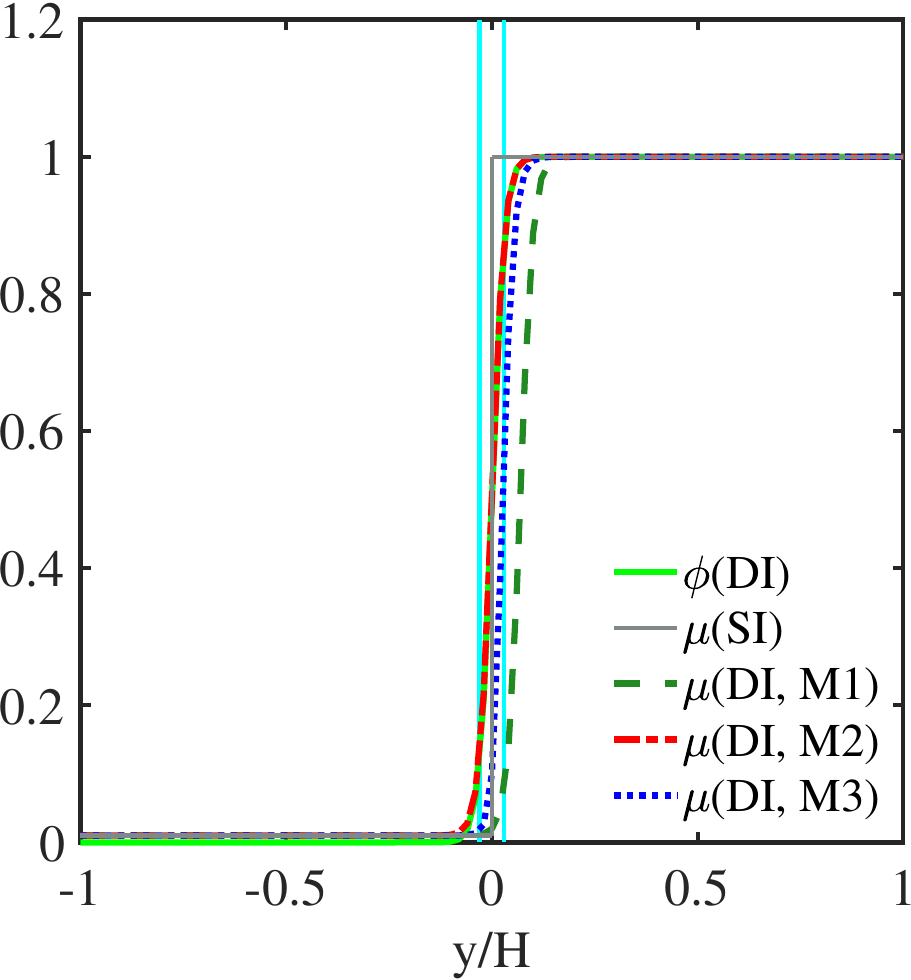}
			\label{fig:3modelsmu100W006mu}
		\end{minipage}
	}
	\subfigure[The profiles of $\phi$ and $\mu$ at $W^*=0.03$]{
		\begin{minipage}[t]{0.48\linewidth}
			\centering
			\includegraphics[width=0.85\columnwidth,trim={0cm 0cm 0cm 0cm},clip]{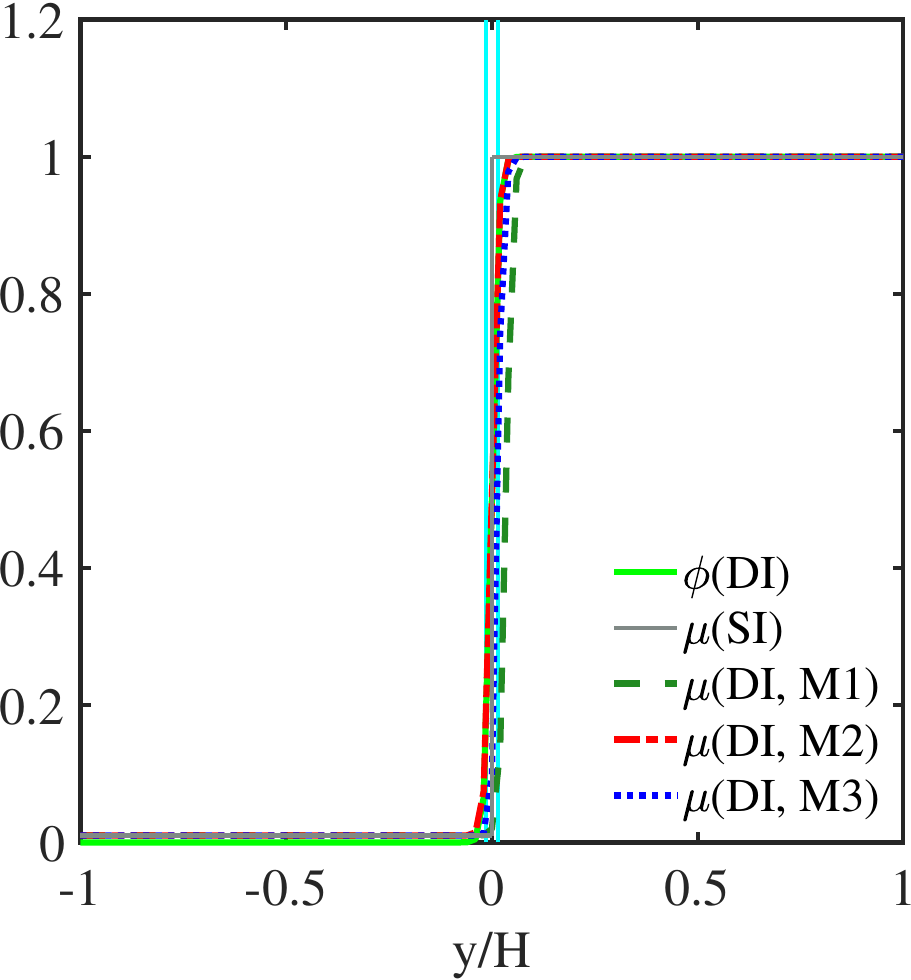}
			\label{fig:3modelsmu100W003mu}
		\end{minipage}
	}\\
	\subfigure[The profiles of $u^*$ at $W^*=0.06$]{
		\begin{minipage}[t]{0.48\linewidth}
			\centering
			\includegraphics[width=0.9\columnwidth,trim={0cm 0cm 0cm 0cm},clip]{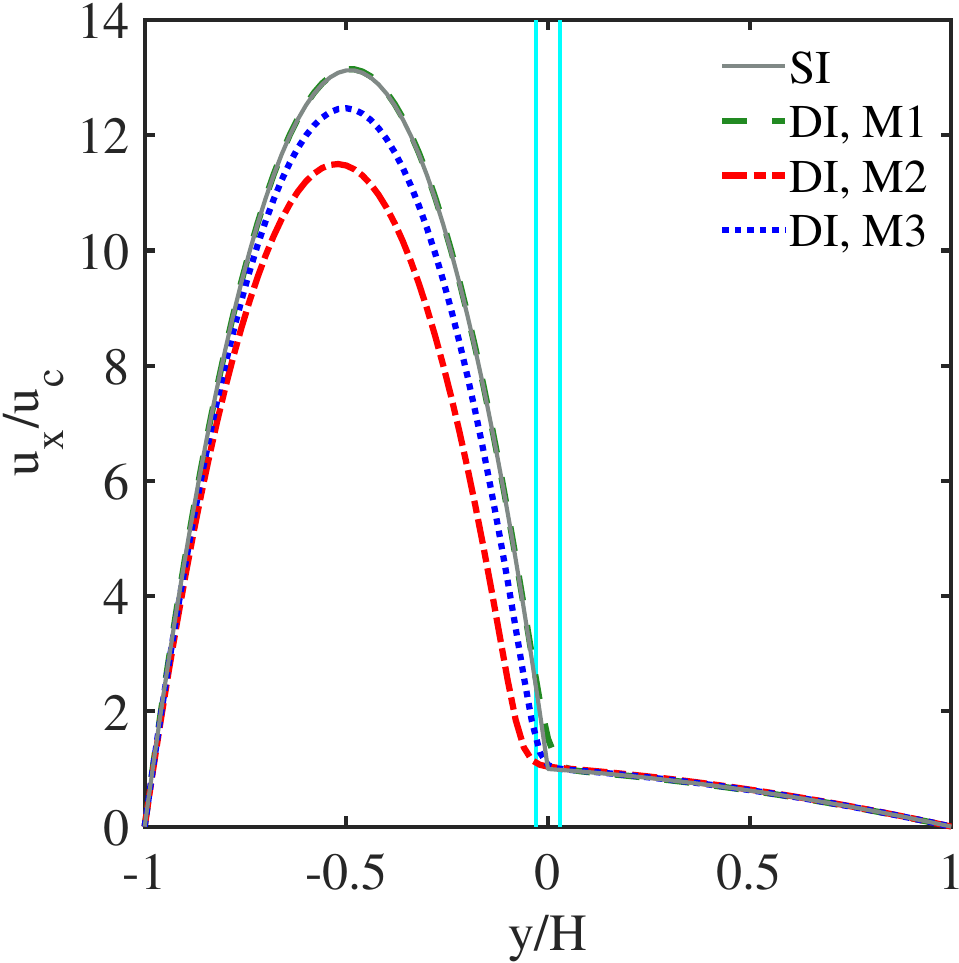}
			\label{fig:3modelsmu100W006u}
		\end{minipage}
	}
	\subfigure[The profiles of $u^*$ at $W^*=0.03$]{
		\begin{minipage}[t]{0.48\linewidth}
			\centering
			\includegraphics[width=0.9\columnwidth,trim={0cm 0cm 0cm 0cm},clip]{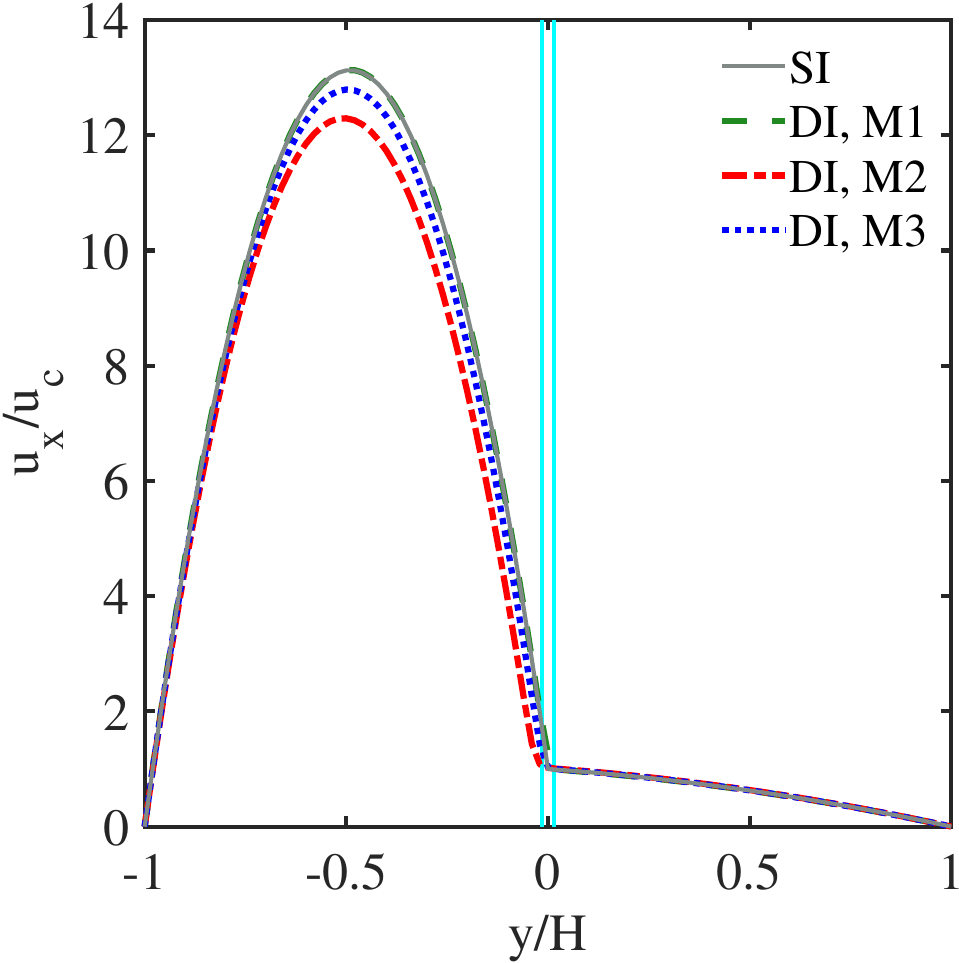}
			\label{fig:3modelsmu100W003u}
		\end{minipage}
	}
	\centering
	\caption{(a) and (b) are the profiles of $\phi$, $\mu$ at different $W^*$. (c) and (d) are the corresponding analytical velocity profiles. The width of the two light blue lines represents the value of $W^*$. $\mu_{A}$ is set to $1$ without losing generality. $\mu^*=100$ in this case.}
	\label{fig:3modelsmu100}
\end{figure}

\begin{figure}
	\centering    
	\subfigure[The profiles of $\phi$ and $\mu$ at $W^*=0.06$]{
		\begin{minipage}[t]{0.48\linewidth}
			\centering
			\includegraphics[width=0.85\columnwidth,trim={0cm 0cm 0cm 0cm},clip]{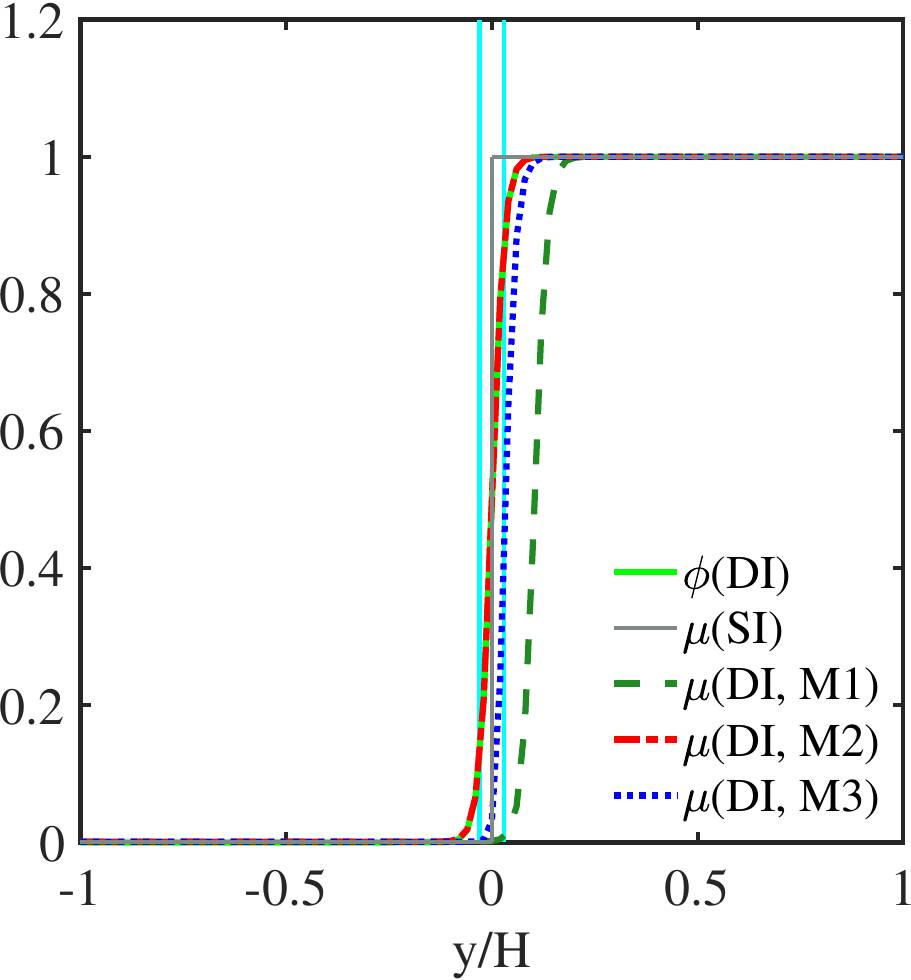}
			\label{fig:3modelsmu1000W006mu}
		\end{minipage}
	}
	\subfigure[The profiles of $\phi$ and $\mu$ at $W^*=0.03$]{
		\begin{minipage}[t]{0.48\linewidth}
			\centering
			\includegraphics[width=0.85\columnwidth,trim={0cm 0cm 0cm 0cm},clip]{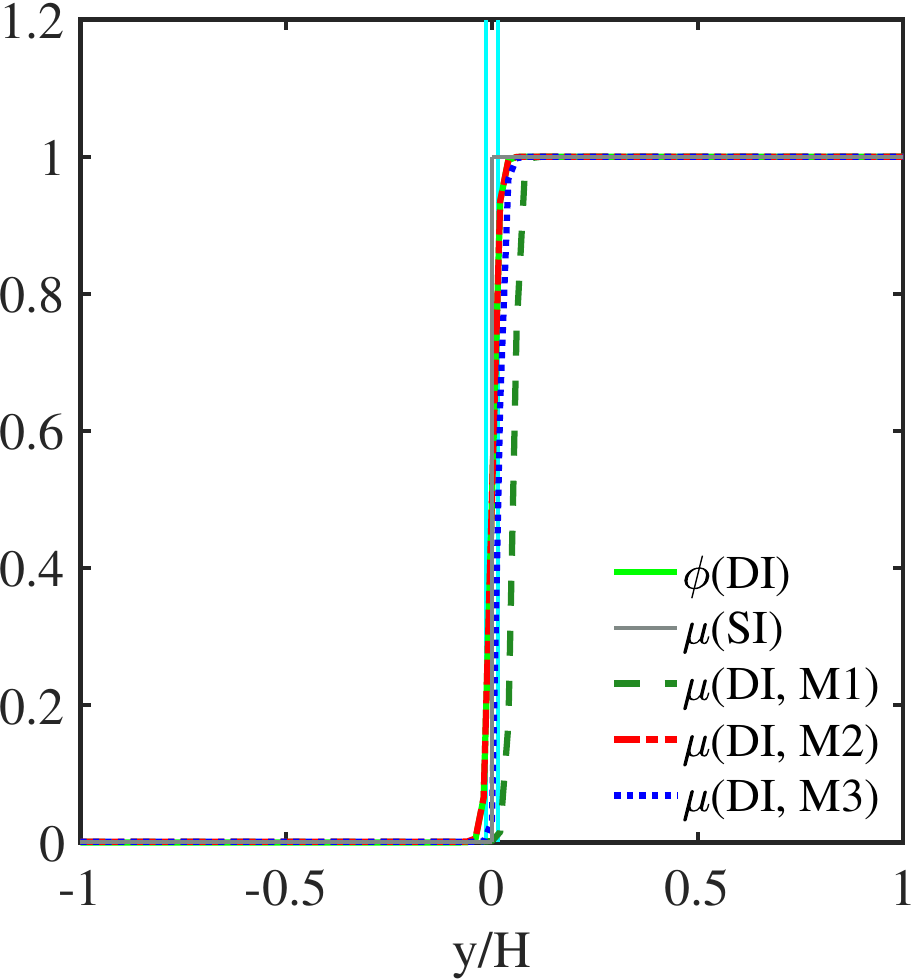}
			\label{fig:3modelsmu1000W003mu}
		\end{minipage}
	}\\
	\subfigure[The profiles of $u^*$ at $W^*=0.06$]{
		\begin{minipage}[t]{0.48\linewidth}
			\centering
			\includegraphics[width=0.9\columnwidth,trim={0cm 0cm 0cm 0cm},clip]{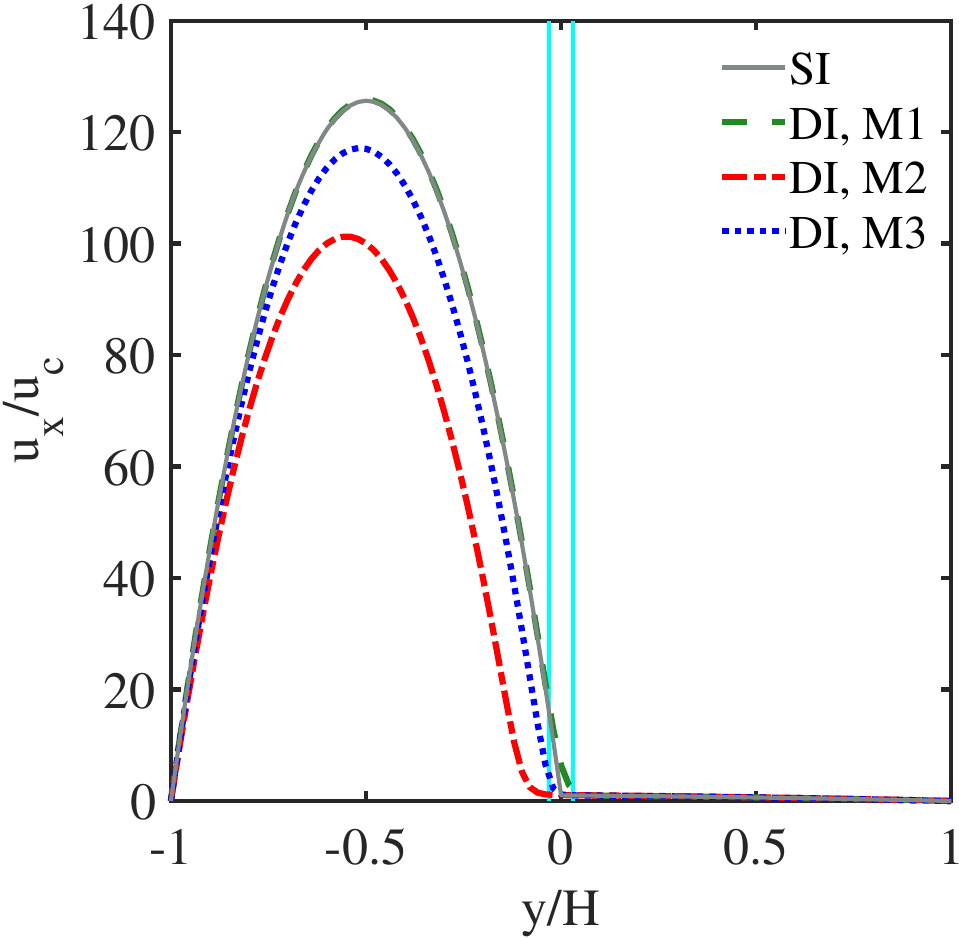}
			\label{fig:3modelsmu1000W006u}
		\end{minipage}
	}
	\subfigure[The profiles of $u^*$ at $W^*=0.03$]{
		\begin{minipage}[t]{0.48\linewidth}
			\centering
			\includegraphics[width=0.9\columnwidth,trim={0cm 0cm 0cm 0cm},clip]{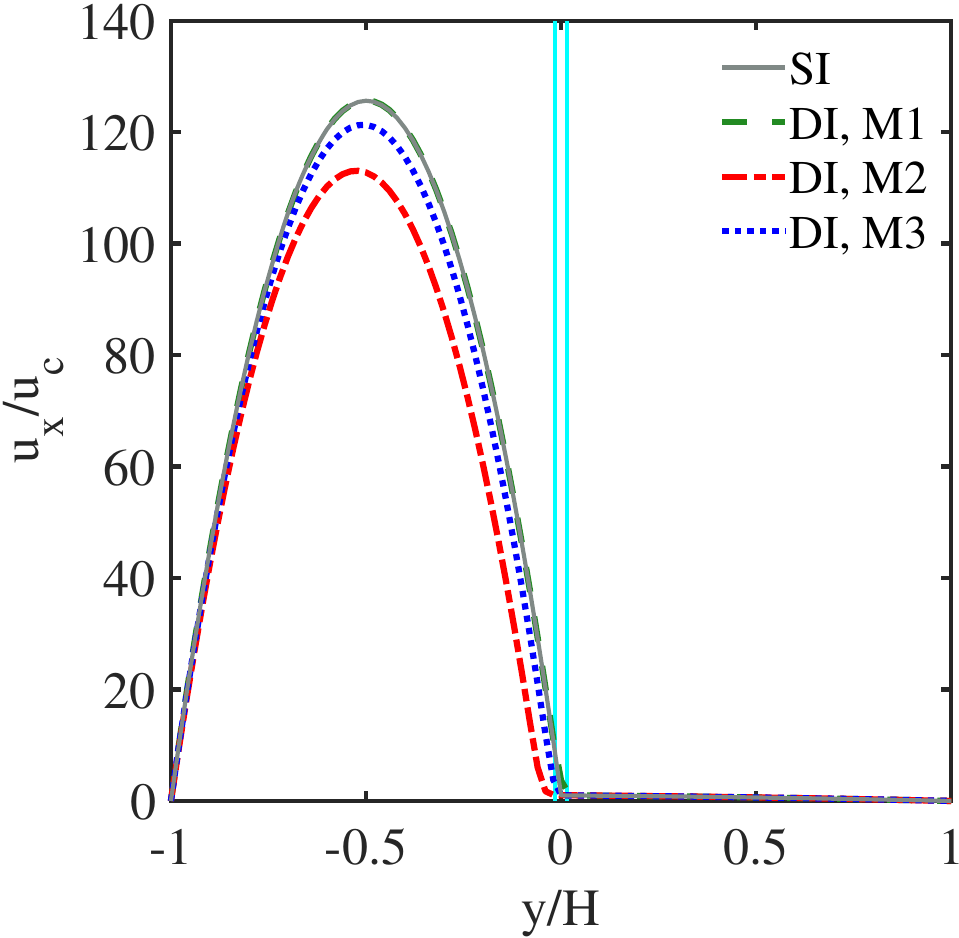}
			\label{fig:3modelsmu1000W003u}
		\end{minipage}
	}
	\centering
	\caption{(a) and (b) are the profiles of $\phi$, $\mu$ at different $W^*$. (c) and (d) are the corresponding analytical velocity profiles. The width of the two light blue lines represents the value of $W^*$. $\mu_{A}$ is set to $1$ without losing generality. $\mu^*=1000$ in this case.}
	\label{fig:3modelsmu1000}
\end{figure}

\begin{table*}[]
	\centering
	\caption{The normalized maximum velocity at different $\mu^*$, $W^*$, and viscosity models}
	\label{TabMu3}
	\begin{tabular}{ccccccccc}
		\toprule
		$\mu^*$ \multirow{2}{*}{} & \multicolumn{2}{c}{$3$} & \multicolumn{2}{c}{$10$} & \multicolumn{2}{c}{$100$} & \multicolumn{2}{c}{$1000$}\\
		\cmidrule(r){2-3} \cmidrule(r){4-5} \cmidrule(r){6-7} \cmidrule(r){8-9}
		$u^*_{SI,\max}$ \multirow{2}{*}{} & \multicolumn{2}{c}{$1.125$} & \multicolumn{2}{c}{$1.920$} & \multicolumn{2}{c}{$13.13$} & \multicolumn{2}{c}{$125.6$}\\
		\cmidrule(r){2-3} \cmidrule(r){4-5} \cmidrule(r){6-7} \cmidrule(r){8-9}
		$W^*$ \multirow{2}{*}{} & \multicolumn{1}{c}{$0.4$} & \multicolumn{1}{c}{$0.2$} & \multicolumn{1}{c}{$0.12$} & \multicolumn{1}{c}{$0.06$} & \multicolumn{1}{c}{$0.06$} & \multicolumn{1}{c}{$0.03$} & \multicolumn{1}{c}{$0.06$} & \multicolumn{1}{c}{$0.03$}\\
		\midrule
		$u_{M1,\max}^* $             & $1.147$                          & $1.131$                    & $1.928$                   & $1.922$           & $13.14$           & $13.13$          & $125.8$          & $125.7$                     \\
		$u_{M2,\max}^* $             &$1.111$                          & $1.112$                    & $1.785$                   & $1.849$          & $11.50$           & $12.29$          & $101.2$          & $113.1$          \\
		$u_{M3,\max}^* $             &$1.127$                         & $1.121$                  & $1.859$                   & $1.887$           &$12.47$           & $12.79$          & $117.1$           & $121.3$          \\
		$\left\{\frac{u_{M1,\max}^* }{ u_{SI,\max}^*}-1\right\} $             &$2.0\%$                          & $0.55\%$                    & $0.42\%$                   & $0.10\%$           & $0.14\%$          & $0.035\%$          & $0.15\%$           & $0.037\%$          \\
		$\left\{\frac{u_{M2,\max}^* }{ u_{SI,\max}^*}-1\right\} $             &$-1.2\%$                          & $-1.1\%$                    & $-7.0\%$                   & $-3.7\%$           & $-12\%$           & $-6.3\%$          & $-19\%$           & $-10\%$          \\
		$\left\{\frac{u_{M3,\max}^* }{ u_{SI,\max}^*}-1\right\} $             &$0.23\%$                          & $-0.33\%$                    & $-3.2\%$                  & $-1.7\%$           & $-5.0\%$          & $-2.6\%$         & $-6.7\%$          & $-3.4\%$          \\
		\bottomrule
	\end{tabular}\\
	{$\mu^*=\mu_{A}/\mu_{B}$ is the dynamic viscosity ratio, $W^*=W/H$ is the normalized interfacial thickness, $u^*=(\mu_{A}+\mu_{B})u_x/(GH^2)$ is the normalized velocity magnitude, $u^*_{\cdot,\max}$ denotes the analytical results of the maximum normalized velocity magnitude in the specific viscosity model.
	}\\
\end{table*}

\section{The choice of mixture dynamic viscosity models and $W^*$ at different viscosity ratios}\label{sec: Guidelines}

In this section, we apply the above analytical results to discuss a broader parameter range in the layered Poiseuille flow problem, and determine the selection requirements for the normalized interface thickness $W^*$ under different circumstances.
Since the density does not affect the analytical results in theory, we only focus on the dynamic viscosity ratio.
If $\mu^*=1$, all the results with different DI models are the same as that with SI model, for the layered Poiseuille flow, as we discuss before (For example, Eq.~\eqref{momsamemu} in Subsection~\ref{subsec: Governing equation}, and Eqs.~\eqref{usamemu} in Appendix~\ref{sec: App Analytical solutions}).
Hence $W^*$ can be any positive value when $\mu^*=1$.
Here we only concern about the cases $\mu^*\neq 1$.
Specifically, $\mu^*=3, 10, 100, 1000$ are selected, which are some of the commonly used viscosity ratios in the literature~\cite{2007Lattice,2008Lattice,LECLAIRE20122237,zu2013phase,WANG2015404,2017Improved,PhysRevE.97.033309,zhang2018discrete}.

Fig.~\ref{fig:3modelsmu3} shows the dynamic viscosity profiles and the corresponding velocity profiles at $\mu^*=3$.
The specific viscosity of each phase would not affect the velocity profiles, here $\mu_{A}$ is set to $1$ for plotting.
When $\mu^*=1$, $W^*$ would not affect the results.
When $\mu^*>1$, the influence of $W^*$ on the analytical velocity profiles increases with the increase of $\mu^*$.
$W^*$ is normally chosen less than $0.2$ in the previous simulations~\cite{zu2013phase,2017Improved,PhysRevE.97.033309,zhang2018discrete}.
However, since $\mu^*=3$ is a relatively small viscosity ratio, $W^*$ can be chosen as a relatively large value.
$W^*=0.4$ a very thick interface compared to the channel width (Fig.~\ref{fig:3modelsmu3W04mu}),
but the analytical velocity profiles of the DI model are not far from that of the SI model (Fig.~\ref{fig:3modelsmu3W04u}).
If the interface thickness decreases to a normal large value, $W^*=0.2$ (Fig.~\ref{fig:3modelsmu3W02mu}), then the analytical results of the DI model are much closer to that of the SI model (Fig.~\ref{fig:3modelsmu3W02u}).
Quantitatively, we can compare the relative error between the maximum velocity calculated by the DI model and that calculated by the SI model, which are shown in Table~\ref{TabMu3}.
When $W^*=0.4$, the result of M1 has the largest relative error ($2.0\%$).
When $W^*=0.2$, the results of all the three models are reasonable, if we want to limit the absolute value of the relative error close to or less than $1.0\%$ ($0.55\%, -1.1\%, -0.33\%$).

When $\mu^*=10$, $W^*$ should be smaller.
$W^*=0.12, 0.06$ are chosen, as shown in Fig.~\ref{fig:3modelsmu10}.
It can be seen that the M1 model and the SI model agree very well, while both the M2 model and the M3 model have deviations, especially in the regions with large velocities.
For $W^*=0.12, 0.06$, the relative errors is $0.42\%, 0.10\%$ for M1, respectively.
However, the absolute values of the relative errors are all greater than $1.0\%$ for M2 and M3,
even go up to $7.0\%$, as shown in Table~\ref{TabMu3}.
Compared to the cases $\mu^*=3$, the absolute values of the relative errors have increased a lot for M2 and M3 models, although the interface thickness has decreased (only $0.3$ times that of the cases $\mu^*=3$).

For much larger $\mu^*$, $W^*$ should be much smaller.
However, we have not seen $W^*<0.03$ in the previous studies~\cite{zu2013phase,2017Improved,PhysRevE.97.033309,zhang2018discrete},
because the researchers are reluctant to spend too much computing resources on such a simple physical problem.
We choose $W^*=0.06, 0.03$ for both $\mu^*=100$ and $\mu^*=1000$ to analyze, as shown in Figs.~\ref{fig:3modelsmu100}-\ref{fig:3modelsmu1000}.
When $W^*$ becomes smaller, the viscosity in the DI model tends to that in the SI model (Figs.~\ref{fig:3modelsmu100W006mu},~\ref{fig:3modelsmu100W003mu},~\ref{fig:3modelsmu1000W006mu},~\ref{fig:3modelsmu1000W003mu}),
then the velocity profile in the DI model is also close to that in the SI model (Figs.~\ref{fig:3modelsmu100W006u},~\ref{fig:3modelsmu100W003u},~\ref{fig:3modelsmu1000W006u},~\ref{fig:3modelsmu1000W003u}).
For the same $W^*$ and the same DI model, the absolute values of the relative errors are larger in the $\mu^*=1000$ cases than those in the $\mu^*=100$ cases, as shown in Table~\ref{TabMu3}.
All these qualitative results are expected.
Furthermore, Table~\ref{TabMu3} also shows that for the large $\mu^*$, we can only use M1 among the three mixture dynamic viscosity models, otherwise the errors would be larger than we can accept.

In general, the M1 model produces the best results compared to the SI results, the error is the smallest in most of the cases.
In the layered Poiseuille flow problem, it is better to choose $W^*\le 0.06$, so that the errors are small for the range of $1\le \mu^*\le 1000$.


\begin{figure}
	\centering    
	\subfigure[The first set of dynamic viscosity models]{
		\begin{minipage}[t]{0.48\linewidth}
			\centering
			\includegraphics[width=0.9\columnwidth,trim={0cm 0cm 0cm 0cm},clip]{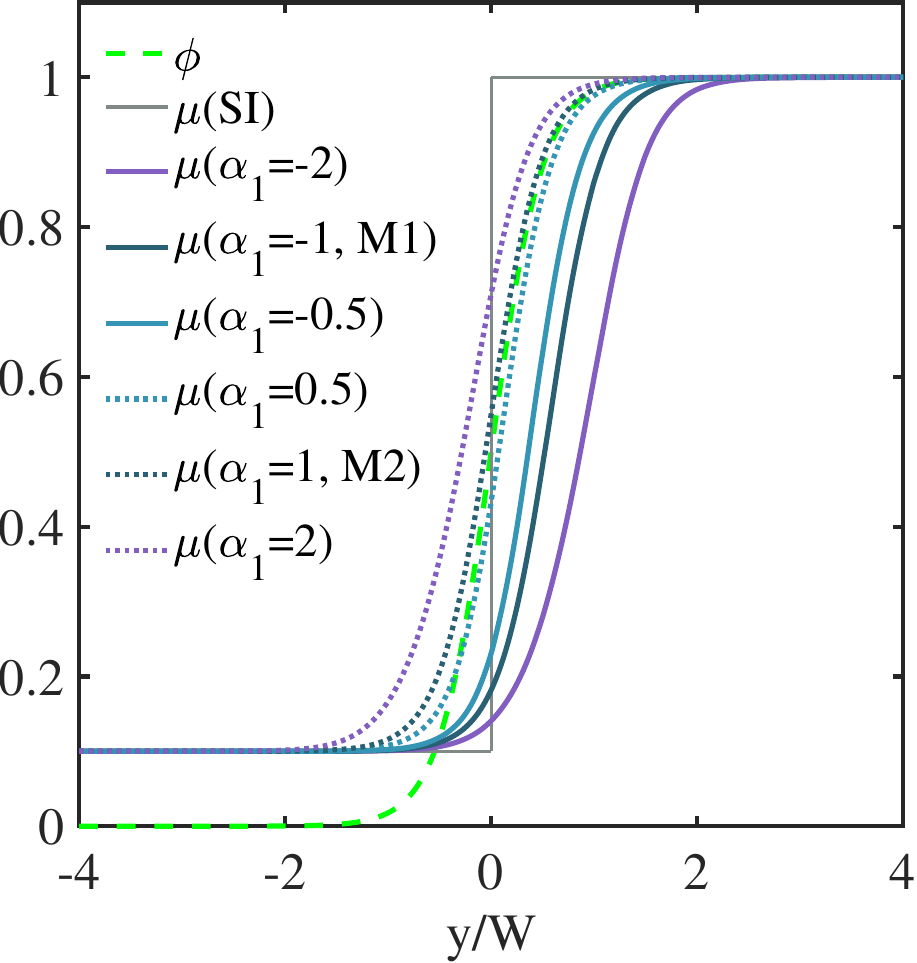}
			\label{fig:alpha1}
		\end{minipage}
	}
	\subfigure[The second set of dynamic viscosity models]{
		\begin{minipage}[t]{0.48\linewidth}
			\centering
			\includegraphics[width=0.9\columnwidth,trim={0cm 0cm 0cm 0cm},clip]{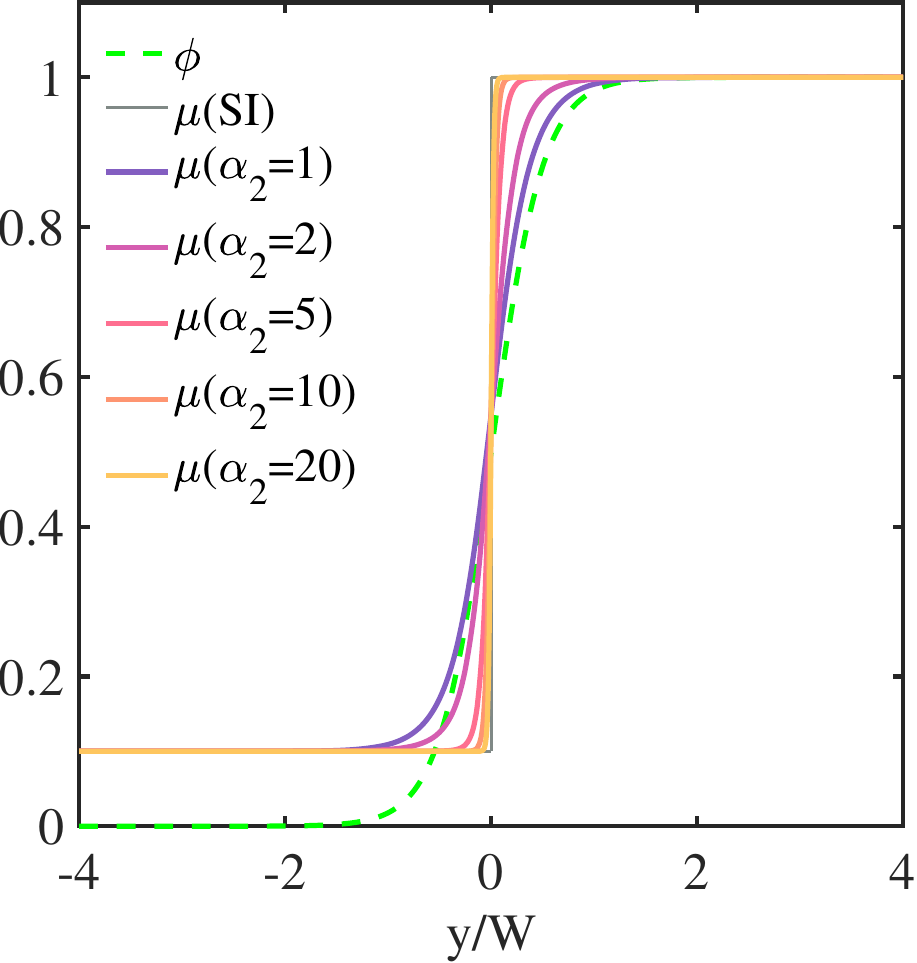}
			\label{fig:alpha2}
		\end{minipage}
	}
	\centering
	\caption{The new designed dynamic viscosity models. Here the fixed dynamic viscosities for two-phase flow are $0.1$ and $1$, respectively, for ease of plotting.}
	\label{fig:alpha}
\end{figure}

\section{The generalized mixture dynamic viscosity models and the corresponding velocity profiles}\label{sec: NewMu}

\begin{figure}
	\centering    
	\subfigure[$W^*=0.24$]{
		\begin{minipage}[t]{0.31\linewidth}
			\centering
			\includegraphics[width=0.95\columnwidth,trim={0cm 0cm 0cm 0cm},clip]{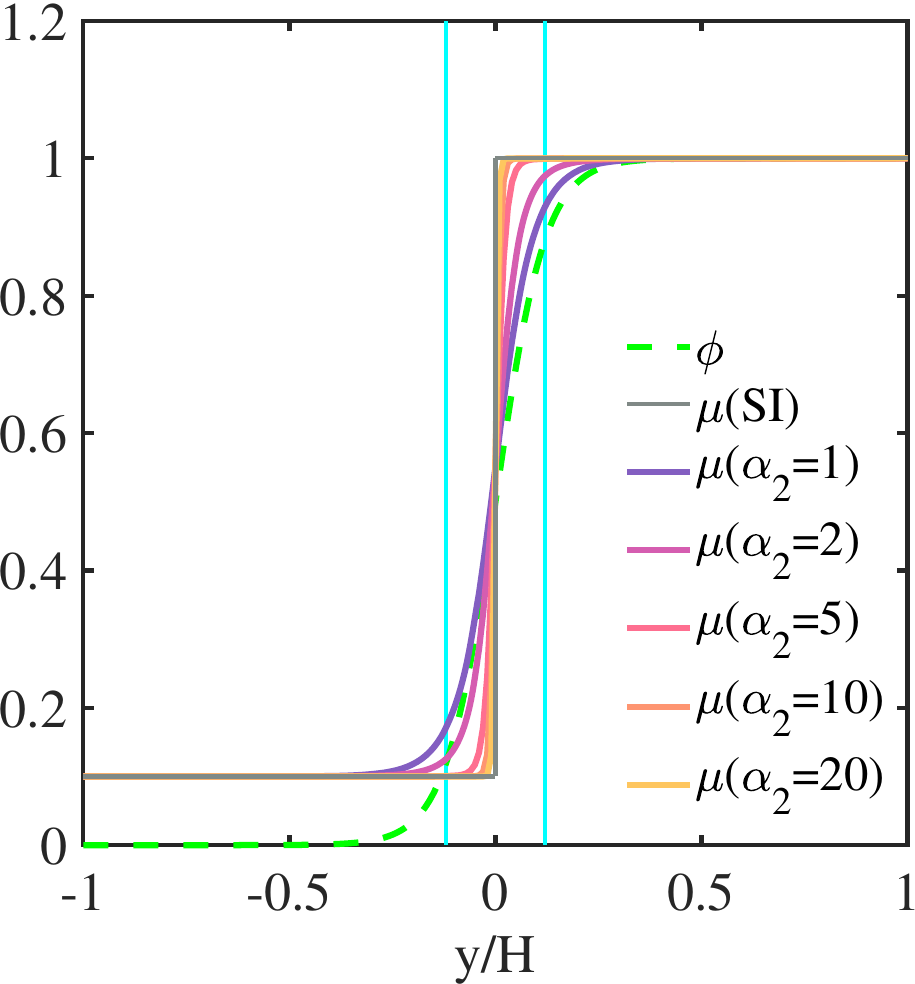}
			\label{fig:nmodelsmu10W024mu}
		\end{minipage}
	}
	\subfigure[$W^*=0.12$]{
		\begin{minipage}[t]{0.31\linewidth}
			\centering
			\includegraphics[width=0.95\columnwidth,trim={0cm 0cm 0cm 0cm},clip]{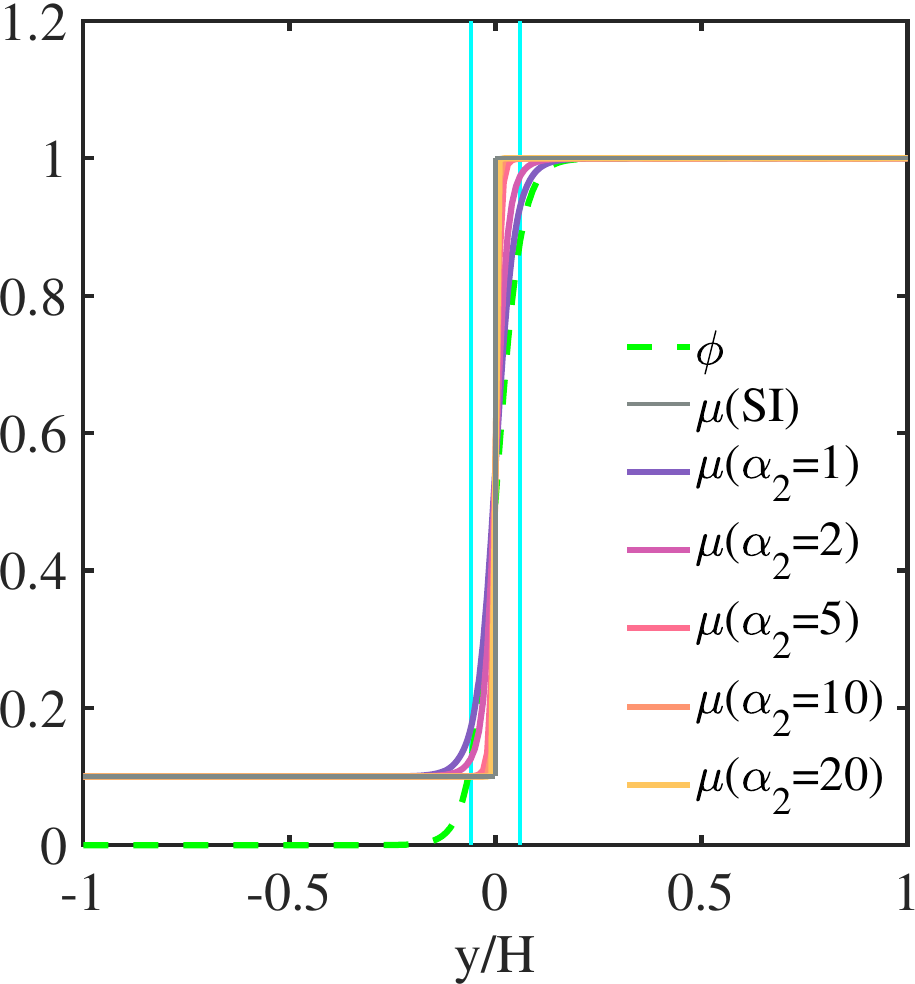}
			\label{fig:nmodelsmu10W012mu}
		\end{minipage}
	}
	\subfigure[$W^*=0.06$]{
		\begin{minipage}[t]{0.31\linewidth}
			\centering
			\includegraphics[width=0.95\columnwidth,trim={0cm 0cm 0cm 0cm},clip]{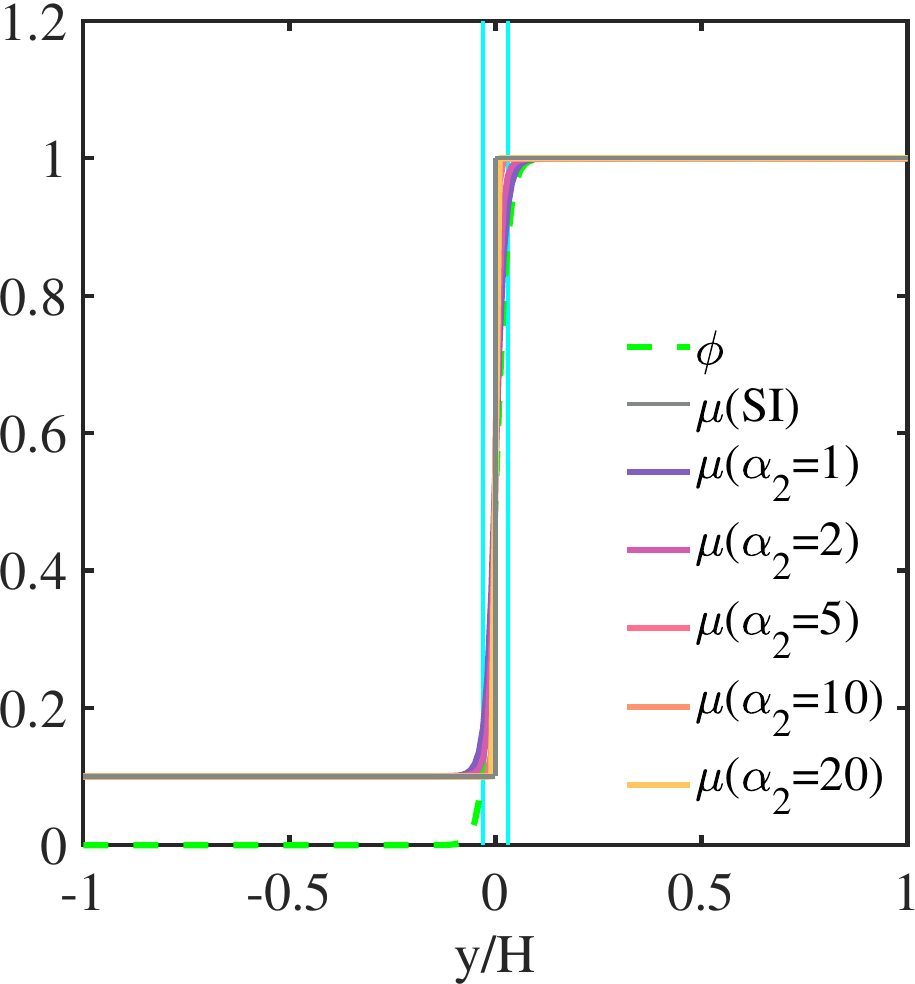}
			\label{fig:nmodelsmu10W006mu}
		\end{minipage}
	}\\
	\subfigure[$W^*=0.24$]{
		\begin{minipage}[t]{0.31\linewidth}
			\centering
			\includegraphics[width=0.99\columnwidth,trim={0cm 0cm 0cm 0cm},clip]{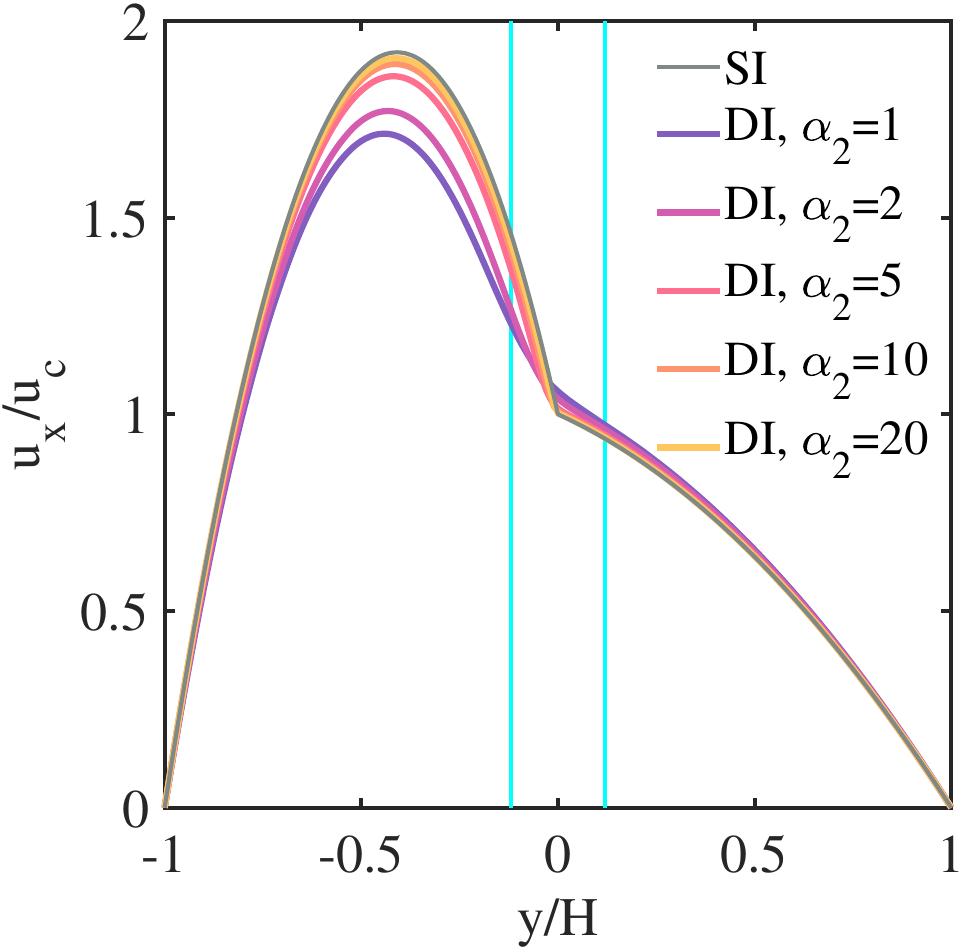}
			\label{fig:nmodelsmu10W024u}
		\end{minipage}
	}
	\subfigure[$W^*=0.12$]{
		\begin{minipage}[t]{0.31\linewidth}
			\centering
			\includegraphics[width=0.99\columnwidth,trim={0cm 0cm 0cm 0cm},clip]{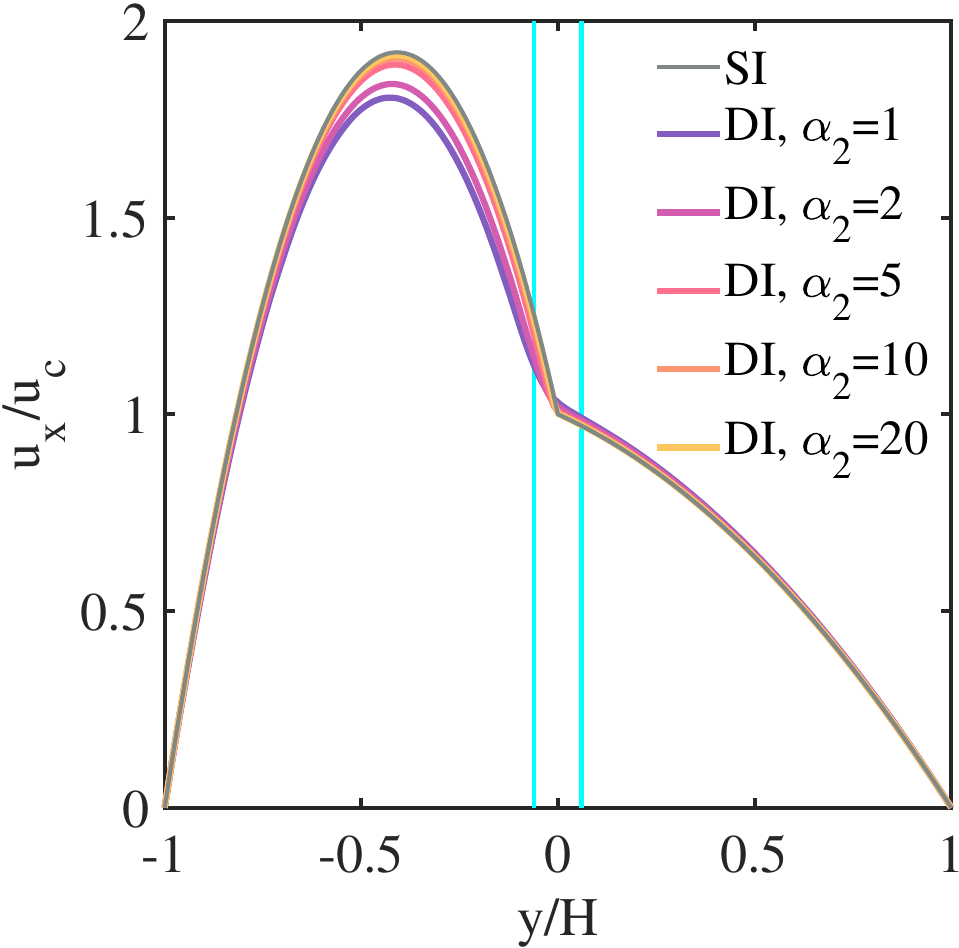}
			\label{fig:nmodelsmu10W012u}
		\end{minipage}
	}
	\subfigure[$W^*=0.06$]{
		\begin{minipage}[t]{0.31\linewidth}
			\centering
			\includegraphics[width=0.99\columnwidth,trim={0cm 0cm 0cm 0cm},clip]{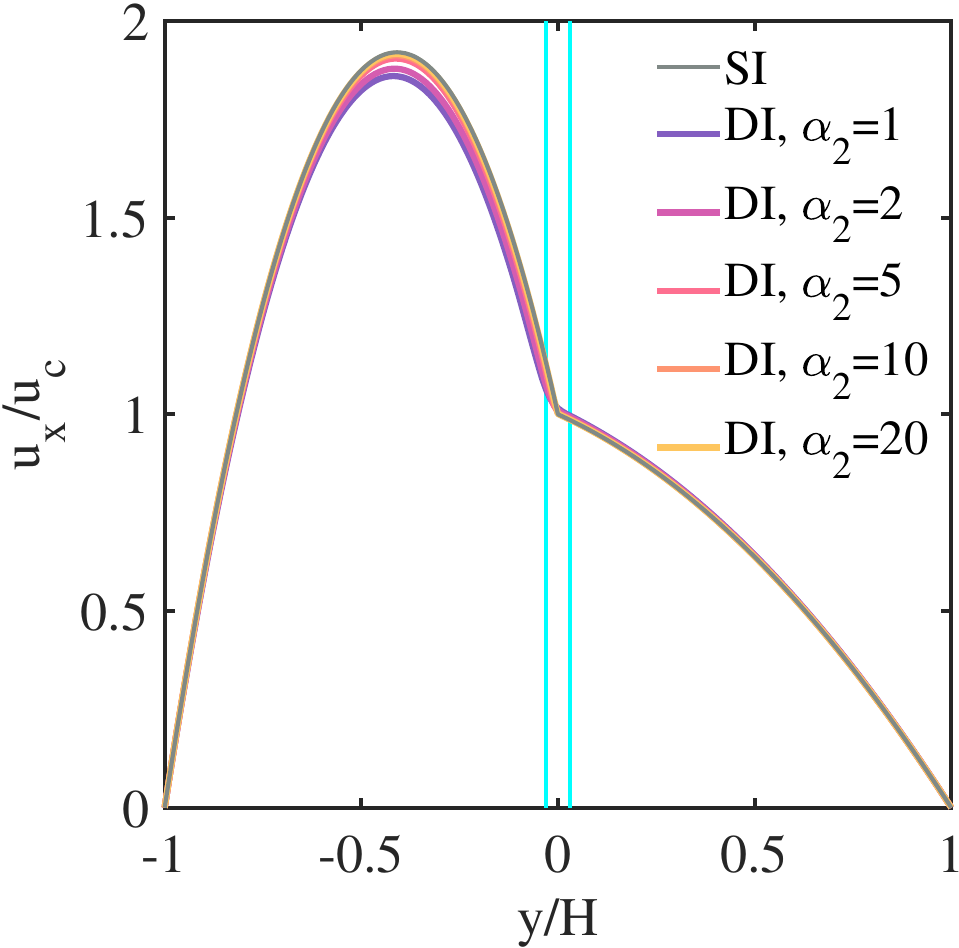}
			\label{fig:nmodelsmu10W006u}
		\end{minipage}
	}
	\centering
	\caption{(a)-(c) are the profiles of $\phi$, $\mu$ at different $W^*$. (d)-(f) are the corresponding analytical velocity profiles. The width of the two light blue lines represents the value of $W^*$. $\mu_{A}$ is set to $1$ without losing generality. $\mu^*=10$ in this case.}
	\label{fig:nmodelsmu10}
\end{figure}

\begin{figure}
	\centering    
	\subfigure[$W^*=0.12$]{
		\begin{minipage}[t]{0.31\linewidth}
			\centering
			\includegraphics[width=0.95\columnwidth,trim={0cm 0cm 0cm 0cm},clip]{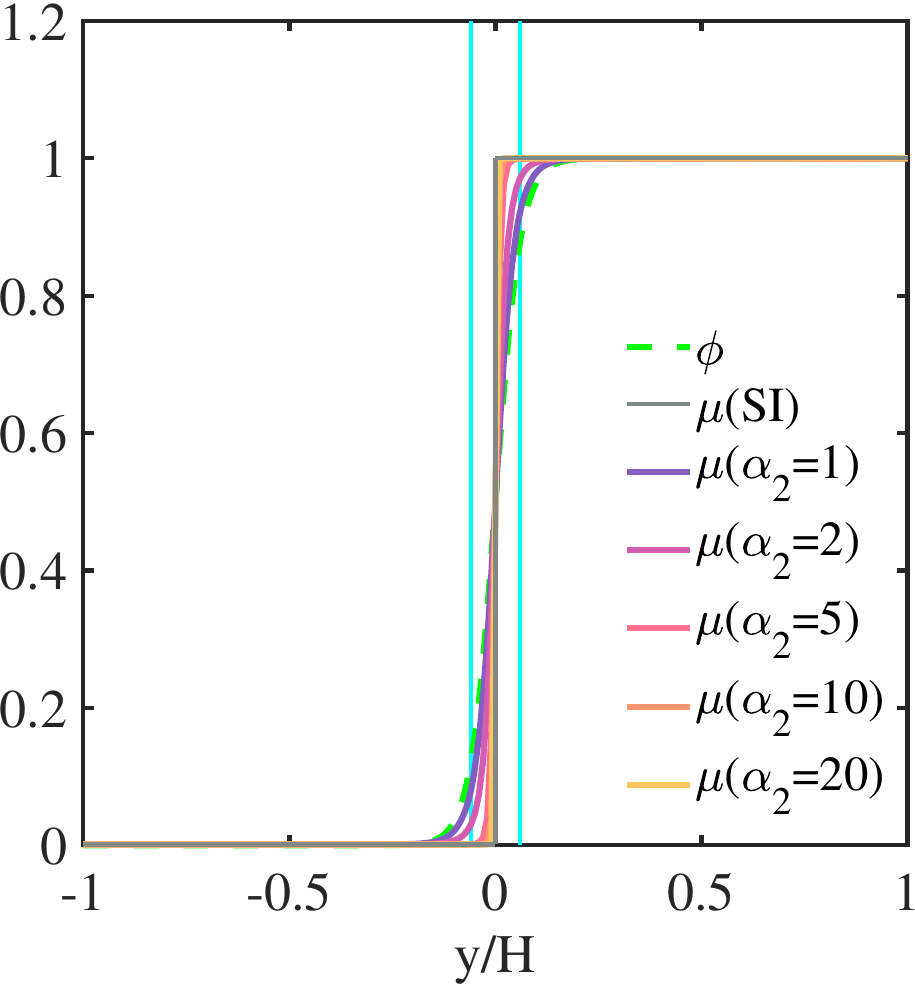}
			\label{fig:nmodelsmu1000W012mu}
		\end{minipage}
	}
	\subfigure[$W^*=0.06$]{
		\begin{minipage}[t]{0.31\linewidth}
			\centering
			\includegraphics[width=0.95\columnwidth,trim={0cm 0cm 0cm 0cm},clip]{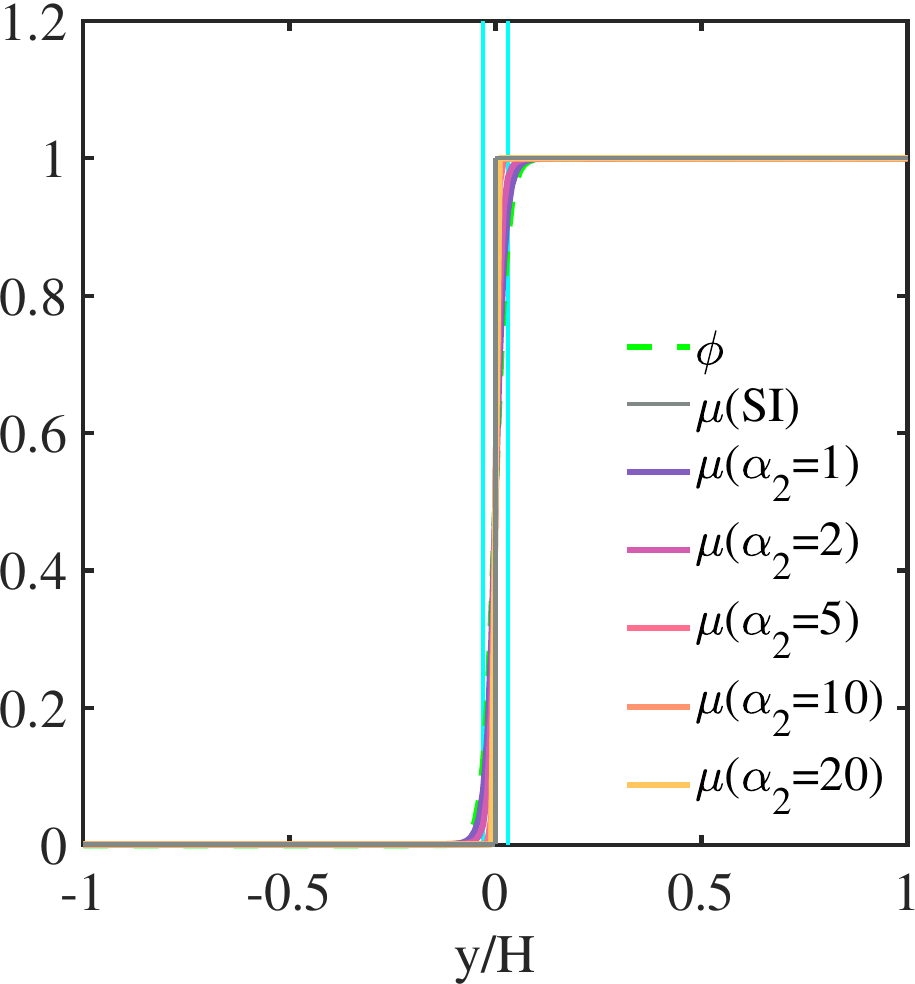}
			\label{fig:nmodelsmu1000W006mu}
		\end{minipage}
	}
	\subfigure[$W^*=0.03$]{
		\begin{minipage}[t]{0.31\linewidth}
			\centering
			\includegraphics[width=0.95\columnwidth,trim={0cm 0cm 0cm 0cm},clip]{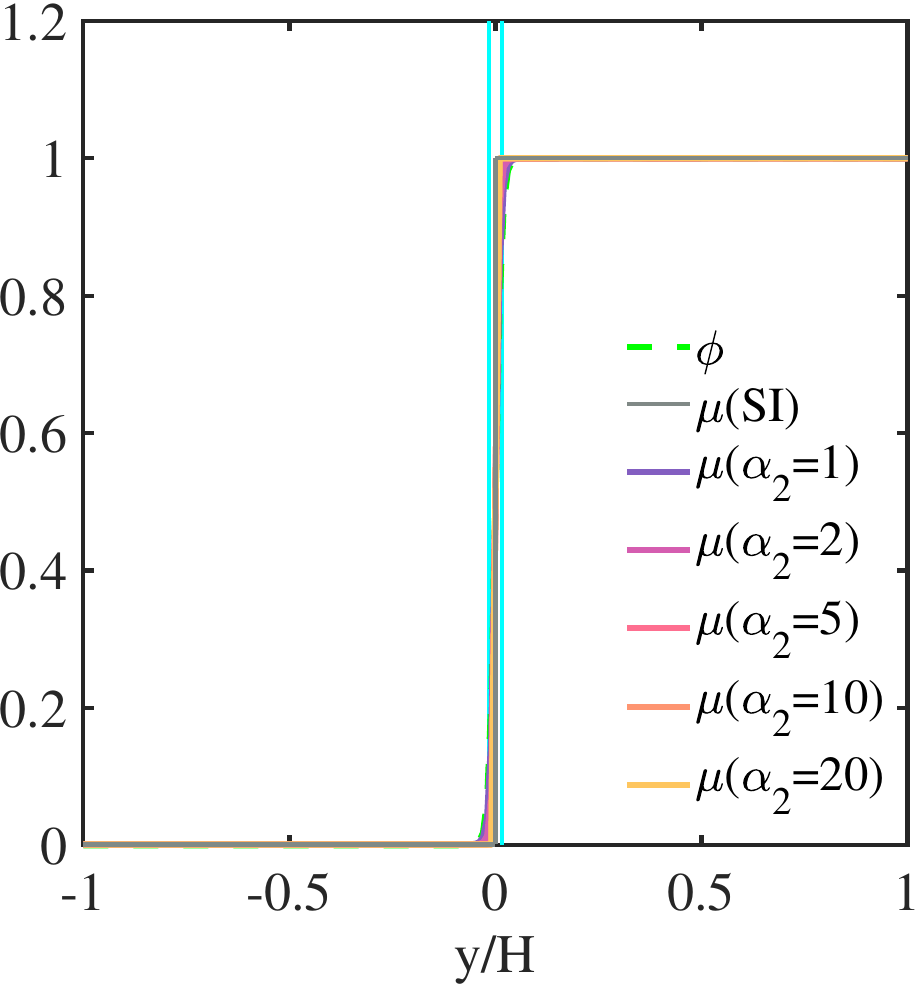}
			\label{fig:nmodelsmu1000W003mu}
		\end{minipage}
	}\\
	\subfigure[$W^*=0.12$]{
		\begin{minipage}[t]{0.31\linewidth}
			\centering
			\includegraphics[width=0.99\columnwidth,trim={0cm 0cm 0cm 0cm},clip]{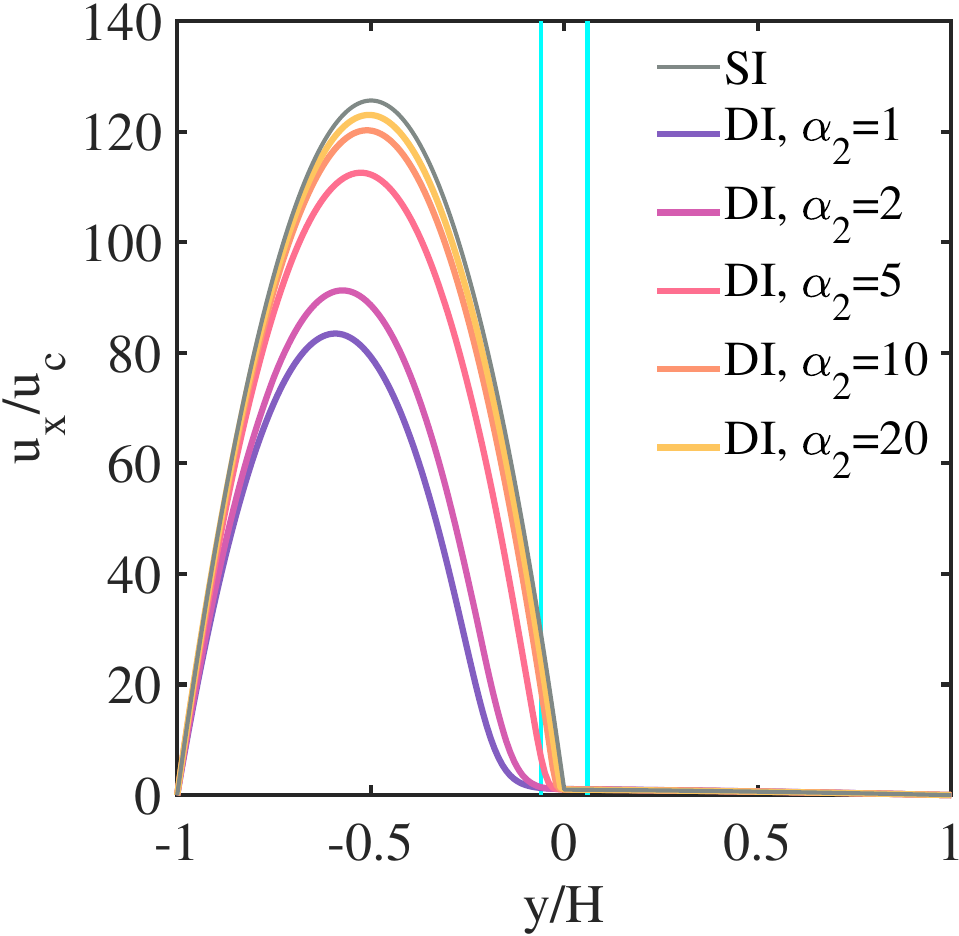}
			\label{fig:nmodelsmu1000W012u}
		\end{minipage}
	}
	\subfigure[$W^*=0.06$]{
		\begin{minipage}[t]{0.31\linewidth}
			\centering
			\includegraphics[width=0.99\columnwidth,trim={0cm 0cm 0cm 0cm},clip]{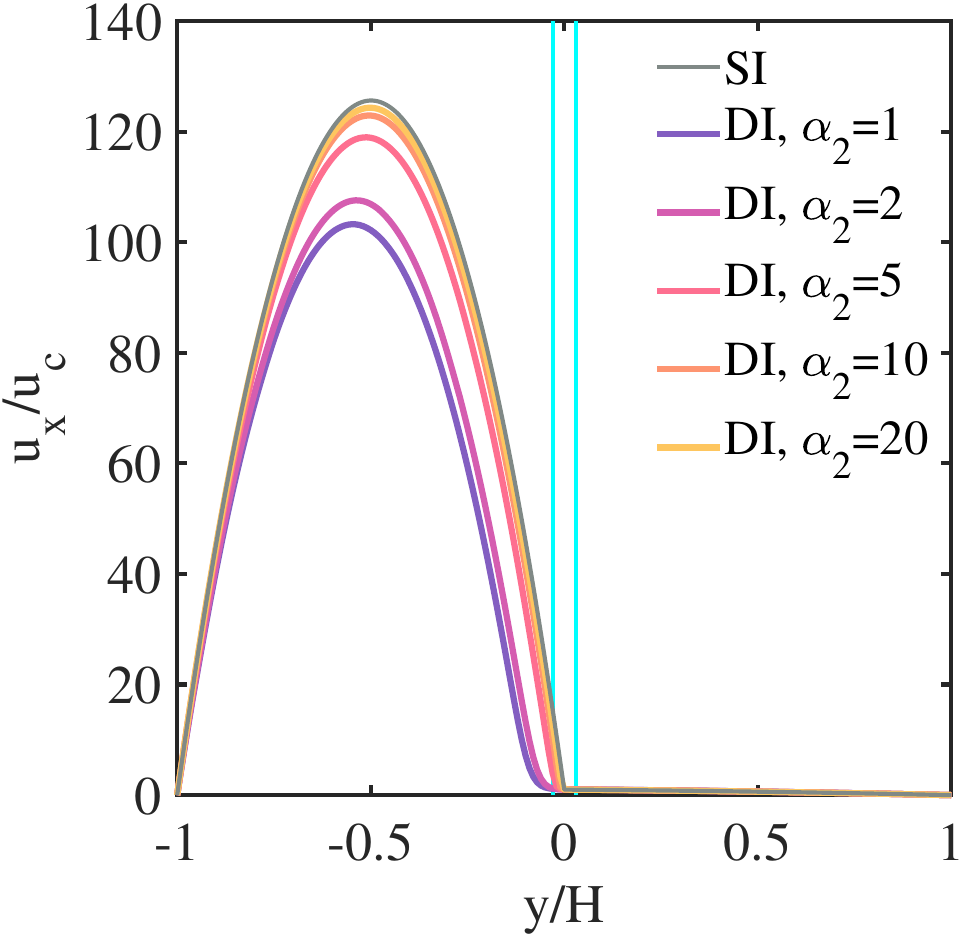}
			\label{fig:nmodelsmu1000W006u}
		\end{minipage}
	}
	\subfigure[$W^*=0.03$]{
		\begin{minipage}[t]{0.31\linewidth}
			\centering
			\includegraphics[width=0.99\columnwidth,trim={0cm 0cm 0cm 0cm},clip]{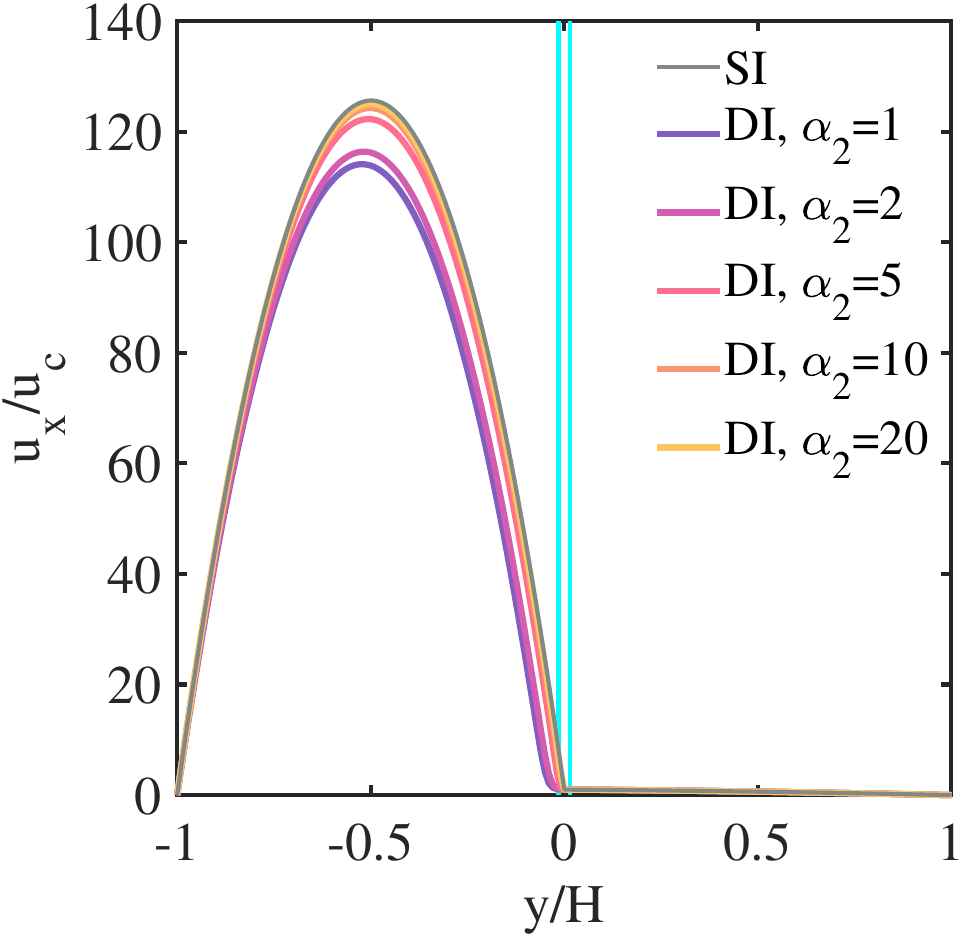}
			\label{fig:nmodelsmu1000W003u}
		\end{minipage}
	}
	\centering
	\caption{(a)-(c) are the profiles of $\phi$, $\mu$ at different $W^*$. (d)-(f) are the corresponding analytical velocity profiles. The width of the two light blue lines represents the value of $W^*$. $\mu_{A}$ is set to $1$ without losing generality. $\mu^*=1000$ in this case.}
	\label{fig:nmodelsmu1000}
\end{figure}

Although we choose M1 in Section~\ref{sec: Guidelines} for less errors in the velocity profiles, the result is not perfect.
In fact, the dynamic viscosity profile of M1 may deviate greatly from that of SI, clearly shown in Fig.~\ref{fig:3modelsmu1000W006mu}, 
which is not acceptable.
In addition, the viscosity profile of M1 is asymmetric with respect to the two-phase interface.
In this section, we want to propose new mixture dynamic viscosity models to better describe the viscosity variation at the two-phase interface.

Based on the viscosity-properties discussions in Section~\ref{Vis}, other new viscosity models can be designed, such as
\begin{subequations}\label{newmumodel}
	\begin{equation}\label{newmumodel1}
	\mu=\left( \mu_{A}^{\alpha_1} \frac{\phi-\phi_{B}}{\phi_{A}-\phi_{B}}+\mu_{B}^{\alpha_1} \frac{\phi-\phi_{A}}{\phi_{B}-\phi_{A}}\right) ^{1/\alpha_1},
	\end{equation}
	and
	\begin{equation}\label{newmumodel2}
	\mu=\frac{\mu_{A}+\mu_{B}}{2}+\frac{\mu_{A}-\mu_{B}}{2} \frac{\operatorname{tanh}\left[ \alpha_2\left(\frac{2\phi-\phi_{A}-\phi_{B}}{\phi_{A}-\phi_{B}}\right)\right] }{\operatorname{tanh}\alpha_2}.
	\end{equation}
\end{subequations}
We call Eqs.~\eqref{newmumodel1} and~\eqref{newmumodel2} the first and second set of dynamic viscosity models.
It is obvious that the commonly used models M1 and M2 are the special cases of the first set of models ($\alpha_1=-1,1$).
For the second set of dynamic viscosity models, we only need to consider $\alpha_2>0$, based on the first property in Appendix~\ref{sec: property of the second set}.
The profiles of the two sets of viscosity models near the interface are shown in Fig.~\ref{fig:alpha}.
It can be seen that the viscosity profiles of the first set are difficult to control (Fig.~\ref{fig:alpha1}).
The viscosity profiles of the second set are symmetrical with respect to the interface $y/W=0$ (Fig.~\ref{fig:alpha2}). 
The second property in Appendix~\ref{sec: property of the second set} also shows this result.
In particular, when $\alpha_2$ is selected a larger value, the result is closer to the SI model, as expected, due to the third property in Appendix~\ref{sec: property of the second set}.
Therefore, the second set is better to describe the dynamic viscosity at the interface.

\begin{table*}[]
	\centering
	\caption{The normalized maximum velocity at different $\mu^*$, $W^*$, and new viscosity models}
	\label{TabMuNew}
	\begin{tabular}{ccccccc}
		\toprule
		$\mu^*$ \multirow{2}{*}{} & \multicolumn{3}{c}{$10$} & \multicolumn{3}{c}{$1000$}\\
		\cmidrule(r){2-4} \cmidrule(r){5-7}
		$u^*_{SI,\max}$ \multirow{2}{*}{} & \multicolumn{3}{c}{$1.920$} & \multicolumn{3}{c}{$125.6$}\\
		\cmidrule(r){2-4} \cmidrule(r){5-7}
		$W^*$ \multirow{2}{*}{} & \multicolumn{1}{c}{$0.24$} & \multicolumn{1}{c}{$0.12$} & \multicolumn{1}{c}{$0.06$} & \multicolumn{1}{c}{$0.12$} & \multicolumn{1}{c}{$0.06$} & \multicolumn{1}{c}{$0.03$}\\
		\midrule
		$u^*_{\max}(\alpha_2=1) $  & $1.713$  & $1.805$ & $1.860$  & $83.48$  & $103.2$ & $114.1$  \\
		$u^*_{\max}(\alpha_2=2) $  & $1.771$  & $1.840$ & $1.879$  & $91.28$  & $107.6$ & $116.4$  \\
		$u^*_{\max}(\alpha_2=5) $  & $1.860$  & $1.889$ & $1.905$  & $112.6$  & $119.0$ & $122.3$  \\
		$u^*_{\max}(\alpha_2=10) $  & $1.891$  & $1.905$ & $1.913$  & $120.2$  & $122.9$ & $124.3$  \\
		$u^*_{\max}(\alpha_2=20) $  & $1.905$  & $1.913$ & $1.914$  & $123.0$  & $124.3$ & $124.3$  \\
		$\left\{\frac{u^*_{\max}(\alpha_2=1)}{ u_{SI,\max}^*}-1\right\} $ & $-11\%$  & $-6.0\%$ & $-3.1\%$  & $-34\%$  & $-18\%$ & $-9.2\%$                           \\
		$\left\{\frac{u^*_{\max}(\alpha_2=2)}{ u_{SI,\max}^*}-1\right\} $ & $-7.8\%$  & $-4.2\%$ & $-2.2\%$  & $-27\%$  & $-14\%$ & $-7.4\%$                           \\
		$\left\{\frac{u^*_{\max}(\alpha_2=5)}{ u_{SI,\max}^*}-1\right\} $ & $-3.2\%$  & $-1.6\%$ & $-0.83\%$  & $-10\%$  & $-5.3\%$ & $-2.7\%$                           \\
		$\left\{\frac{u^*_{\max}(\alpha_2=10)}{ u_{SI,\max}^*}-1\right\} $ & $-1.6\%$  & $-0.79\%$ & $-0.40\%$  & $-4.3\%$  & $-2.2\%$ & $-1.1\%$                           \\
		$\left\{\frac{u^*_{\max}(\alpha_2=20)}{ u_{SI,\max}^*}-1\right\} $ & $-0.78\%$  & $-0.39\%$ & $-0.36\%$  & $-2.1\%$  & $-1.0\%$ & $-1.0\%$                           \\
		\bottomrule
	\end{tabular}\\
	{$\mu^*=\mu_{A}/\mu_{B}$ is the dynamic viscosity ratio, $W^*=W/H$ is the normalized interfacial thickness, $u^*=(\mu_{A}+\mu_{B})u_x/(GH^2)$ is the normalized velocity magnitude, $u^*_{\cdot,\max}$ denotes the analytical results of the maximum normalized velocity magnitude in the specific viscosity model.
	}\\
\end{table*}

Now we check the velocity profiles of the second set.
$\mu^*=10, 1000$ are chosen for these cases.
$W^*=0.24, 0.12, 0.06$ and $W^*= 0.12, 0.06, 0.03$ for $\mu^*=10$ and $\mu^*=1000$, respectively, as shown in Fig.~\ref{fig:nmodelsmu10} and Fig.~\ref{fig:nmodelsmu1000}.
For any $\mu^*$, any $W^*$, and any $\alpha_2$, the dynamic viscosity profile is always symmetric about the two-phase interface.
For the same $\mu^*$ and the same $W^*$, the dynamic viscosity profile in the DI model is closer to that in the SI model, when $\alpha_2$ increase.
These results are all shown in Figs.~\ref{fig:nmodelsmu10W024mu}-\ref{fig:nmodelsmu10W006mu} and Figs.~\ref{fig:nmodelsmu1000W012mu}-\ref{fig:nmodelsmu1000W003mu}.
These again confirm the properties demonstrated in Appendix~\ref{sec: property of the second set}.
In the layered Poiseuille flows problem, the velocity field is determined by the dynamic viscosity field. Therefore, we believe that the larger $\alpha_2$ is, the closer the velocity profile in the DI model is to that in the SI model.
Figs.~\ref{fig:nmodelsmu10W024u}-\ref{fig:nmodelsmu10W006u} and Figs.~\ref{fig:nmodelsmu1000W012u}-\ref{fig:nmodelsmu1000W003u} confirm our expectation.
In addition, when $\mu^*$ is larger, $W^*$ should be smaller, in order to obtain an accurate velocity profile.

Table~\ref{TabMuNew} gives the normalized maximum velocity for these cases.
Again, we apply the absolute values of the relative errors of the maximum velocities to judge the choice of parameters.
For the case $\mu^*=10$, $\alpha_2$ should be chosen at least $5$.
If $\alpha_2=5$, $W^*$ should be smaller than $0.12$, then the relative errors would be smaller than $1.6\%$.
If $\alpha_2=10$, $W^*$ should be smaller than $0.24$, then the relative errors would be smaller than $1.6\%$.
If $\alpha_2=5$, $W^*$ can be larger.
For the case $\mu^*=1000$, $\alpha_2$ should be chosen at least $10$.
If $\alpha_2=10$, $W^*$ should be smaller than $0.03$, then the relative errors would be smaller than $1.1\%$.
If $\alpha_2=20$, $W^*$ should be smaller than $0.06$, then the relative errors would be smaller than $1.0\%$.

\section{Summary and Conclusions}\label{Conclusion}

In this paper,
the general form of macroscopic governing equation of two-phase flow is simplified, to obtain the equation of layered Poiseuille flows.
Then the general analytic solution of the velocity profile of layered Poiseuille flows is obtained by solving this equation.
The dynamic viscosity is an essential factor affecting the velocity profiles for the two-phase Poiseuille flow, while the density has no effect on this problem.
For example, the analytical solution of the velocity profile is always the same as the analytical solution of the single-phase flow with the same dynamic viscosity, as long as the dynamic viscosities of the two phases are the same, regardless of the difference in the density of the two phases.

The profiles of dynamic viscosities are different when the dynamic viscosity models are different, hence the specific analytical solutions need to be determined according to the dynamic viscosity models.
Three different velocity profiles are obtained by substituting the three dynamic viscosity models commonly used in literature into the general form of the analytical solutions.
The PF(AC)-DUGKS algorithm and PF(CH)-DUGKS algorithm are applied to simulate the layered Poiseuille flows with the same parameters, and compared with the analytical solutions. 
It is found that the results of different numerical algorithms and analytic solutions agree well, for all three viscosity models.
For the three dynamic viscosity models, the velocity profiles are different, indicating that the choice of viscosity model has an important influence on the flow field.

In addition to the dynamic viscosity model, the interfacial thickness parameter also affects the results since it would affect the profile of dynamic viscosity. The thinner the interface thickness of the DI model, the closer the flow velocity profile is to the SI model.
The interface thickness should be much thiner for the high viscosity ratio cases.

In order to test which viscosity model among the three models can yield better velocity profile, we use the analytic solution of velocity to calculate the results of different viscosity ratios and different interface thicknesses. The profile figures and the maximum velocity comparison tables show,
the M1 model should be chosen. It is better to choose $W^*\le 0.06$, so that the absolute values of the relative errors are small (less than $0.15\%$) for the $1\le \mu^*\le 1000$ cases in the layered Poiseuille flow problem.

However, the M1 model only ensures that the velocity profiles of the DI model are reasonable compared to that of the SI model,
but the dynamic viscosity profiles of M1 would deviate from that of SI.
We propose a set of viscosity models which can ensure the rationality of both viscosity profiles and velocity profiles.
This set of viscosity models ensures that the viscosity profiles are symmetric with respect to the two-phase interface.
At the same time, there is a tunable positive parameter $\alpha_2$ in this set of models. When $\alpha_2$ is larger, the viscosity profiles and velocity profiles in the DI model are more similar to those in the SI model.
$\alpha_2$ should be chosen at least $10$ for the $1\le \mu^*\le 1000$ cases.
For ($\alpha_2=10$, $W^*\le 0.03$) and ($\alpha_2=20$, $W^*\le 0.06$),
the absolute values of the relative errors are sufficient small
(less than $1.1\%$).
We suggest to use this new set of mixture dynamic viscosity models to simulate the two-phase flow in the DI model.

The DI model was used to simulate the layered Poiseuille flow, but the results were compared with the analytical solution of SI model in the previous studies.
Our results show that, when the viscosity ratio is large, the analytical solutions of DI model and SI model are quite different.
Therefore, the simulation results by the DI model should be compared with the analytical solution of the DI model introduced in this work, in order to better quantify the numerical error.
The model error is the error between the DI model and the SI model, which cannot be eliminated by the numerical algorithms.
The two errors should be separated out, and a new set of mixture dynamic viscosity models should be used.


\begin{acknowledgments}
	This work has been supported by the National Numerical Wind Tunnel program, 
	the National Natural Science Foundation of China (NSFC award numbers T2250710183 and U2241269), NSFC Basic Science Center Program (Award number 11988102), 
	Guangdong Provincial Key Laboratory of Turbulence Research and
	Applications (2019B21203001), Guangdong-Hong Kong-Macao Joint Laboratory for Data-Driven Fluid Mechanics and Engineering Applications (2020B1212030001), Shenzhen Science and Technology Program (Grant No. KQTD20180411143441009),
	Start-up Fund for Talent Introduction at Guangzhou Jiaotong University (K42024035),
	and Tertiary Education Scientific research project of Guangzhou Municipal Education Bureau (2024312534). 
	Computing resources are provided by
	the Center for Computational Science and Engineering of Southern University
	of Science and Technology. The authors also wish to thank Dr. Tao Chen, Dr. Shengqi Zhang, Mr. Mingyu Su 
	for helpful discussions.
\end{acknowledgments}

\appendix

\section{Discussion on the convective term in the interface equation in phase field model} \label{sec: convective}

In phase field model, the interface equation should show the evolution of order parameter under the action of driving factors.
The evolution of order parameter is described by the material derivative of order parameter, $D\phi/Dt$.
The action of driving factors can be represented by $J$ in general.
Hence,
\begin{equation}\label{Dphi}
\frac{D \phi}{D t}=J
\quad\text{or}\quad 
\frac{\partial \phi}{\partial t}+\boldsymbol{u}\cdot\nabla \phi=J.
\end{equation}
The term $\boldsymbol{u}\cdot\nabla \phi$ can be regard as the physical convective term in the interface equation.
However, this is a non-conserved form and is sometimes not benefit from numerical simulation and research.
We have two ways to transform $\boldsymbol{u}\cdot\nabla \phi$ to the conserved form $\nabla\cdot(\phi\boldsymbol{u}) $.
One is to assume velocity divergence free in the whole domain,
$\nabla\cdot\boldsymbol{u}=0 $.
The other one is to focus on the material derivative of $\phi/\rho$, but not $\phi$.
Based on
\begin{equation}\label{phirho}
\frac{\partial \phi}{\partial t}+\nabla \cdot(\phi \boldsymbol{u})=J
\quad\text{and}\quad 
\frac{\partial \rho}{\partial t}+\nabla \cdot(\rho \boldsymbol{u})=0,
\end{equation}
it is easy to prove
\begin{equation}
\frac{D}{Dt}\left(\frac{\phi}{\rho} \right) =\frac{J}{\rho}.
\end{equation}
Therefore, the convection terms $\boldsymbol{u}\cdot\nabla \phi$ and $\nabla\cdot(\phi\boldsymbol{u}) $ is corresponding to the material derivatives of $\phi$ and $\phi/\rho$, respectively. They are equivalent to each other if $\nabla\cdot\boldsymbol{u}=0 $.

\section{The relationship between continuity equation and velocity nondivergence condition in phase field model} \label{sec: conti}

In a single phase flow, the continuity equation can be consistent with the incompressible condition.
In a two-phase flow based on the phase field model,
the main physical quantity of the continuity equation is density, which is completely determined by the order parameter, $\rho\left(\phi \right)$.
The linear function is used in the literature, and the continuity equation is always contradictory to the solenoidal velocity field.
Is it possible to design a functional relation $\rho\left(\phi \right)$ that satisfies both the continuity equation and the velocity divergence-free condition?
We apply Theorems~\ref{mass and div} and~\ref{contrad} to give the conclusion.
It is worth noting that the bulk region of each phase is almost single-phase flow, so the region that does not meet the conditions is mainly near the interface.

\begin{theorem}\label{mass and div} 
	
	In the phase field model, the evolution of order parameter $\phi$ satisfies
	\begin{equation}\label{phieqs}
	\frac{\partial \phi}{\partial t}+\boldsymbol{u}\cdot\nabla \phi=J
	\quad\text{or}\quad 
	\frac{\partial \phi}{\partial t}+\nabla \cdot(\phi \boldsymbol{u})=J.
	\end{equation}
	The density $\rho$ is the function of $\phi$,
	\begin{equation}
	\rho=\rho\left(\phi \right) .
	\end{equation}
	Then
	\begin{enumerate}
		\item If $\nabla \cdot\boldsymbol{u}=0$ and $\partial_t \rho+\nabla \cdot(\rho \boldsymbol{u})=0$, then $\rho=\text{const.}$ or $J =0$.
		\item If $\rho=\text{const.}$, then $\nabla \cdot\boldsymbol{u}=0$ and $\partial_t \rho+\nabla \cdot(\rho \boldsymbol{u})=0$
		are the sufficient and necessary conditions for each other.
		\item Under the evolution equation $\partial_t \phi+\boldsymbol{u}\cdot\nabla \phi=J$, if $J =0$, then $\nabla \cdot\boldsymbol{u}=0$ and $\partial_t \rho+\nabla \cdot(\rho \boldsymbol{u})=0$ are the sufficient and necessary conditions for each other;
		\\
		Under the evolution equation $\partial_t \phi+\nabla \cdot(\phi \boldsymbol{u})=J$,
		if $J =0$ and ${\phi}/{\rho}\neq const.$, then $\nabla \cdot\boldsymbol{u}=0$ and $\partial_t \rho+\nabla \cdot(\rho \boldsymbol{u})=0$ are the sufficient and necessary conditions for each other.
	\end{enumerate}
	
\end{theorem}

\begin{proof} 
	
	1. If
	\begin{equation}
	\nabla \cdot\boldsymbol{u}=0,
	\quad
	\frac{\partial \rho}{\partial t}+\nabla \cdot(\rho \boldsymbol{u})=0.
	\end{equation}
	Then no matter which evolution equation in Eq.~\eqref{phieqs} the order parameter satisfies, there is always
	\begin{equation}
	\frac{d \rho}{d \phi}J =\frac{d \rho}{d \phi}\left(\frac{\partial \phi}{\partial t}+\boldsymbol{u}\cdot\nabla \phi \right) =\frac{\partial \rho}{\partial t}+\boldsymbol{u}\cdot\nabla \rho =0.
	\end{equation}
	Therefore,
	\begin{equation}
	\rho\left(\phi \right)=\text{const.}\quad\text{or}\quad J =0.
	\end{equation}
	
	Comment: If the densities of the two phases are not the same, then $\rho\left(\phi \right)\neq\text{const.}$ near the two-phase interface;
	In addition, near the interface, $J =0$ only in some special cases (such as a flat interface at equilibrium state), generally $J \neq0$.
	Therefore, $\nabla \cdot\boldsymbol{u}=0$ and $\partial_t \rho+\nabla \cdot(\rho \boldsymbol{u})=0$ generally cannot be held at the same time when the two phases have different densities, no matter what model $\rho\left(\phi \right)$ takes.

	2. If $\rho=\text{const.}$, then $\nabla \cdot\boldsymbol{u}=0$ and $\frac{\partial \rho}{\partial t}+\nabla \cdot(\rho \boldsymbol{u})=0$ are the sufficient and necessary conditions for each other, obviously.
	
	Comment: If the densities of the two phases are the same, then $\rho=\text{const.}$ due to the Equal viscosity property (2) in subsection~\ref{Vis}. Hence $\nabla \cdot\boldsymbol{u}=0$ and $\partial_t \rho+\nabla \cdot(\rho \boldsymbol{u})=0$ are both satisfied. 
	
	3. If $J =0$, we have
	\begin{equation}\label{J0}
	\frac{\partial \phi}{\partial t}+\nabla \cdot(\phi \boldsymbol{u})=0\quad\text{or}\quad \frac{\partial \phi}{\partial t}+\boldsymbol{u}\cdot\nabla \phi=0.
	\end{equation}    
	(1) If $\nabla \cdot\boldsymbol{u}=0$, then the two equations in \eqref{J0} are the same, and
	\begin{equation}
	\frac{\partial \rho}{\partial t}+\nabla \cdot(\rho \boldsymbol{u})=\frac{d \rho}{d \phi}\left(\frac{\partial \phi}{\partial t}+\boldsymbol{u}\cdot\nabla \phi \right)+\rho\nabla\cdot\boldsymbol{u}=0.
	\end{equation}
	(2) If $\partial_t \rho+\nabla \cdot(\rho \boldsymbol{u})=0$, the two results are different. 
	
	When $\partial_t \phi+\boldsymbol{u}\cdot\nabla \phi=0$, we can obtain
	\begin{equation}
	\rho\nabla\cdot\boldsymbol{u}=
	\frac{\partial \rho}{\partial t}+\nabla \cdot(\rho \boldsymbol{u})-\frac{d \rho}{d \phi}\left(\frac{\partial \phi}{\partial t}+\boldsymbol{u}\cdot\nabla \phi \right)=0.
	\end{equation}
	Since $\rho\neq 0$, we have $\nabla\cdot\boldsymbol{u}=0$.
	
	When $\partial_t \phi+\nabla \cdot(\phi \boldsymbol{u})=0$, we can obtain
	\begin{equation}
	\left( \rho-\phi\frac{d \rho}{d \phi}\right) \nabla\cdot\boldsymbol{u}=
	\frac{\partial \rho}{\partial t}+\nabla \cdot(\rho \boldsymbol{u})-\frac{d \rho}{d \phi}\left[\frac{\partial \phi}{\partial t}+\nabla \cdot\left( \phi \boldsymbol{u}\right)  \right]=0.
	\end{equation}
	Hence the following condition is needed,
	\begin{equation}
	\frac{\phi}{\rho}\neq const.
	\quad i.e. \quad
	\rho\left(\phi \right)\neq C\phi,
	\end{equation}
	to obtain $\nabla\cdot\boldsymbol{u}=0$.
	Here $C$ is any constant.
\end{proof} 
From Theorem~\ref{mass and div}, we can obtain Theorem~\ref{contrad} directly.
\begin{theorem}\label{contrad}     
	In the phase field model ($J\neq 0$), if the densities of the two phases are different, then the continuity equation and divergence-free velocity field are contradictory.
	If the densities of the two phases are the same, then the continuity equation and the divergence-free velocity field are equivalent to each other. This conclusion is independent of the expression for $\rho\left(\phi \right)$.
\end{theorem} 

Comment: Since the relation between the continuity equation and the divergence-free velocity field does not depend on the expression of $\rho\left(\phi \right)$, it is more convenient to select $\rho\left(\phi \right)$ as a linear function from the point of view of mass conservation (the conservation of total $\rho$ is equivalent to the conservation of total $\phi$ due to the linear relation).
Therefore, $\rho\left(\phi \right)$ is suitable for linear functions, and there is no need to consider other models like the way $\mu\left(\phi \right)$ do.

\section{$f^*$ and $g^*$ of three dynamic models in two special cases} \label{sec: App f*g*}

In this appendix, we apply two special cases to further demonstrate our theoretical results in Subsection~\ref{subsec: Governing equation}.
\begin{enumerate}
	\item $\mu^*=1$, {\it i.e.}, the same dynamic viscosity case.
	
	Based on Table~\ref{Tabfgstar},  
	$f^{*},g^{*}$ in all the three models become
	\begin{equation}
	f^{*}=0,\quad g^{*}=-2.
	\end{equation}
	Eq.~\eqref{mom4} becomes
	\begin{equation}
	\left\{
	\begin{aligned}
	&\frac{d^{2} u^{*}}{d y^{* 2}}=-2, & -1< y^{*}< 1, \\
	& u^{*}=0, & y^{*}=\pm 1.
	\end{aligned}
	\right.
	\end{equation}
	It is consistent with Eq.~\eqref{momsamemu} with boundary conditions, which verifies the compatibility and rationality of the formulas in Table~\ref{Tabfgstar}.

	\item $W^*\rightarrow 0^{+}$, {\it i.e.}, the SI limit case.
	
	
	Based on Table~\ref{Tabfgstar}, we can prove that
	\begin{equation}
	\begin{aligned} \lim _{W^{*} \rightarrow 0^{+}} f^{*}
	=\left\{\begin{array}{ccc}{0} & {,} & {y^{*} \neq 0 \quad \text{or} \quad\left(y^{*}=0, \mu^{*}=1\right),} \\ {-\infty} & {,} & {\left(y^{*}=0,\quad \mu^{*}>1\right),} \\ {+\infty} & {,} & {\left(y^{*}=0, \quad 0<\mu^{*}<1\right).}\end{array}\right.
	\end{aligned}
	\end{equation}
	Thus, the SI limit of $f^*$ are the same for all the three different dynamic viscosity models.
	It is observed that $f^*$ is always $0$ in the smooth velocity profiles, while $f^*$ is the singularity where the velocity profiles are not differentiable.

	
	Similarly, $g^*$ in the three models can be solved.
	\begin{equation}
	\lim _{W^{*} \rightarrow 0^{+}} g_{M1}^{*}=\left\{\begin{array}{ccc}{-\mu^{*}-1} & {,} & {y^{*}>0,} \\ {-\frac{\left(\mu^{*}+1\right)^2}{2 \mu^{*}}} & {,} & {y^{*}=0,} \\ {-1-\frac{1}{\mu^{*}}} & {,} & {y^{*}<0.}\end{array}\right.
	\end{equation}
	\begin{equation}
	\lim _{W^{*} \rightarrow 0^{+}} g_{M2}^{*}=\left\{\begin{array}{ccc}{-\mu^{*}-1} & {,} & {y^{*}>0,} \\ {-2} & {,} & {y^{*}=0,} \\ {-1-\frac{1}{\mu^{*}}} & {,} & {y^{*}<0.}\end{array}\right.
	\end{equation}
	\begin{equation}
	\lim _{W^{*} \rightarrow 0^{+}} g_{M3}^{*}=\left\{\begin{array}{ccc}{-\mu^{*}-1} & {,} & {y^{*}>0,} \\ {-\frac{\mu^{*}+1}{\sqrt{\mu^{*}} }} & {,} & {y^{*}=0,} \\ {-1-\frac{1}{\mu^{*}}} & {,} & {y^{*}<0.}\end{array}\right.
	\end{equation}
	Here the subscripts $M1,M2$ and $M3$ represent the corresponding three models, respectively.	
	It can be seen that the $g^*$ of different models are the same inside the bulk region, but different at the two-phase interface ($y^*=0$).

	
	$f^*$ produces a singularity at the interface, and the value of $g^*$ at the interface depends on the viscosity model. If the interface position is excluded, the following momentum equation can be derived from the above models.
	\begin{equation}
	\left\{\begin{aligned}&{\frac{d^{2} u^{*}}{d y^{* 2}}=\left\{\begin{aligned}&{-\mu^{*}-1,} & {0<y^{*}<1,} \\ &{-1-\frac{1}{\mu^{*}},} & {-1<y^{*}<0,}\end{aligned}\right.} \\ &{u^{*}=0},\quad {y^{*}=\pm 1.}\end{aligned}\right.
	\end{equation}
	or
	\begin{equation}
	\left\{\begin{aligned}&{\frac{d^{2} u_x}{d y^{ 2}}=\left\{\begin{aligned}&{-\frac{G}{\mu_A},} & {0<y<H,} \\ &{-\frac{G}{\mu_B},} & {-H<y<0,}\end{aligned}\right.} \\ &{u_x=0},\quad {y=\pm H.}\end{aligned}\right.
	\end{equation}
	This equation fully recovers Eq.~\eqref{PoiseuilleSI} in the bulk region.
	The SI limit case verifies the rationality of the formulas in Table~\ref{Tabfgstar} again.	
	In addition, this limit case also shows that the velocity profiles obtained by different viscosity models may be different, mainly due to the different results at the two-phase interface.

\end{enumerate}

\section{The first solution of layered Poiseuille flow in the DI model} \label{sec: App first solution}
In this appendix, we provide some details for the derivation of Eqs.~\eqref{integrate1} and~\eqref{integrate2}.
First, integrating the momentum equation in Eq.~(\ref{general scalar}) twice, we have
\begin{equation}\label{firstmethod}
u_x=-G\int_a^y \frac{Y}{\mu\left(Y ; W, \mu_{A}, \mu_{B}\right)} dY+c\int_b^y \frac{1}{\mu\left(Y ; W, \mu_{A}, \mu_{B}\right)} dY+d,
\end{equation}
where $a$ and $b$ are any constants between $-H$ and $H$, $c$ and $d$ are obtained by the boundary condition,
\begin{equation}\label{constantc}
c=\frac{G \int_{-H}^{H} \frac{Y}{\mu} d Y
}
{ \int_{-H}^{H} \frac{1}{\mu} d Y},
\end{equation}
\begin{equation}\label{constantd}
d=\frac{G\cdot 
	\left(\int_{a}^{H} \frac{Y}{\mu} d Y \int_{-H}^{H} \frac{1}{\mu} d Y-\int_{-H}^{H} \frac{Y}{\mu} d Y \int_{b}^{H} \frac{1}{\mu} d Y\right) 
}
{ \int_{-H}^{H} \frac{1}{\mu} d Y}.
\end{equation}
Substituting Eqs.~\eqref{constantc} and~\eqref{constantd} into Eq.~\eqref{firstmethod}, the analytical solution is
\begin{equation}\label{first solution}
\begin{aligned}
u_x=&\cfrac{G}{\displaystyle{\int_{-H}^{H} \cfrac{1}{\mu} d Y}}\cdot\left\{\begin{aligned}& 
-\int_{-H}^{H} \frac{1}{\mu} d Y\int_{a}^{y} \frac{Y}{\mu} d Y+ \int_{-H}^{H} \frac{Y}{\mu} d Y\int_{b}^{y} \frac{1}{\mu} d Y
\\&+\int_{a}^{H} \frac{Y}{\mu} d Y \int_{-H}^{H} \frac{1}{\mu} d Y-\int_{-H}^{H} \frac{Y}{\mu} d Y \int_{b}^{H} \frac{1}{\mu} d Y \end{aligned}\right\}
\\=&
\cfrac{G}{\displaystyle{\int_{-H}^{H} \cfrac{1}{\mu} d Y}}\cdot\left\{ 
\int_{-H}^{H} \frac{1}{\mu} d Y\int_{y}^{H} \frac{Y}{\mu} d Y+ \int_{-H}^{H} \frac{Y}{\mu} d Y\int_{H}^{y} \frac{1}{\mu} d Y
\right\}. 
\end{aligned}
\end{equation}  
Eq.~\eqref{first solution} can also be expressed as
\begin{equation}
u_x=\frac{G}{I(H) - I(-H)}\cdot\left\{\begin{array}{l} 
\left[  I(H) - I(-H)\right] \left[I_y(H)-I_y(y) \right]  
\\+\left[  I_y(H) - I_y(-H)\right] \left[I(y)-I(H) \right] \end{array}\right\},
\end{equation}
\begin{equation}
u_x=G\cdot\left\{\begin{array}{l} 
\left[I_y(H)-I_y(y) \right]  
+\cfrac{  I_y(H) - I_y(-H)}{I(H) - I(-H)}\left[I(y)-I(H) \right] \end{array}\right\},
\end{equation}    
\begin{equation}
u_x=\frac{G\cdot\left\{\begin{array}{l} 
	\left[  I(H) - I(-H)\right] \left[I_y(H)-I_y(y) \right]  
	+\left[  I_y(H) - I_y(-H)\right] \left[I(y)-I(H) \right] \end{array}\right\}}{I(H) - I(-H)},
\end{equation}
where $I(y) = \int_a^y \frac{dY}{\mu(Y)}$ is a certain primitive function of $\frac{1}{\mu(y)}$, $I_y(y) = \int_a^y \frac{YdY}{\mu(Y)}$ is a certain primitive function of $\frac{y}{\mu(y)}$.

\section{Expressions of key integrals for velocity profile}\label{sec: int mu}
Eq.~\eqref{uPoiseuilleMethod1} demonstrates that $I(y)$ and $I_y(y)$ are the determinants of $u_x$ profiles. 
In this appendix, both $I(y)$ and $I_y(y)$ in different viscosity models are provided.

SI:
\begin{equation}
\mu 
= \frac{1}{2} \left[ ( 1 + sgn(y) )\mu_A + ( 1 - sgn(y) )\mu_B \right],
\end{equation}
\begin{equation}
\frac{1}{\mu}
= \frac{2}{ \mu_A } \frac{ r }{ r [ 1 + sgn(y) ] + [ 1 - sgn(y) ] },
\end{equation}
\begin{equation}
I(y) 
= \frac{y}{ 2 \mu_A } \left\{ [ 1 + sgn(y) ] + r [ 1 - sgn(y) ] \right\},
\end{equation}
\begin{equation}
\int I(y) dy 
= \frac{y^2}{ 4 \mu_A }  \left\{ [ 1 + sgn(y) ] + r [ 1 - sgn(y) ] \right\},
\end{equation}
\begin{equation}
u(y)= \frac{G}{4\mu_A (r+1)}\left\{ 4rH^2 - y\left[ y(r+1) + H(r-1) \right]\left[ ( 1 + sgn(y) ) + r ( 1 - sgn(y) ) \right] \right\}.
\end{equation}

M1:
\begin{equation}
\int_0^y \frac{Y}{\mu} dY=\frac{y^{2}}{2 \mu_{B}}+\frac{\left(\mu_{B}-\mu_{A}\right)W}{4\mu_{A}\mu_{B}}\left[y \ln \left(1+e^{\frac{4 y}{W}}\right)-I_{1}(y ; W)\right],
\end{equation}
where
\begin{equation}
\begin{aligned}
I_1\left(y;W\right):=&
\int_0^y \ln \left(1+e^{\frac{4Y}{W}}\right) dY
\\=&\left\{\begin{array}{ll}
{\frac{2y^{2}}{W}+\frac{\pi^{2}W}{48}+\frac{W}{4}Li_2\left(-e^{-\frac{4 y}{W}}
	\right),} & {y\ge 0,} \\ {-\frac{\pi^{2}W}{48}-\frac{W}{4}Li_2\left(-e^{\frac{4 y}{W}}
	\right),} & {y\le 0.}\end{array}\right.
\end{aligned}
\end{equation}
\begin{equation}
\int_0^y \frac{1}{\mu} dY=\frac{1}{4\mu_{A}\mu_{B} }\left[2\left(\mu_{B}+\mu_{A}\right)y+\left(\mu_{B}-\mu_{A}\right) W\ln \left(\cosh\left(\frac{2y}{W} \right) \right)\right].
\end{equation}

M2:
\begin{equation}
\int_0^y \frac{Y}{\mu} dY=\frac{y^{2}}{2 \mu_{B}}+\frac{\left(\mu_{B}-\mu_{A}\right)W}{4\mu_{A}\mu_{B}}\left[y \ln \left(1+\frac{\mu_{A}}{\mu_{B}}e^{\frac{4 y}{W}}\right)-I_{2}(y ; W,\mu_{A},\mu_{B})\right],
\end{equation}
where
\begin{equation}
I_2\left(y;W,\mu_{A},\mu_{B}\right):=
\int_{0}^y \ln \left(1+\frac{\mu_A}{\mu_B}e^{\frac{4Y}{W}}\right) dY.
\end{equation}
\begin{equation}
\int_0^y \frac{1}{\mu} dY=\frac{1}{4\mu_{A}\mu_{B} }\left[2\left(\mu_{B}+\mu_{A}\right)y+\left(\mu_{B}-\mu_{A}\right) W\ln \left(\frac{\mu_{A}e^{\frac{2y}{W}}+\mu_{B}e^{-\frac{2y}{W}}}{\mu_{A}+\mu_{B}} \right)\right].
\end{equation}

M3:
\begin{equation}
\int_0^y \frac{Y}{\mu} dY=\frac{1}{\sqrt{\mu_{A} \mu_{B}}} \int_{0}^{y}\left(\frac{\mu_{B}}{\mu_{A}}\right)^{\frac{1}{2}\tanh\left(\frac{2 Y}{W}\right)}  Y d Y,
\end{equation}
\begin{equation}
\int_0^y \frac{1}{\mu} dY=\frac{1}{\sqrt{\mu_{A} \mu_{B}}} \int_{0}^{y}\left(\frac{\mu_{B}}{\mu_{A}}\right)^{\frac{1}{2}\tanh\left(\frac{2 Y}{W}\right)}   d Y.
\end{equation}

\section{Derivation of velocity solutions in different viscosity models} \label{CDI}
The basic idea is to substitute $f^{*}$ and $g^{*}$ of Table~\ref{Tabfgstar} into Eq.~\eqref{analyticalustar}, integrate to get the general solutions, and then substitute the non-slip velocity boundary conditions to find the undetermined constants.

For the first viscosity model M1, there is
\begin{equation}
e^{\int f^{*} d y^{*}}=\frac{1}{\left(\mu^{*}+1\right)+\left(\mu^{*}-1\right) \operatorname{tanh}\left(\frac{2 y^{*}}{W^{*}}\right)},
\end{equation}
then
\begin{equation}\label{uM1starCD}
\begin{aligned}
u_{M1}^{*}=-\left(\mu^{*}+1\right)\cdot\left\{\begin{array}{l} \frac{y^{* 2}}{2\mu^*}+\frac{\left(\mu^{*}-1\right) W^{*} y^{*}}{4 \mu^{*}} \ln \left(1+e^{\frac{4 y^{*}}{W^{*}}}\right)
-\frac{\left(\mu^{*}-1\right) W^{*} }{4 \mu^{*}} I_1\left(y^*;W^*\right)
\\+C_{1}^{*}\left[\left(\mu^{*}+1\right) y^{*}+ \frac{\left(\mu^{*}-1\right)W^*}{2}  \ln \left(\operatorname{cosh}\left(\frac{2 y^{*}}{W^{*}}\right)\right)\right]+D_{1}^{*} \end{array}\right\}
\end{aligned},
\end{equation}
where
\begin{equation}
\begin{aligned}
I_1\left(y;W\right):=&
\int_0^y \ln \left(1+e^{\frac{4x}{W}}\right) dx
\\=&\left\{\begin{array}{ll}
{\frac{2y^{2}}{W}+\frac{\pi^{2}W}{48}+\frac{W}{4}Li_2\left(-e^{-\frac{4 y}{W}}
	\right),} & {y\ge 0,} \\ {-\frac{\pi^{2}W}{48}-\frac{W}{4}Li_2\left(-e^{\frac{4 y}{W}}
	\right),} & {y\le 0.}\end{array}\right.
\end{aligned}
\end{equation}
Here the polylog function is defined by
\begin{equation}
Li_s\left(z
\right):=\sum_{m=1}^{\infty}
\frac{z^m}{m^s}.
\end{equation}

Substituting the boundary condition $u^*(y^*=\pm 1)=0$ into Eq.~\eqref{uM1starCD}, then
\begin{subequations}\label{C1D1}
	\begin{equation}
	C_{1}^{*}=\frac{\left(1-\mu^{*}\right) W^{*}}{8 \mu^{*}\left(\mu^{*}+1\right)}\left[\ln \left(2+2\operatorname{cosh}\left(\frac{4}{W^{*}}\right)\right)+\frac{2}{W^*}-2I_1\left(1;W^{*}\right)
	\right],
	\end{equation}
	\begin{equation}
	D_{1}^{*}=
	-\frac{\mu^*+1}{4\mu^*}
	+
	\left\{\begin{array}{l}
	\frac{(\mu^*-1)^2W^{*2}}{16\mu^*(\mu^*+1)}\ln \left(\operatorname{cosh}\left(\frac{2}{W^{*}}\right)\right)
	\\\cdot
	\left[\ln \left(2+2\operatorname{cosh}\left(\frac{4}{W^{*}}\right)\right)+\frac{2}{W^*}-2I_1\left(1;W^{*}\right)
	\right]
	\end{array}\right\}.
	\end{equation}
\end{subequations}
In the end, the constant coefficient result Eq.~\eqref{C1D1} is substituted back to Eq.~\eqref{uM1starCD}, the final analytical expression of dimensionless velocity can be obtained, that is Eq.~\eqref{uM1star}.

Similarly, for M2, there is
\begin{equation}
e^{\int f^{*} d y^{*}}=\left(\mu^{*}+1\right)-\left(\mu^{*}-1\right) \operatorname{tanh}\left(\frac{2 y^{*}}{W^{*}}\right),
\end{equation}
then
\begin{equation}\label{uM2starCD}
u^{*}_{M2}=-\left(\mu^{*}+1\right)\cdot\left\{\begin{array}{l} \frac{y^{* }\left(y^{*}+C_2^{*}\right)}{2\mu^{*}}+\frac{\left(\mu^{*}-1\right) W^{*} \left(2y^{*}+C_2^{*}\right)}{8 \mu^{*}} \ln \left(1+\frac{1}{\mu^{*}}e^{\frac{4 y^{*}}{W^{*}}}\right)
\\-\frac{\left(\mu^{*}-1\right) W^{*}}{4 \mu^{*}} I_2\left(y^*;W^*,\mu^*\right)+D_{2}^{*} \end{array}\right\},
\end{equation}
where
\begin{equation}
\begin{aligned}
&I_2\left(y;W,\mu\right):=
\int_{-1}^y \ln \left(1+\frac{1}{\mu}e^{\frac{4x}{W}}\right) dx
\\=&\left\{\begin{array}{ll}
{-\frac{W}{4}\left[Li_2\left(-\frac{e^{\frac{4 y}{W}}}{\mu}
	\right)-Li_2\left(-\frac{e^{-\frac{4 }{W}}}{\mu}
	\right)\right],} & {
	\left\{\begin{array}{l}\frac{W}{4}\ln\mu\ge 1 \text { or }\\ -1\le y\le\frac{W}{4}\ln\mu<1,\end{array}\right\}} \\ 
{\left\{\begin{array}{l}\frac{2y^{2}}{W}-y\ln\mu+\frac{W}{8}\ln^2\mu+\frac{\pi^2 W}{24}\\
	+\frac{W}{4}
	\left[
	Li_2\left(-\mu e^{-\frac{4 y}{W}}
	\right)
	+Li_2\left(-\frac{e^{-\frac{4 }{W}}}{\mu}
	\right)
	\right]
	\end{array}\right\}
	,} & { -1<\frac{W}{4}\ln\mu\le y\le 1,}
\\ {\left\{\begin{array}{l}\frac{2}{W}\left(y^{2}-1\right)-\left(y+1\right)\ln\mu\\
	+\frac{W}{4}
	\left[
	Li_2\left(-\mu e^{-\frac{4 y}{W}}
	\right)-Li_2\left(-\mu e^{\frac{4 }{W}}
	\right)
	\right]
	\end{array}\right\}
	,} & { \frac{W}{4}\ln\mu\le - 1.}\end{array}\right.
\end{aligned}
\end{equation}

Substituting the boundary condition
$u^*(y^*=\pm 1)=0$ into Eq.~\eqref{uM2starCD}, we have
\begin{subequations}\label{C2D2}
	\begin{equation}
	C_{2}^{*}=\frac{2\left(1-\mu^{*}\right) W^{*}\cdot
		\left\{\begin{array}{l}
		\ln \left[1+\frac{1}{\mu^{*2}}+\frac{2}{\mu^*}\operatorname{cosh}\left(\frac{4}{W^{*}}\right)\right]
		-I_2\left(1;W^*,\mu^*\right)
		\end{array}\right\}}
	{8+(\mu^*-1) W^* \ln \left(\frac{\mu^*+e^{\frac{4}{W^*}}}{\mu^*+e^{-\frac{4}{W^*}}}\right)},
	\end{equation}
	\begin{equation}
	D_{2}^{*}=\frac{
		\left\{\begin{array}{l}
		16+6\left(\mu^*-1\right)W^*\ln\left(\frac{\mu^*+e^\frac{4}{W^*}}{\mu^*+e^{-\frac{4}{W^*}}}\right)
		\\
		-4\left(\mu^*-1\right)W^*I_2\left(1;W^*,\mu^*\right) 
		\\+\left(\mu^*-1\right)^2 W^{*2}\ln\left(1+\frac{e^{-\frac{4}{W^*}}}{\mu^*}\right)\cdot
		\left\{\begin{array}{l}
		-2\ln\left(1+\frac{e^\frac{4}{W^*}}{\mu^*}\right)
		\\+I_2\left(1;W^*,\mu^*\right)
		\end{array}\right\}
		\end{array}\right\}}
	{-4\mu^*\left[8+(\mu^*-1) W^* \ln \left(\frac{\mu^*+e^{\frac{4}{W^*}}}{\mu^*+e^{-\frac{4}{W^*}}}\right)\right]
	}.
	\end{equation}
\end{subequations}
The constant coefficient result Eq.~\eqref{C2D2} is substituted back to Eq.~\eqref{uM2starCD}, the final analytical expression of dimensionless velocity can be obtained, that is Eq.~\eqref{uM2star}.

For M3, there is
\begin{equation}
e^{\int f^{*} d y^{*}}=\left(\mu^{*}\right)^{-\frac{1}{2}\tanh\left(\frac{2 y^{*}}{W^{*}}\right)},
\end{equation}
then 
\begin{equation}\label{uM3starCD}
u^{*}_{M3}=-\frac{\mu^{*}+1}{\sqrt{\mu^{*}}} \int_{0}^{y^*}\left(\mu^{*}\right)^{\frac{1}{2}\tanh\left(\frac{2 Y^{*}}{W^{*}}\right)}  Y^{*} d Y^{*}
+C_3^*\int_{0}^{y^*}\left(\mu^{*}\right)^{\frac{1}{2}\tanh\left(\frac{2 Y^{*}}{W^{*}}\right)}   d Y^{*}+D_3^*.
\end{equation}

Substituting the boundary condition
$u^*(y^*=\pm 1)=0$ into Eq.~\eqref{uM3starCD}, then
\begin{subequations}\label{C3D3}
	\begin{equation}
	C_{3}^*=\frac{\mu^{*}+1}{\sqrt{\mu^{*}}}\cdot
	\frac{\int_{-1}^{1}\left(\mu^{*}\right)^{\frac{1}{2}\tanh\left(\frac{2 Y^{*}}{W^{*}}\right)}  Y^{*} d Y^{*}}{\int_{-1}^{1}\left(\mu^{*}\right)^{\frac{1}{2}\tanh\left(\frac{2 Y^{*}}{W^{*}}\right)}   d Y^{*}},
	\end{equation}
	\begin{equation}
	D_{3}^*=\frac{\mu^{*}+1}{\sqrt{\mu^{*}}}\cdot
	\frac{
		\left\{\begin{array}{l} 
		\int_{0}^{1}\left(\mu^{*}\right)^{\frac{1}{2}\tanh\left(\frac{2 Y^{*}}{W^{*}}\right)} Y^{*}  d Y^{*} \cdot
		\int_{-1}^{1}\left(\mu^{*}\right)^{\frac{1}{2}\tanh\left(\frac{2 Y^{*}}{W^{*}}\right)}   d Y^{*}
		\\-\int_{-1}^{1}\left(\mu^{*}\right)^{\frac{1}{2}\tanh\left(\frac{2 Y^{*}}{W^{*}}\right)} Y^{*}  d Y^{*} \cdot
		\int_{0}^{1}\left(\mu^{*}\right)^{\frac{1}{2}\tanh\left(\frac{2 Y^{*}}{W^{*}}\right)}   d Y^{*}
		\end{array}\right\}
	}
	{\int_{-1}^{1}\left(\mu^{*}\right)^{\frac{1}{2}\tanh\left(\frac{2 Y^{*}}{W^{*}}\right)}   d Y^{*}}.
	\end{equation}
\end{subequations}
In the end, the constant coefficient result Eq.~\eqref{C3D3} is substituted back to Eq.~\eqref{uM3starCD}, the final analytical expression of dimensionless velocity can be obtained, that is Eq.~\eqref{uM3star}.

Based on the general solution expressions, Eqs.~\eqref{uM1starCD}, \eqref{uM2starCD}, and~\eqref{uM3starCD}, we can observe that there is a basic physical commonality for these three viscosity models.
If $\mu^*=1$, then $W^*$ will disappear and the solution becomes simple.
This is the case of the same viscosity, as we discussed earlier. At this time, the interfacial thickness has no influence on the solution, and the flow velocity profile is consistent with the single-phase flow with the same parameters.
In general, $\mu^*\neq 1$, then $u^*$ is complicated and depends on $W^*$.

\section{Analytical solutions of three dynamic models in two special cases} \label{sec: App Analytical solutions}

In this appendix, we apply two special cases to further demonstrate our theoretical results in Subsection~\ref{subsec: Analytical solutions}.

\begin{enumerate}
	\item If the dynamic viscosity of the two phases are the same ($\mu^*=1$ or $\mu_A=\mu_B=\mu_0$),
	then the three analytical solutions (Eqs.~\eqref{uM1star}, \eqref{uM2star}, and~\eqref{uM3star}) can obtain the same result, {\it i.e.},
	\begin{subequations}\label{usamemu}
		\begin{equation}
		u^*\left( \mu^*=1\right) =1-y^{*2}
		\end{equation}
		or
		\begin{equation}
		u_x\left(\mu_A=\mu_B=\mu_0 \right) 
		=\frac{GH^2}{2\mu_0}\left[1-\left(\frac{y}{H}\right)^{2}\right].
		\end{equation}
	\end{subequations}
	This is the analytic solution of single-phase flow velocity, which is consistent with the previous results (For example, the discussions in Subsection~\ref{subsec: Governing equation}).
	
	\item When the interface thickness tends to $0$, the velocity results of all models are also the same, {\it i.e.},
	\begin{subequations}
		\begin{equation} 
		\lim _{W^{*} \rightarrow 0^{+}} u^{*}=\left\{\begin{aligned}
		&\frac{1-y^*}{2}\left[2+\left(\mu^*+1\right)y^*\right], & {0<y^{*}\le 1,} \\ 
		& 1, & {y^{*}=0,} \\ &\frac{1+y^*}{2\mu^*}\left[2\mu^*-\left(\mu^*+1\right)y^*\right], & {-1\le y^{*}<0.}\end{aligned}\right.
		\end{equation}
		or
		\begin{equation} 
		\lim _{W \rightarrow 0^{+}} u_x=\left\{\begin{aligned}
		& -\frac{GH^2}{2\mu_{A}}\left(\frac{y}{H}-1\right)\left(\frac{y}{H}+\frac{2\mu_A}{\mu_A+\mu_B}\right), & {0<y\le H,} \\ 
		& \frac{GH^2}{\mu_A+\mu_B}, & {y=0,} \\ 
		& -\frac{GH^2}{2\mu_{B}} \left(\frac{y}{H}+1\right)\left(\frac{y}{H}-\frac{2\mu_B}{\mu_A+\mu_B}\right), & {-H\le y<0.}
		\end{aligned}\right.
		\end{equation}
	\end{subequations}
	This result is consistent with the solution of SI model (Eqs.~\eqref{uSInonD} and~\eqref{uSI})! Notice that the velocity at the interface is also correct.
	This shows that, in theory, the velocity profiles obtained by these viscosity models can fully recover the results in the SI model as long as the interface thickness is sufficiently small, which is expected.
\end{enumerate}

\section{Viscouse stress of layered Poiseuille flow in the DI model}\label{sec: App Viscouse stress}
According to Eq.~\eqref{SIBCtau}, the viscous stress is an important physical quantity that determines the boundary conditions in the SI model.
Therefore, it is necessary to investigate the results of viscous stress.
The viscous stress is dimensionless as
\begin{equation}
\tau^*=\frac{\tau}{GH},
\end{equation}
then the analytical results of
\begin{equation}
\tau=\mu\frac{du_x}{dy}
\end{equation}
in different dynamic viscosity models are as follows.

\begin{equation}
\tau_{M1}^*=\frac{2\mu^*}{\left(\mu^*+1\right)\left[\left(\mu^*+1\right)+\left(\mu^*-1\right)\operatorname{tanh}\left(\frac{2y^*}{W^*}\right)\right]}\frac{du_{M1}^*}{dy^*},
\end{equation}
where
\begin{equation}
\begin{aligned}
\frac{du_{M1}^*}{dy^*}
=\left\{
\begin{array}{l}
-\left(\mu^{*}+1\right)y^*
+\frac{(\mu^{*2}-1) y^{*}}{\mu^*\left(1+e^{\frac{4y^*}{W^*}}\right)} 
\\+
\left\{
\begin{array}{l}
\frac{(\mu^*-1)W^*}{8\mu^*}
\left[(\mu^*+1) +(\mu^*-1) \operatorname{tanh}\left(\frac{2y^* }{W^*}\right)\right]
\\
\cdot\left[\ln \left(2+2\operatorname{cosh}\left(\frac{4}{W^{*}}\right)\right)+\frac{2}{W^*}-2I_1\left(1;W^{*}\right)
\right]
\end{array}
\right\}
\end{array}
\right\}
\end{aligned}.
\end{equation}

\begin{equation}
\tau_{M2}^*=\frac{\left[\left(\mu^*+1\right)-\left(\mu^*-1\right)\operatorname{tanh}\left(\frac{2y^*}{W^*}\right)\right]}{2\left(\mu^*+1\right)}\frac{du_{M2}^*}{dy^*},
\end{equation}
where
\begin{equation}
\begin{aligned}
\frac{du_{M2}^*}{dy^*}
=\frac{16y^*-2\left(\mu^{*}-1\right) W^{*}\cdot
	\left\{\begin{array}{l}
	\ln \left[1+\frac{1}{\mu^{*2}}+\frac{2}{\mu^*}\operatorname{cosh}\left(\frac{4}{W^{*}}\right)\right]
	\\-y^*\ln \left(\frac{\mu^*+e^{\frac{4}{W^*}}}{\mu^*+e^{-\frac{4}{W^*}}}\right)
	-I_2\left(1;W^*,\mu^*\right)
	\end{array}\right\}}
{
	\left[-1+\frac{\mu^*-1}{\mu^*+1}\operatorname{tanh}\left(\frac{2y^*}{W^*}\right)\right]\cdot
	\left[8+(\mu^*-1) W^* \ln \left(\frac{\mu^*+e^{\frac{4}{W^*}}}{\mu^*+e^{-\frac{4}{W^*}}}\right)\right]}.
\end{aligned}
\end{equation}

\begin{equation}
\tau_{M3}^*=\frac{1}{\mu^*+1}\cdot
\left(\mu^*\right)^{\frac{1}{2}\left[1-\operatorname{tanh}\left(\frac{2 y^*}{W^*}\right)\right]}\cdot
\frac{du_{M3}^*}{dy^*},
\end{equation}
where
\begin{equation}
\begin{aligned}
\frac{du_{M3}^*}{dy^*}
=\frac{\mu^{*}+1}{\sqrt{\mu^{*}}}\cdot
\frac{
	\left\{\begin{array}{l} 
	-\int_{-1}^{1}\left(\mu^{*}\right)^{\frac{1}{2}\tanh\left(\frac{2 Y^{*}}{W^{*}}\right)}   d Y^{*}\cdot
	\left(\mu^{*}\right)^{\frac{1}{2}\tanh\left(\frac{2 y^{*}}{W^{*}}\right)}  y^{*} 
	\\
	+\int_{-1}^{1}\left(\mu^{*}\right)^{\frac{1}{2}\tanh\left(\frac{2 Y^{*}}{W^{*}}\right)}  Y^{*} d Y^{*}\cdot
	\left(\mu^{*}\right)^{\frac{1}{2}\tanh\left(\frac{2 y^{*}}{W^{*}}\right)}   
	\end{array}\right\}
}
{\int_{-1}^{1}\left(\mu^{*}\right)^{\frac{1}{2}\tanh\left(\frac{2 Y^{*}}{W^{*}}\right)}   d Y^{*}}.
\end{aligned}
\end{equation}

Other related physical quantities, such as flow rate, average velocity, maximum velocity, can be calculated based on the expression of velocity.
But these quantities are not directly related to analytical solutions, so we will not discuss them here.

\section{The information at the interface of layered Poiseuille flow in the DI model}\label{sec: App Information at the interface}
In this appendix, we show the results of the velocity and viscous stress at the interface of layered Poiseuille flow. This key information can be used to detect the relationship between the DI model and the SI model.

The expressions of velocity field at the two-phase interface ($y^*=0$) are, respectively,

\begin{subequations}
	
	\begin{equation}
	u^*_{M1}(y^*=0)=
	\frac{\left(\mu^*+1\right)^2}{4\mu^*}
	-\left\{
	\begin{array}{l}
	\frac{(\mu^*-1)^2W^{*2}}{16\mu^*}
	\ln \left(\operatorname{cosh}\left(\frac{2 }{W^*}\right)\right)
	\\\cdot
	\left[\ln \left(2+2\operatorname{cosh}\left(\frac{4}{W^{*}}\right)\right)+\frac{2}{W^*}-2I_1\left(1;W^{*}\right)
	\right]
	\end{array}\right\},
	\end{equation}
	
	\begin{equation}
	u^*_{M2}(y^*=0)=\frac{
		\left\{\begin{array}{l}
		-16
		\\+2\left(\mu^*-1\right) W^{*}\cdot
		\left\{\begin{array}{l}
		-3\ln\left(1+\frac{e^{\frac{4}{W^*}}}{\mu^*}\right)
		+3\ln\left(1+\frac{e^{-\frac{4}{W^*}}}{\mu^*}\right)
		\\
		+2I_2\left(1;W^*,\mu^*\right)
		-4I_2\left(0;W^*,\mu^*\right)
		\end{array}\right\}
		\\
		+\left(\mu^*-1\right)^2 W^{*2}\cdot
		\left\{\begin{array}{l}
		-
		\ln\left(1+\frac{e^\frac{4}{W^*}}{\mu^*}\right)
		\ln\left(1+\frac{1}{\mu^*}\right)
		\\-
		\ln\left(1+\frac{e^{-\frac{4}{W^*}}}{\mu^*}\right)
		\ln\left(1+\frac{1}{\mu^*}\right)
		\\
		+2\ln\left(1+\frac{e^\frac{4}{W^*}}{\mu^*}\right)
		\ln\left(1+\frac{e^{-\frac{4}{W^*}}}{\mu^*}\right)
		\\
		+I_2\left(1;W^*,\mu^*\right)\ln\left(1+\frac{1}{\mu^*}\right)
		\\
		-I_2\left(0;W^*,\mu^*\right)\ln\left(1+\frac{e^{\frac{4}{W^*}}}{\mu^*}\right)
		\\
		+\left\{\begin{array}{l}
		I_2\left(0;W^*,\mu^*\right)
		\\-I_2\left(1;W^*,\mu^*\right)
		\end{array}\right\}
		\ln\left(1+\frac{e^{-\frac{4}{W^*}}}{\mu^*}\right)
		\end{array}\right\}
		\end{array}\right\}}
	{-\frac{4\mu^*}{\mu^*+1}\left[8+(\mu^*-1) W^* \ln \left(\frac{\mu^*+e^{\frac{4}{W^*}}}{\mu^*+e^{-\frac{4}{W^*}}}\right)\right]
	},
	\end{equation}
	
	\begin{equation}
	u_{M3}^*(y^*=0)=\frac{\mu^{*}+1}{\sqrt{\mu^{*}}}\cdot
	\frac{
		\left\{\begin{array}{l} 
		\int_{0}^{1}\left(\mu^{*}\right)^{\frac{1}{2}\tanh\left(\frac{2 Y^{*}}{W^{*}}\right)} Y^{*}  d Y^{*} \cdot
		\int_{-1}^{1}\left(\mu^{*}\right)^{\frac{1}{2}\tanh\left(\frac{2 Y^{*}}{W^{*}}\right)}   d Y^{*}
		\\-\int_{-1}^{1}\left(\mu^{*}\right)^{\frac{1}{2}\tanh\left(\frac{2 Y^{*}}{W^{*}}\right)} Y^{*}  d Y^{*} \cdot
		\int_{0}^{1}\left(\mu^{*}\right)^{\frac{1}{2}\tanh\left(\frac{2 Y^{*}}{W^{*}}\right)}   d Y^{*}
		\end{array}\right\}
	}
	{\int_{-1}^{1}\left(\mu^{*}\right)^{\frac{1}{2}\tanh\left(\frac{2 Y^{*}}{W^{*}}\right)}   d Y^{*}}.
	\end{equation}
	
\end{subequations}

When the interface thickness tends to $0$, the velocity results of all models are consistent, {\it i.e.},
\begin{subequations}
	\begin{equation}
	\lim _{W^* \rightarrow 0^{+}}u^*(y^*=0)
	=1
	\end{equation}
	or
	\begin{equation}
	\lim _{W \rightarrow 0^{+}}u_x(y=0)=\frac{GH^2}{\mu_{A}+\mu_{B}}.
	\end{equation}
\end{subequations}
These results are consistent with that of the velocity at the two-phase interface in the SI model (see Eq.~\eqref{uSI}).

Next, the results of the viscous stress at the interface are investigated.
\begin{subequations}
	
	
	\begin{equation}
	\tau_{M1}^*\left(y^*=0 \right) 
	=
	\frac{(\mu^*-1)W^*}{4(\mu^*+1)}\left[\ln \left(2+2\operatorname{cosh}\left(\frac{4}{W^{*}}\right)\right)+\frac{2}{W^*}-2I_1\left(1;W^{*}\right)
	\right],
	\end{equation}
	
	\begin{equation}
	\tau_{M2}^*\left(y^*=0 \right)=\frac{\left(\mu^{*}-1\right) W^{*}\cdot
		\left\{\begin{array}{l}
		\ln \left[1+\frac{1}{\mu^{*2}}+\frac{2}{\mu^*}\operatorname{cosh}\left(\frac{4}{W^{*}}\right)\right]
		-I_2\left(1;W^*,\mu^*\right)
		\end{array}\right\}}
	{8+(\mu^*-1) W^* \ln \left(\frac{\mu^*+e^{\frac{4}{W^*}}}{\mu^*+e^{-\frac{4}{W^*}}}\right)},
	\end{equation}
	
	\begin{equation}
	\tau_{M3}^*\left(y^*=0 \right)
	=
	\frac{\int_{-1}^{1}\left(\mu^{*}\right)^{\frac{1}{2}\tanh\left(\frac{2 Y^{*}}{W^{*}}\right)}  Y^{*} d Y^{*}}{\int_{-1}^{1}\left(\mu^{*}\right)^{\frac{1}{2}\tanh\left(\frac{2 Y^{*}}{W^{*}}\right)}   d Y^{*}}.
	\end{equation}
\end{subequations}

When the interface thickness tends to $0$, the viscous stress results of all models are consistent, {\it i.e.},
\begin{subequations}
	\begin{equation}
	\lim _{W^* \rightarrow 0^{+}}\tau^*\left(y^*=0 \right)
	=
	\frac{\mu^*-1}{2\left(\mu^*+1\right)}
	\end{equation}
	or
	\begin{equation}\label{tauW0y0}
	\lim _{W \rightarrow 0^{+}}\tau \left(y=0 \right)
	=
	\frac{GH}{2}
	\frac{\mu_{B}-\mu_{A}}{\mu_{B}+\mu_{A}}.
	\end{equation}
\end{subequations}
Substituting the analytical solution of velocity in SI model, Eq.~\eqref{uSI}, into the expression of viscous stress, Eq.~\eqref{SIBCtau}, it is found that the results are the same as Eq.~\eqref{tauW0y0}.

%

In summary, for all these three viscosity models, M1, M2, and M3, when the interfacial thickness tends to $0$, the velocity and viscous stress at the interface of the two phases are exactly the same as the results in the SI model. So these DI models are compatible with the SI model.

\section{Special properties of the second set of dynamic viscosity models}\label{sec: property of the second set}
In this appendix, three special properties of the second set of dynamic viscosity models are given, as follows.

\begin{enumerate}
	\item $\mu$ is even with respect to $\alpha_2$.
	
	Proof: 
	Since $\mu$ can be viewed as a function of $\alpha_2$,
	\begin{equation}\label{newmumodel2alpha}
	\mu\left(\alpha_2\right)=\frac{\mu_{A}+\mu_{B}}{2}+\frac{\mu_{A}-\mu_{B}}{2} \frac{\operatorname{tanh}\left[ \alpha_2\left(\frac{2\phi-\phi_{A}-\phi_{B}}{\phi_{A}-\phi_{B}}\right)\right] }{\operatorname{tanh}\alpha_2},
	\end{equation}
	we have
	\begin{equation}\label{newmumodel2alpha=}
	\mu\left(-\alpha_2\right)=\mu\left(\alpha_2\right),
	\end{equation}
	meaning $\mu$ is even with respect to $\alpha_2$.
	Furthermore, $\alpha_2$ cannot be $0$ because $\operatorname{tanh}\alpha_2$ is in the denominator.
	Therefore, we only need to consider the case of $\alpha_2>0$ in the discussion of this set of models.
	
	\item $\mu$ is odd with respect to the interface $\phi=(\phi_{A}+\phi_{B})/2$.
	
	Proof: 
	$\mu$ is a function of $\phi$,
	\begin{equation}\label{newmumodel2phi}
	\mu\left(\phi\right)=\frac{\mu_{A}+\mu_{B}}{2}+\frac{\mu_{A}-\mu_{B}}{2} \frac{\operatorname{tanh}\left[ \alpha_2\left(\frac{2\phi-\phi_{A}-\phi_{B}}{\phi_{A}-\phi_{B}}\right)\right] }{\operatorname{tanh}\alpha_2},
	\end{equation}
	then
	\begin{equation}\label{newmumodel2phi=}
	\mu\left(\frac{\phi_{A}+\phi_{B}}{2}+\delta\phi\right)
	+\mu\left(\frac{\phi_{A}+\phi_{B}}{2}-\delta\phi\right)
	=\mu_{A}+\mu_{B}.
	\end{equation}
	Since $\phi$ is odd with respect to the two-phase interface $(\phi_{A}+\phi_{B})/2$,
	$\mu$ is also an odd function with respect to the interface due to Eq.~\eqref{newmumodel2phi=}.
	This shows that the profiles of $\mu$ should be symmetric with respect to the two-phase interface.
	
	\item When $\alpha_2>0$ increases, $|d\mu/dy|$ increases at the interface.
	
	Proof: 
	$\mu$ can be also viewed as a function of $y$,
	\begin{equation}\label{newmumodel2phiy}
	\mu\left(\phi\left(y\right)\right)=\frac{\mu_{A}+\mu_{B}}{2}+\frac{\mu_{A}-\mu_{B}}{2} \frac{\operatorname{tanh}\left[ \alpha_2\left(\frac{2\phi\left(y\right)-\phi_{A}-\phi_{B}}{\phi_{A}-\phi_{B}}\right)\right] }{\operatorname{tanh}\alpha_2},
	\end{equation}
	where $\phi\left(y\right)$ is Eq.~\eqref{phi}.
	The two-phase interface is at $\phi=(\phi_{A}+\phi_{B})/2$ or $y=0$.
	We have
	\begin{equation}
	\frac{d \mu}{d \phi}=\frac{\alpha_2}{\tanh \alpha_2} \frac{\mu_A-\mu_B}{\phi_A-\phi_B} \operatorname{sech}^2\left(\alpha_2 \frac{2 \phi-\phi_A-\phi_B}{\phi_A-\phi_B}\right),
	\end{equation}
	then, 
	\begin{equation}
	\frac{d \mu}{d y}\left(y=0\right)=\frac{\alpha_2}{\tanh \alpha_2} \frac{\mu_A-\mu_B}{\phi_A-\phi_B} 
	\frac{d \phi}{d y}\left(y=0\right).
	\end{equation}
	The key is
	\begin{equation}
	K\left(\alpha_2\right)= \frac{\alpha_2}{\tanh \alpha_2} 
	\end{equation}
	since the other terms are constant with respect to $\alpha_2$.
	It is easy to prove that
	\begin{equation}
	\lim _{\alpha_2 \rightarrow 0^{+}}K\left(\alpha_2\right)
	=
	1
	\end{equation}
	and $K\left(\alpha_2\right)$ increases when $\alpha_2>0$ increases.
	Therefore, $|d\mu/dy|$ increases at the interface.
\end{enumerate}

\bibliography{apssamp}

\end{document}